\documentclass[twocolumn]{aastex63}

\newcommand{\cii}{[C{~\small{II}}]$_{158\, \mu\rm{m}}$}
\newcommand{\nii}{[N{~\small{II}}]$_{205\, \mu\rm{m}}$}
\newcommand{\oi}{[O{~\small{I}}]$_{146\, \mu\rm{m}}$}
\newcommand{\ci}{[C{~\small{I}}]$_{369\, \mu\rm{m}}$}
\newcommand{\kms}{\rm{km \,s}^{-1}}

\usepackage[normalem]{ulem}
\usepackage{multirow}

\received{August 16, 2021}
\revised{October 6, 2021}
\accepted{January 20, 2022}

\submitjournal{ApJ}

\shorttitle{A fine structure line study of the ISM of J1148+5251}
\shortauthors{Meyer et al.}

\graphicspath{{./}{figures/}}

\begin{document}

\title{Physical constraints on the extended interstellar medium of the $z=6.42$ quasar J1148+5251: \cii, \nii\ and \oi\ observations}

\correspondingauthor{Romain A. Meyer}
\email{meyer@mpia.de}

\author[0000-0001-5492-4522]{Romain A. Meyer}
\affiliation{Max Planck Institute for Astronomy, Königstuhl 17, D-69117, Heidelberg, Germany}
\author[0000-0003-4793-7880]{Fabian Walter}
\affiliation{Max Planck Institute for Astronomy, Königstuhl 17, D-69117, Heidelberg, Germany}
\author[0000-0003-0522-6941]{Claudia Cicone}
\affiliation{Institute of Theoretical Astrophysics, University of Oslo, P.O. Box 1029, Blindern, 0315 Oslo, Norway}
\author[0000-0003-2027-8221]{Pierre Cox}
\affiliation{Sorbonne Universit\'e, UPMC Universit\'e Paris 6 and CNRS, UMR 7095, Institut d’Astrophysique de Paris, 98bis Boulevard Arago,75014 Paris, France}
\author[0000-0002-2662-8803]{Roberto Decarli}
\affiliation{INAF – Osservatorio di Astrofisica e Scienza dello Spazio di Bologna, via Gobetti 93/3, I-40129, Bologna, Italy}
\author[0000-0002-7176-4046]{Roberto Neri}
\affiliation{Institut de Radioastronomie Millimétrique (IRAM), 300 Rue de la Piscine, 38400 Saint-Martin-d’H\'eres, France}
\author[0000-0001-8695-825X]{Mladen Novak}
\affiliation{Max Planck Institute for Astronomy, Königstuhl 17, D-69117, Heidelberg, Germany}
\author[0000-0001-9815-4953]{Antonio Pensabene}
\affiliation{INAF – Osservatorio di Astrofisica e Scienza dello Spazio di Bologna, via Gobetti 93/3, I-40129, Bologna, Italy}
\author[0000-0001-9585-1462]{Dominik Riechers}
\affiliation{I. Physikalisches Institut, Universit\"at zu K\"oln, Z\"ulpicher Strasse 77, D-50937 K\"oln, Germany}
\author[0000-0003-4678-3939]{Axel Weiss}
\affiliation{Max-Planck-Institut f\"ur Radioastronomie, Auf dem H\"ugel 71, 53121 Bonn, Germany}

\begin{abstract}
We report new Northern Extended Millimeter Array (NOEMA) observations of the \cii, \nii\ and \oi\ atomic fine structure lines and dust continuum emission of J1148+5251, a $z=6.42$ quasar, that probe the physical properties of its interstellar medium (ISM). 
The radially-averaged \cii\ and dust continuum emission have similar extensions (up to $\theta = 2.51^{+0.46}_{-0.25}\ \rm{arcsec}$, corresponding to $r= 9.8^{+3.3}_{-2.1}\ \rm{kpc}$ accounting for beam-convolution), confirming that J1148+5251 is the quasar with the largest \cii\ --emitting has reservoir known at these epochs.Moreover, if the \cii\ emission is examined only along its NE-SW axis, a significant excess ($>5.8\sigma$) of \cii\ emission (with respect to the dust) is detected.
The new wide--bandwidth observations enable us to accurately constrain the continuum emission, and do not statistically require the presence of broad \cii\ line wings that were reported in previous studies.
We also report the first detection of the \oi\  and (tentatively) \nii\ emission lines in J1148+5251. Using Fine Structure Lines (FSL) ratios of the  \cii, \nii, \oi\ and previously measured \ci\ emission lines, we show that J1148+5251 has similar ISM conditions compared to lower--redshift (ultra)-luminous infrared galaxies. CLOUDY modelling of the FSL ratios exclude X--ray dominated regions (XDR) and favours photodissociation regions (PDR) as the origin of the FSL emission. We find that a high radiation field ($10^{3.5-4.5}\,G_0$), high gas density ($n  \simeq 10^{3.5-4.5}\, \rm{cm}^{-3}$) and HI column density of $10^{23} \,\rm{cm^{-2}}$ reproduce the observed FSL ratios well.
\end{abstract}

\section{Introduction} \label{sec:intro}
Luminous quasar activity is a key process of galaxy evolution. Indeed, massive outflows driven by the radiation pressure generated by the accretion of gas onto the central supermassive black hole (SMBH), or so--called “Active Galactic Nuclei (AGN) feedback", are invoked in most models of galaxy formation to clear massive galaxies of their gas and quench star formation \citep[e.g.,][]{Silk1998,DiMatteo2005,Springel2005,King2010,Costa2014,Ishibashi2015,Richardson2016,Negri2017,Oppenheimer2020,Koudmani2021}. Luminous quasars are most interesting at high--redshift in particular, when they probe the early phase of co--evolution between the first galaxies and their central black hole. The advent of large optical and infrared surveys has enabled the discovery of quasars up to $z\sim 7.5$ \citep{Banados2018,Yang2020,Wang2021}, with several hundreds at $z>6$ \citep[see][for an up--to--date list]{Bosman2020_qsolist}. These early quasars harbour SMBHs with $M_\bullet \gtrsim 10^8\, \rm{M}_\odot$ and accrete gas at or near the Eddington limit for most of their life, challenging models of SMBH formation and growth \citep[e.g.,][]{DeRosa2014,Banados2018,Mazzucchelli2017,Wang2021}. Because the bright quasar light outshines that of the host in the optical and near--infrared, the galaxies hosting early luminous quasars have remained relatively mysterious until the advent of modern (sub--)millimeter observatories.

Numerous observations of $z>6$ quasars targeting the bright Far--Infrared (FIR) \cii\ emission line have revealed their host galaxies to be infrared luminous, dusty and actively forming stars with estimated rates of $10^{2}-10^{3} M_\odot\ \rm{yr}^{-1}$ \citep[e.g.,][]{Walter2003,Walter2009, Maiolino2005,Maiolino2012, Banados2015a, Decarli2018,Venemans2018,Venemans2020,Novak2019,Novak2020,Wang2013,RWang2019, Yang2020}. \cii\ kinematics also show that most quasar hosts are massive galaxies ($M_*\sim10^{10}\,M_\odot$) displaying a variety of morphologies such as stable disks, bulge--dominated galaxies and mergers with nearby companions \citep{Shao2017,Shao2019,Wang2013,RWang2019,Decarli2019,Decarli2019a,Neeleman2019a, Neeleman2021}. The reports of broad \cii
line wings in the z=6.42 quasar J1148+5251 \citep[][]{Maiolino2012,Cicone2015} spurred the use of the \cii\ emission line to identify quasar outflow signatures in the early Universe. Such features have however remained rare, and stacking analyses have led to contradictory results \citep[][]{Bischetti2019,Novak2020}. Recently, \citet{Izumi2021,Izumi2021a} reported broad \cii\ line wings in two low-luminosity quasars at $z=6.72$ and $z=7.07$.

Two decades of \cii\ and CO studies have shown that early luminous quasars provide an unparalleled observational window into the physics of the earliest (and most massive) galaxies in the Universe. However, multi--line studies using lines other than \cii\ and CO transitions have been much rarer until now. Since different fine structure lines (FSL) trace different gas densities and excitation levels, only in combination can they probe the ionized and neutral atomic gas phases, and the excitation source(s) of the gas \citep[e.g.,][for a review]{Carilli2013}. Besides \cii, potential atomic fine structure lines of interest include [N{~\small{II}}]$_{122\, \mu\rm{m}}$, [N{~\small{II}}]$_{205\, \mu\rm{m}}$, [O{~\small{I}}]$_{63\, \mu\rm{m}}$, [O{~\small{I}}]$_{145\, \mu\rm{m}}$ [O{~\small{III}}]$_{88\, \mu\rm{m}}$ and [C{~\small{I}}]$_{369\, \mu\rm{m}}$, all accessible at $z\sim 6$ with (sub--)millimeter arrays such as the Atacama Large Millimeter Array (ALMA) or the Northern Extended Millimeter Array (NOEMA). Moreover, these lines have been observed with \textit{Herschel} in large samples of local (ultra)-luminous infrared galaxies -(U)LIRGs, which can which can be readily compared to $z\sim6$ quasars hosts \citep[e.g.,][]{Diaz-Santos2017,Herrera-Camus2018}. 

Detections of FSL other than \cii\ in $z>6$ quasars are still relatively recent \citep[][]{Walter2018,Hashimoto2019, Novak2019,Li2020b}, and a complex picture is emerging from these first results. Emission lines probing the neutral phase ([O{~\small{I}}]$_{145\, \mu\rm{m}}$ and [C{~\small{I}}]$_{369\, \mu\rm{m}}$) show good agreement between the line ratios and line--to--far--infrared (FIR) ratios of distant quasars and local (U)LIRGs \citep{Novak2019,Li2020b}. Whilst \citet[][]{Novak2019} and \citet[][]{Li2020b} report potentially high [O{~\small{III}}]$_{88\, \mu\rm{m}}$ to \cii\ ratios ($\sim 2$) in J1342+0928 and J2310+1855, respectively, this is not the case for the quasar J2100--1715 and its companion galaxy \citep[][]{Walter2018} which have ratios similar to the average of the local population of LIRGS and AGNs \citep[$\sim0.1-1$;][]{Herrera-Camus2018}. These first results are however difficult to interpret since the origin of FSL is not always clearly determined and can be linked to different phases (as is the case for the \cii\ emission line). Clearly, more FIR multi--line studies of $z>6$ quasars are needed to understand the ISM of their host galaxies. 

SDSS J1148+5251 is one of the earliest high--redshift quasars discovered in the SDSS survey \citep[][z=6.4189]{Fan2003}, and harbours a $3\times10^{9}M_\odot$ SMBH \citep[][]{Willott2003}. Being the redshift record--holder for many years after its discovery, it was extensively observed  with the Very Large Array and the IRAM Plateau de Bure Interferometer (PdBI) and was the first object detected in CO and \cii\ at $z>5$ \citep{Walter2003,Walter2004,Bertoldi2003,Bertoldi2003a,Maiolino2005,Riechers2009,Walter2009,Walter2009a}. These pioneering studies probed the host galaxy Star Formation Rate (SFR), dust and ISM properties. Additionally, \citet{Maiolino2012} and \citet{Cicone2015} reported the presence of a broad \cii\ emission ($\sigma_v = 900\, \kms$) component in the PdBI data, suggesting the presence of an outflow as well as spatially extended \cii\ emission (up to $r\sim30\,$ kpc). In this paper, we return to J1148+5251 with a new set of NOEMA observations, capitalizing on larger bandwidths and more antennas, thus improving on the image fidelity as compared to earlier PdBI observations. The new observations targeted atomic fine structure emission lines (\oi, \nii) and other molecular (CO, H$_2$O) rotational transitions. The aim of the observations was to dissect the ISM phases without relying on assumptions about the origin of \cii\ which can come from both the ionized and neutral phases. Indeed, \oi\ traces exclusively the neutral phase/ photo--dissociated regions (PDRs), whereas \nii\ traces the ionized/H{~\small II} regions. This set of observations is complemented with earlier \ci\ data \citep[][]{Riechers2009} that trace the neutral/molecular gas. Thanks to the wide spectral coverage of the new NOEMA correlator \textit{PolyFix} and the upgraded NOEMA array, these observations achieved a high fidelity that resulted in deep \cii\ observations and tight constraints on the underlying dust continuum. 

The structure of this paper is as follows. We present in Section \ref{sec:obs} and \ref{sec:results} the new observations of the \cii, \nii\  and \oi\ emission lines in J1148+5251 as well as the FIR continuum observed between $200$ and $280$ GHz. We focus on the \cii\ emission line to re-assess the evidence for a broad velocity component and investigate its spatial extension in Section \ref{sec:cii}. In Section \ref{sec:ism}, we derive ISM properties from the strength of the atomic fine-structure emission lines  observed, before concluding our study in Section \ref{sec:conclusion}. Throughout this paper, we assume a concordance cosmology with $H_0=70\, \rm{km\, s}^{-1} \rm{Mpc}^{-1}$, $\Omega_M = 0.3, \Omega_\Lambda=0.7$. At the redshift of the target ($z=6.42$), $1"$ corresponds to $5.62$ proper kpc. 

\section{Observations and data reduction}
\label{sec:obs}

We have observed the z=6.4189 quasar J1148+5251 using NOEMA (Project ID: w17ex001/w17ex001, PI: F. Walter). The pointing and phase center of our observations were chosen to correspond to the quasar position in the optical SDSS imaging (RA = 11:48:16.64, DEC = +52:51:50.32). The observations included two spectral setups taking advantage of the new \textit{PolyFix} correlator covering simultaneously two $7.744$\,GHz--wide sidebands \footnote{\url{https://www.iram-institute.org/EN/content-page-96-7-56-96-0-0.html}}. The first spectral setup was centered at $267$ GHz such that the lower sideband ($255-263$ GHz) covers the redshifted \cii\ emission with one sideband whilst the upper sideband ($271-279$ GHz) covers the \oi\ emission. The second setup was centered at $208$ GHz to cover the \nii\ in the lower sideband ($196-204$ GHz). The setups also covered two high--J CO (14-13 and 13-12) and H$_2$O rotational transitions ($5_{23}-4_{32}$ and $3_{22}- 3_{13}$), which will be discussed in future works.
The observations were executed between December 2017 and May 2018. The \cii\ and \oi\ setup was mostly observed in configuration 9D except for two tracks using 8 antennas (with baselines ranging from $24$m to $176$m), for a total observing time of $18.7$h. Data was (remotely) reduced at IRAM Grenoble using the \emph{CLIC} package within the GILDAS framework (jan2021a version)\footnote{\url{https://www.iram.fr/~gildas/dist/}}. We reach a rms noise of $0.64(0.88)$ mJy beam$^{-1}$ in $50\, \kms$ channels, and the synthesized beam FWHM size is $1.97"\times1.59"$($1.83"\times1.51"$) for the \cii(\oi) line observations. The \nii\ and CO lines were observed with 8 antennas for a total of $15.7$h with baselines ranging from $24$m to $176$m. For the \nii\ line the noise rms is $0.78$ mJy beam$^{-1}$ in $50\, \kms$ channels and the synthesized beam size $1.79"\times1.51"$. For the continuum, the synthesized beam size achieved is $1.78"\times1.51"$,  $1.65"\times1.46"$, $1.91"\times1.71"$ and $1.88"\times1.52"$ at $200,212,259$ and $272$ GHz. 

Imaging and cleaning was performed using the latest version of MAPPING/GILDAS (jan2021a). The dirty maps were obtained from the visibilities without tapering and using natural weighting. The data were not primary beam corrected \footnote{As the source is located at the phase center, even at the edge of the aperture considered in the analysis ($r=3"$) the correction is minimal ($1.06$) and does not impact the results.}. Cleaning was performed down to $2\sigma$ (where $\sigma$ is the rms noise in the dirty map) using a circular clean region of radius $r=5"$. The reason for choosing such a wide radius are the earlier reports of extended \cii\ emission \citep{Maiolino2012,Cicone2015}. An additional clean region with radius $r=2"$ was added on the NW source reported by \citet{Leipski2010} and \citet{Cicone2015} which is also detected in the NOEMA data \footnote{We found no emission lines for this source in any of the spectral setups used in this work.}. 
The final products were created using the following procedure. Firstly, data cubes with $50\, \kms$ channels were produced to search for significant emission lines. The \cii, \nii, \oi, H$_2$O ($5_{23}-4_{32}$ and $3_{22}- 3_{13}$, $\nu_{\rm{rest}}=258.7$ GHz), CO(14-13) and CO(13-12) emission lines were fitted with a single Gaussian profile in order to estimate their FWHM. Continuum maps were created using all channels at least $1.25\times$FWHM away from each emission line. The continuum, determined from the line--free channels using an order 1 interpolation (GILDAS \emph{UV\_BASELINE} routine) over the $\sim 7.6$ GHz sidebands, was subtracted in the \textit{uv} plane to create continuum--subtracted cubes. 

In order to determine the significance of emission lines, velocity--integrated emission line maps (“line maps") were created by integrating channels over $1.2$ times the FWHM of the \cii\ line (i.e., $482\,\kms$) at the redshifted frequency of the line. We use the \cii\ redshift ($z=6.4189$) to determine redshifted frequency of all lines. Such maps, assuming the line is Gaussian, contain by definition $84\%$ of the total flux\footnote{We have checked that this correction holds for the [{C~\scriptsize II}]$_{158\mu \rm{m}}$ emission where the flux recovered in a $(-1500,+1500)\,\kms$ velocity-integrated map is $S_{\rm{line}}\Delta v = 10.5\pm0.92 \,\rm{Jy}\,\kms$ and the $1/0.84$--corrected flux in the nominal $1.2\times$FWHM$_{\rm{[C~II]}}=482\, \kms$ channel is $S_{\rm{line}}\Delta v = 10.4\pm0.5 \,\rm{Jy}\,\kms$, where $S_{\rm{line}}$ is integrated in a circular aperture with $r=3"$. Indeed, the choice of a $1.2\times$FWHM--wide channel maximises the SNR without flux losses.}  \citep[see for a short derivation appendix A of][]{Novak2020} . All total line fluxes measured from the line maps and reported in this paper are accounting for this effect. Additionally, all continuum and line fluxes in this paper were computed using the residual scaling method \citep[e.g.,][]{Jorsater1995,Walter1999,Walter2008, Novak2019}. 

In order to determine the aperture needed to recover most of the flux of the emission lines and the dust continuum, a curve of growth approach was adopted. We show in Appendix \ref{sec:curve_growth} that all line and continuum fluxes reach a maximum or plateau at an aperture radius $r=3"$, which corresponds to $16.9$ kpc at $z=6.42$. A nominal aperture radius of $3"$ is thus adopted throughout the paper. The line subtraction procedure described above was repeated using the $r=3"$ aperture to obtain final datacubes. Additionally, we have investigated the use of Multiscale cleaning in appendix \ref{sec:cleaning}. We conclude that the \cii\ emission and the $259-272$ GHz continuum are better recovered using Multiscale cleaning, and we therefore use Multiscale cleaning with Gildas/MAPPING for these data throughout the paper.

\section{Results}
\label{sec:results}
\subsection{Dust continuum emission}
\label{sec:dust_cont}
\begin{figure*}
    \centering
    \includegraphics[height = 0.23\textheight]{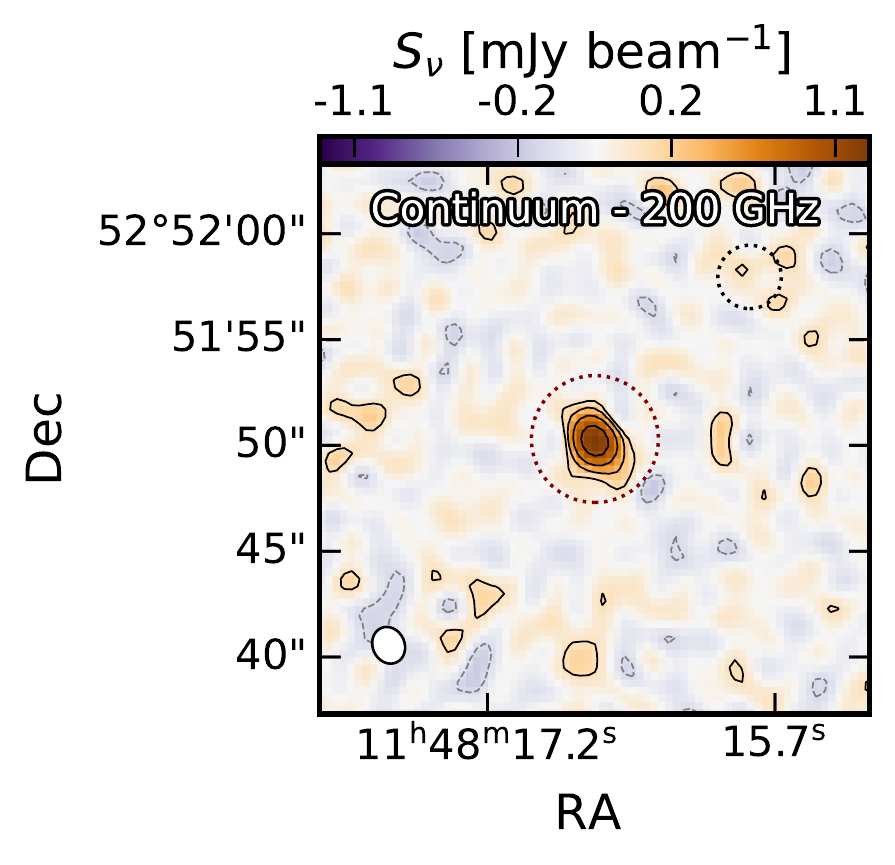} 
    \includegraphics[height= 0.23\textheight,trim={3.2cm 0 0 0},clip]{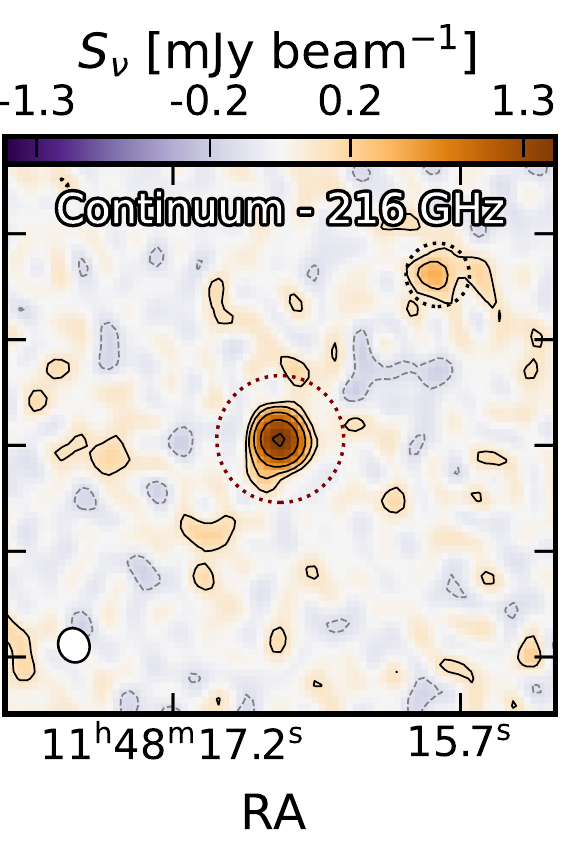}
    \includegraphics[height = 0.23\textheight,trim={3.2cm 0 0 0},clip]{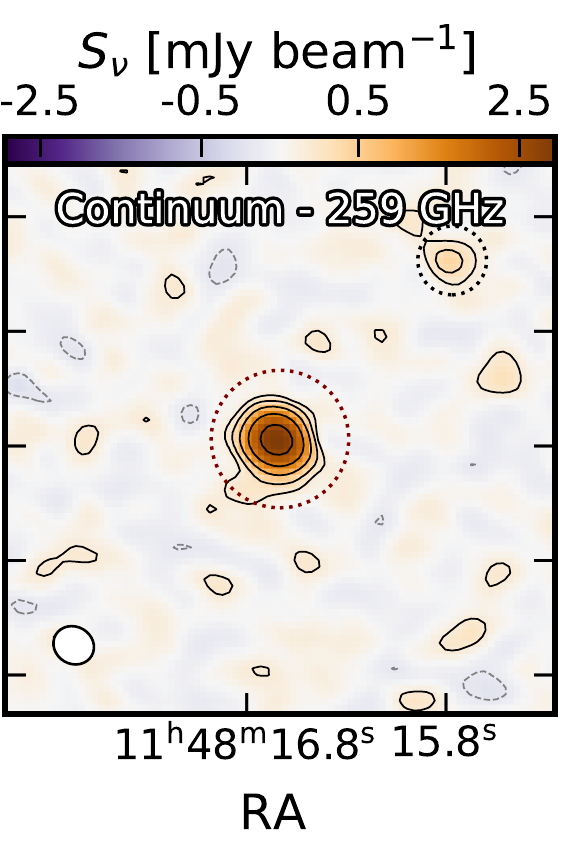}
    \includegraphics[height = 0.23\textheight,trim={3.2cm 0 0 0},clip]{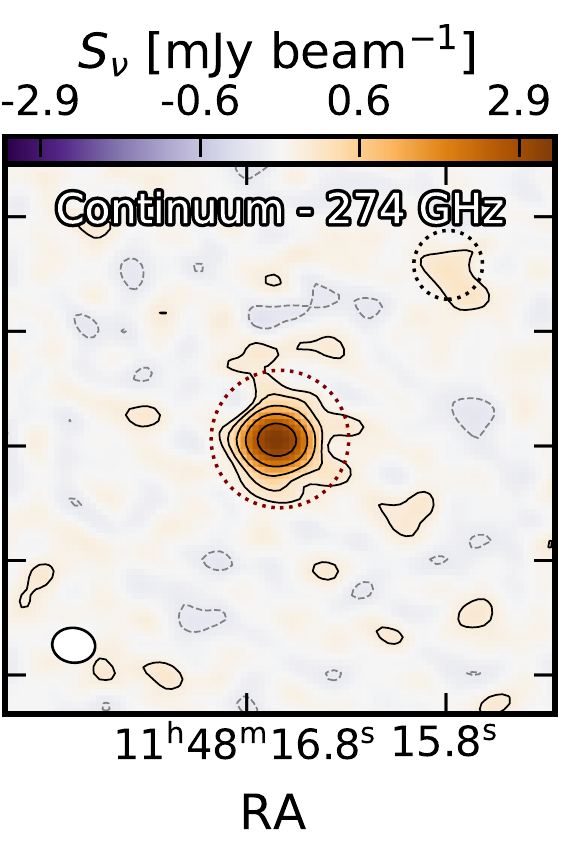}
    \caption{Dust continuum emission maps integrated over the $\sim 7.68$ GHz of the \textit{PolyFix} correlator effective bandwidth for each sideband with central frequency $200,216,259$ and $274$ GHz. The dotted circle centred on J1148+5251 indicates the aperture radius of $3"$ adopted throughout this paper for flux density estimates. An additional source, reported previously by \citet[][]{Leipski2010} and \citet{Cicone2015} is visible in the three higher--frequency bands at $\sim 10"$ to the north--west (black dotted circle). The contours are logarithmic $(-4,-2,2,4,8,16,32)\sigma$ (rms). The colour scaling is log--linear, the threshold being at $3\sigma$. }
    \label{fig:cont_maps}
\end{figure*}

We first present the FIR continuum maps in Figure \ref{fig:cont_maps}. The FIR continuum is clearly detected in all four sidebands. The measured continuum flux densities are tabulated in Table \ref{tab:continuum}. Due to the upgraded bandwidth of the NOEMA \textit{PolyFix} correlator, these continuum measurements have higher sensitivity than previous observations at $\sim 260$ GHz \citep[][]{Walter2009,Maiolino2012,Cicone2015}. The new continuum flux density at $259$ GHz ($4.64\pm0.26 \, \rm{mJy}$) is in good agreement\footnote{We have checked that there is no continuum offset between the two side--bands by imaging the calibrators (1150+497 and 1216+487) for every track and the stacked data.} with the earlier PdBI measurement from \citet[][]{Walter2009} and the 1.2 mm continuum measurement \citep[$5.0\pm 0.6\, \rm{mJy}$,][]{Bertoldi2003a}. We combine our new continuum measurements with previous literature results at different frequencies \citep[][]{Bertoldi2003,Walter2003,Robson2004,Riechers2009,Leipski2010,Gallerani2014} to fit the FIR spectral energy distribution. 

\begin{table}
   \centering
    \begin{tabular}{c|cccc}
    \hline \hline
$\nu_{\rm{obs}}$ [GHz] & $S_\nu$  [mJy] & rms [mJy] & Beam size   \\ \hline
200 & $1.78\pm0.23$ & $0.06$ & $1.78"\times1.51"$\\ 
216 &  $2.33\pm0.21$ & $0.05$ &  $1.65"\times1.46"$\\ 
259 &  $4.64\pm0.26$ & $0.08$ &  $1.91"\times1.71"$ \\ 
274 &  $5.85\pm0.29$ & $0.08$ & $1.88"\times1.52"$\\ 
\hline
    \end{tabular}
\caption{Continuum measurements at $200-274$ GHz extracted from the continuum images (Figure \ref{fig:cont_maps}, effective bandwidth $\simeq 7.68$ GHz (masking the edges and central baseband gap) and integrated in a circular aperture with radius $r=3"$.  }
    \label{tab:continuum}
\end{table}

\begin{figure}
    \hspace{-0.5cm}
    \includegraphics[width=0.5\textwidth]{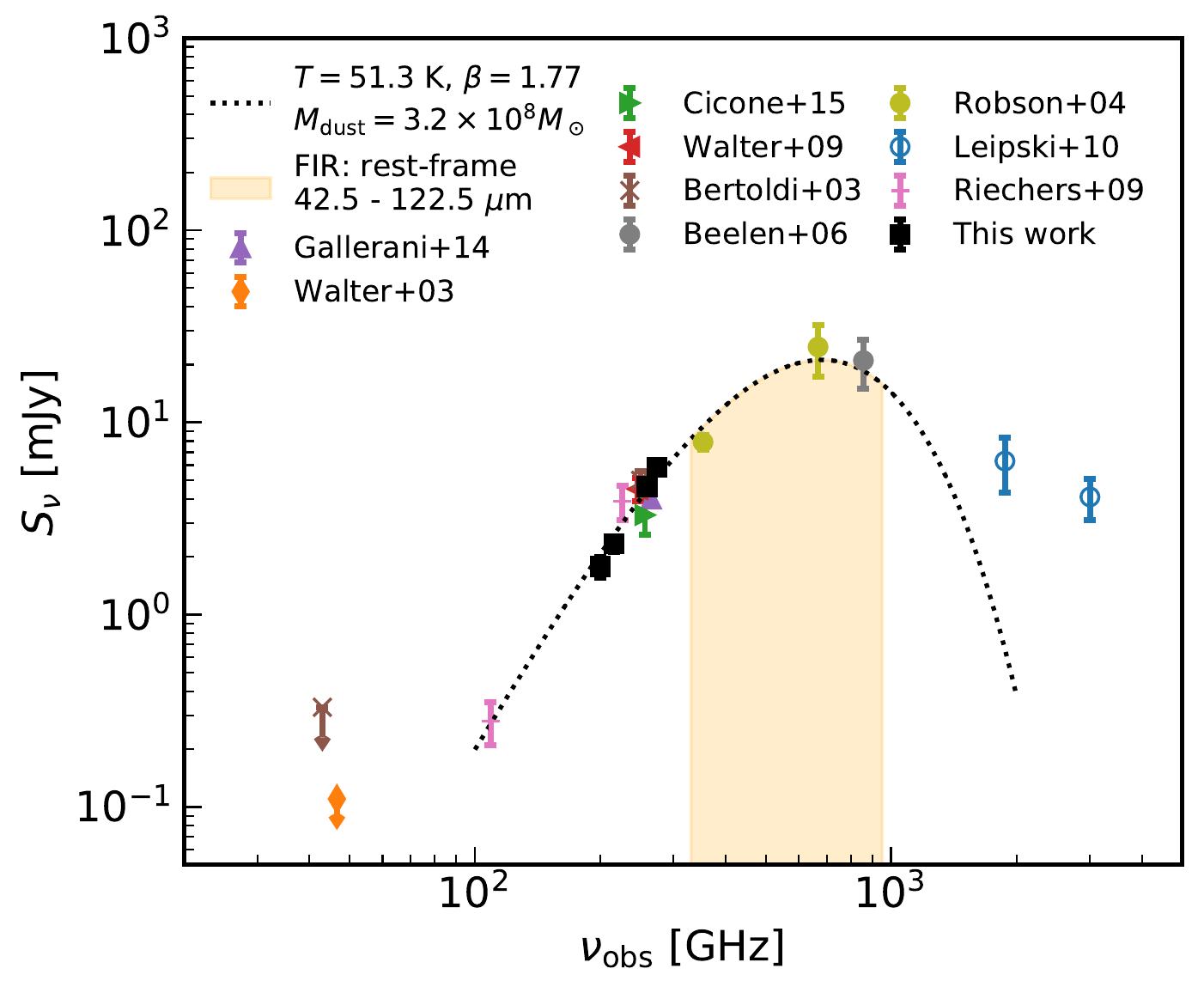}
        \caption{Continuum measurements from the literature and this work with the best--fit modified black body dust emission model. The effect of the CMB is accounted for as prescribed by \citet[][]{DaCunha2013} and the data points at $\nu>10^3$ GHz are not used for the fit. The FIR luminosity, integrated over the shaded orange area, is not significantly affected by this choice. 
        }
    \label{fig:dust_sed}
\end{figure}

Assuming optically thin dust emission at $\lambda>40\mu\rm{m}$ \citep[e.g.,][]{Beelen2006}, we use a modified black body model for the dust emission and correct both for contrast and CMB heating as prescribed by \citet[][]{DaCunha2013}. The dust mass is derived assuming an opacity $\kappa_{\nu_{rest}} = \kappa_{\nu_0}(\nu_{rest}/\nu_0)^\beta$ with $\nu_0=c/(125\mu\rm{m})$ and $\kappa_0=2.64 \rm{m}^{2}\rm{kg}^{-1}$ following \citet{Dunne2003} with $\beta$ being the dust spectral emissivity index. Our purpose is primarily to measure the FIR luminosity to constrain the SFR in J1148+5251. Therefore, we omit data points at $\nu_{\rm{rest}} > 1000\, \rm{GHz}\ (\lambda_{\rm{rest}} \lesssim 125 \,\mu\rm{m})$ where contamination by the quasar non--thermal and torus emission becomes significant \citep[e.g.,][]{Leipski2010, Leipski2014}. The dust SED model uses three free parameters (total dust mass $M_d$, dust emissivity index $\beta$, dust temperature $T_d$) and is fitted using MCMC with the \emph{emcee} package \citep[][]{Foreman-Mackey2013} The resulting best--fit and observational constraints are shown in Figure \ref{fig:dust_sed} and the posterior probability distribution of the dust SED parameters is displayed in Appendix \ref{app:dust_sed_parameters}. The median dust mass is $3.2\times10^{8} M_\odot$, the median dust SED index is $\beta=1.77^{+0.30}_{-0.26}$, and the median dust temperature is moderately high ($T_d=51.3$ K), in agreement with earlier studies \citep[e.g.,][]{Beelen2006,Leipski2010,Cicone2015}, which is not surprising considering that most of the constraining power comes from observations at lower and higher frequencies than those reported in this work.

We integrate the modified black body to derive the total infrared (IR, $8-1000\,\mu$m) and far--infrared \citep[FIR, $42.5-122.5 \mu$m, e.g.,][]{Helou1985} luminosities. The total infrared luminosity is $L_{\rm{IR}} = (20.9\pm 6.8)\times 10^{12} L_\odot$ and the far--infrared luminosity $L_{\rm{FIR}}=(13.4\pm 2.4)\times 10^{12} L_\odot$, in agreement with earlier studies of J1148+5251.  Assuming that dust heating is dominated by young stars, the infrared luminosity can be converted to a SFR using the \citet[][]{Kennicutt1998} and \citet{Kennicutt2012} conversions, giving SFR$=(1830\pm 595 )-(2090\pm 680) M_\odot\,\rm{yr}^{-1}$, respectively. This is in agreement with earlier studies which found that J1148+5251 is in an intense starburst phase \citep[][]{Maiolino2005,Walter2009,Maiolino2012,Cicone2015}.

\subsection{Fine structure line detections}

\begin{figure*}
    \centering
    \includegraphics[height=0.22\textheight,trim={0 0 0.5cm 0},clip]{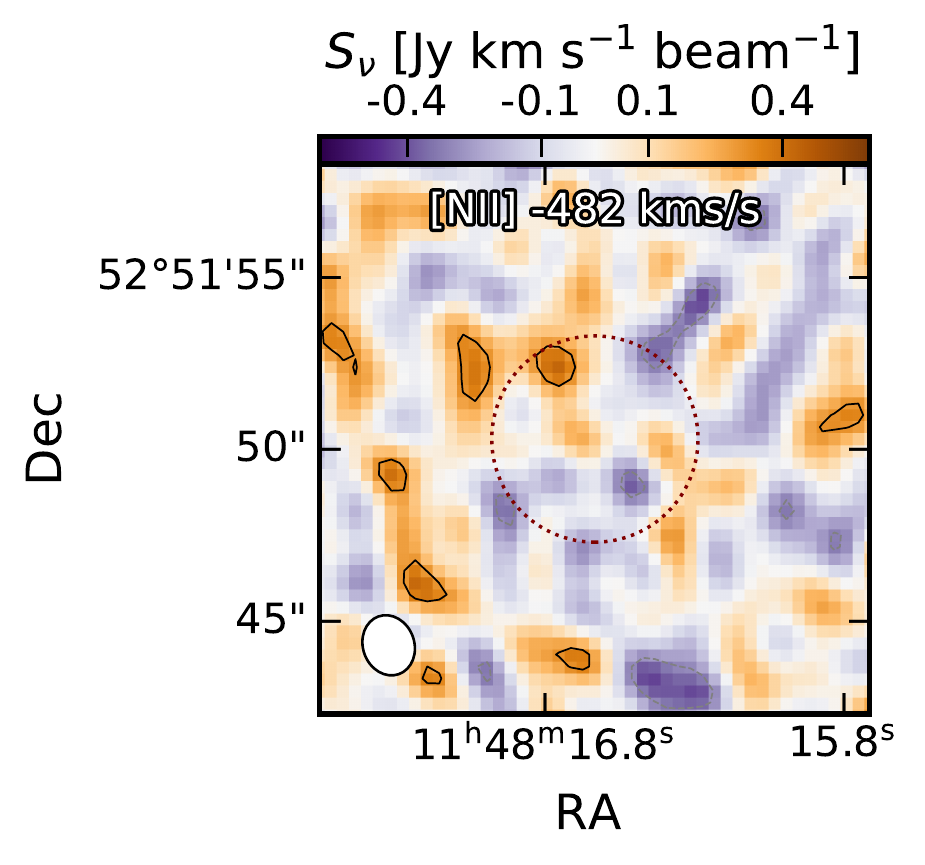}
    \includegraphics[height=0.22\textheight,trim={3.25cm 0 0.5cm 0},clip]{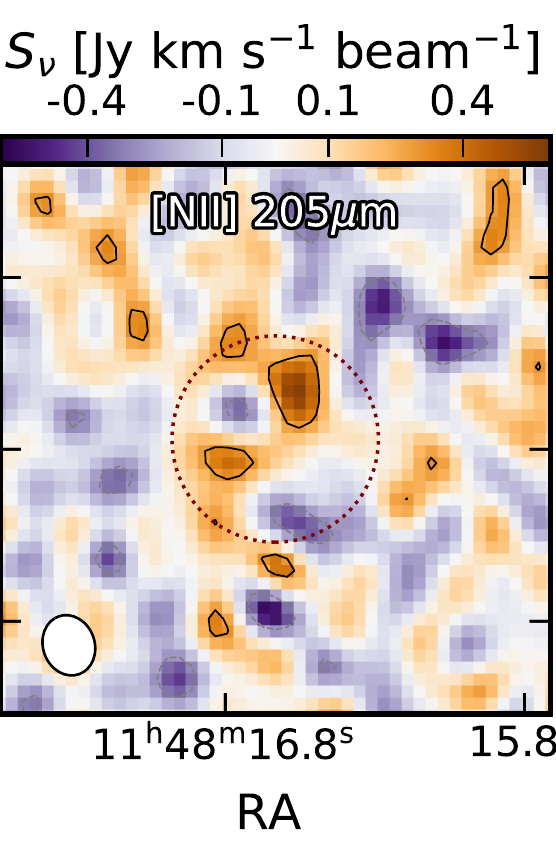}
    \includegraphics[height=0.22\textheight,trim={3.25cm 0 0.5cm 0},clip]{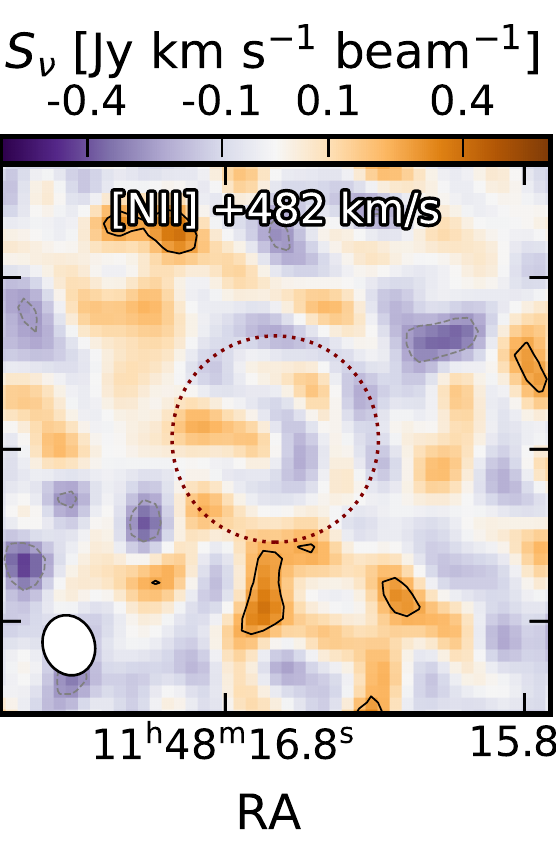} \\
    
    \includegraphics[height=0.22\textheight,trim={0 0 0.5cm 0},clip]{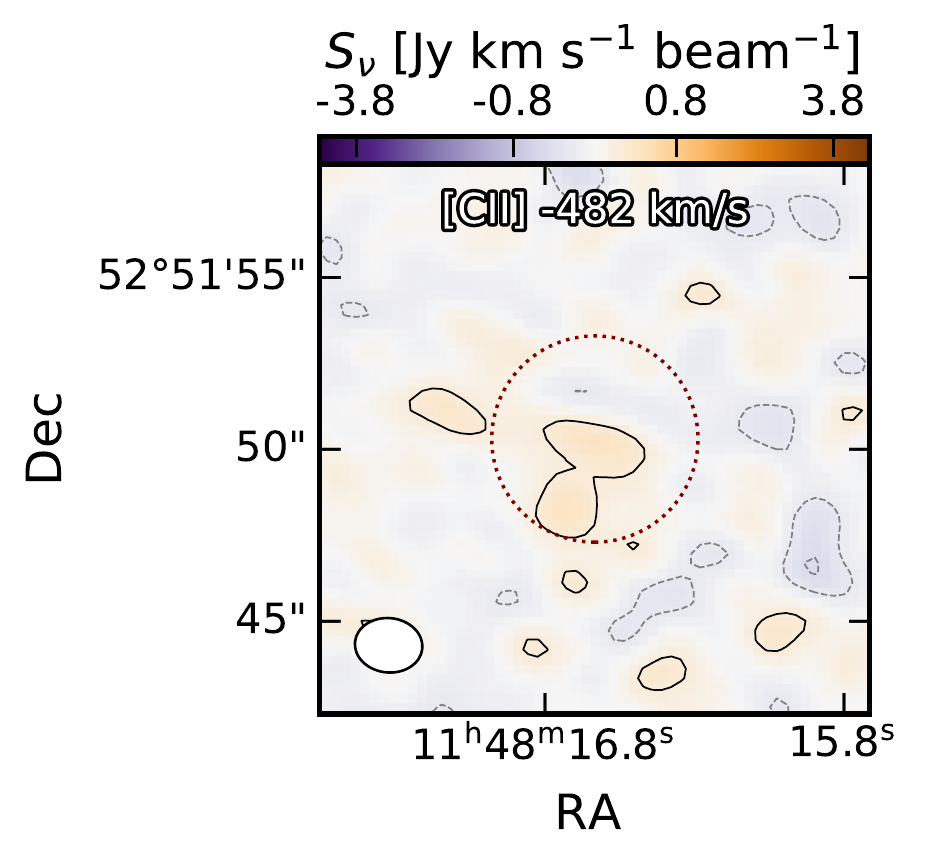}
    \includegraphics[height=0.22\textheight,trim={3.25cm 0 0.5cm 0},clip]{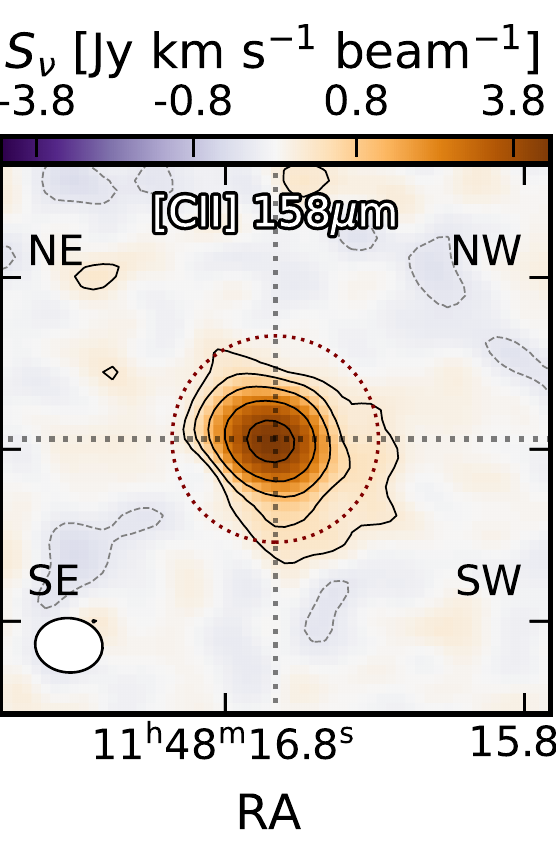}
    \includegraphics[height=0.22\textheight,trim={3.25cm 0 0.5cm 0},clip]{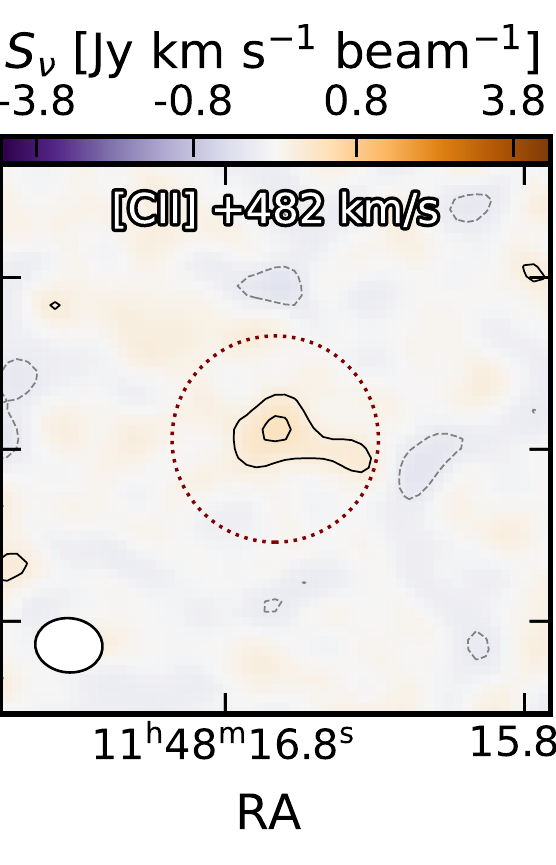} \\
    
    \includegraphics[height=0.22\textheight,trim={0 0 0.5cm 0},clip]{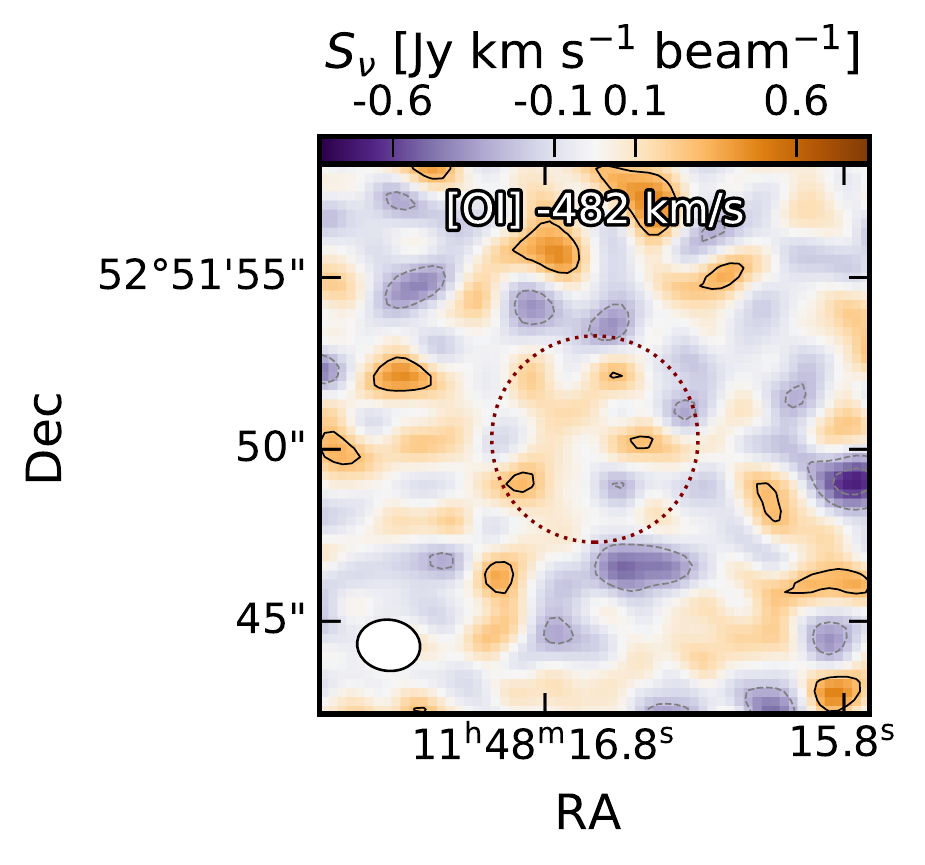}
    \includegraphics[height=0.22\textheight,trim={3.25cm 0 0.5cm 0},clip]{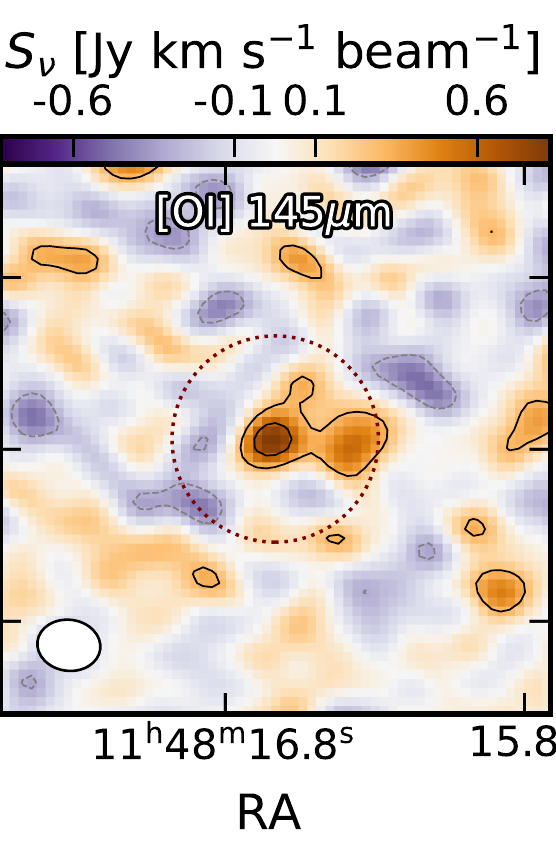}
    \includegraphics[height=0.22\textheight,trim={3.25cm 0 0.5cm 0},clip]{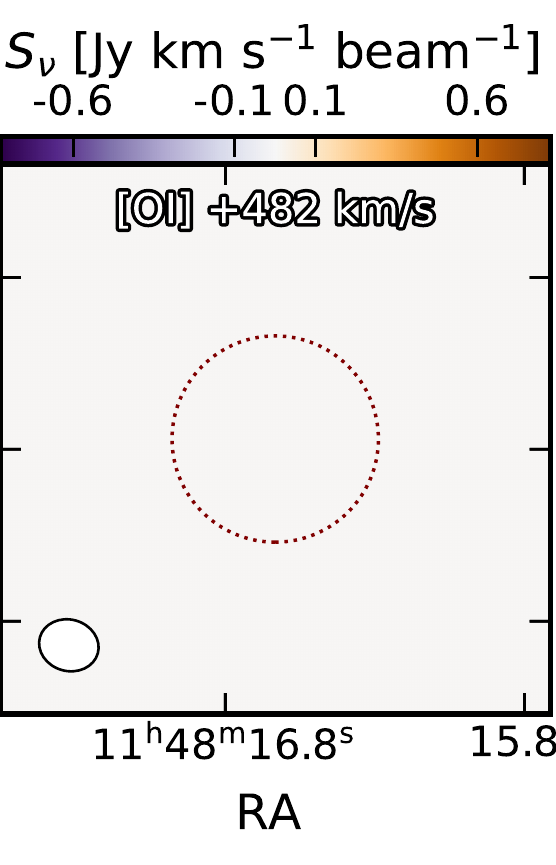}

    \caption{Line maps, velocity--integrated over $1.2\times \rm{FWHM}_{\rm{[C~II]}}=482\,\kms$ at the expected frequency of \cii, \nii\ and \oi\ and  assuming the \cii\ redshift. For each line, we also plot two additional collapsed maps centered at $\pm 482 \,\kms$ away from the line emission. For \oi\ only one of the adjacent maps is empty as the emission is at the edge of the band. On the central \cii\ map we plot the quadrants used for the spatial extension analysis in Section \ref{sec:cii_extension}. The contours are logarithmic $(-4,-2,2,4,8,16,32)\sigma$ (rms). }
    \label{fig:lines_maps}
\end{figure*}

\begin{figure}
    \includegraphics[width=0.46\textwidth,trim={0 0 0 0cm},clip]{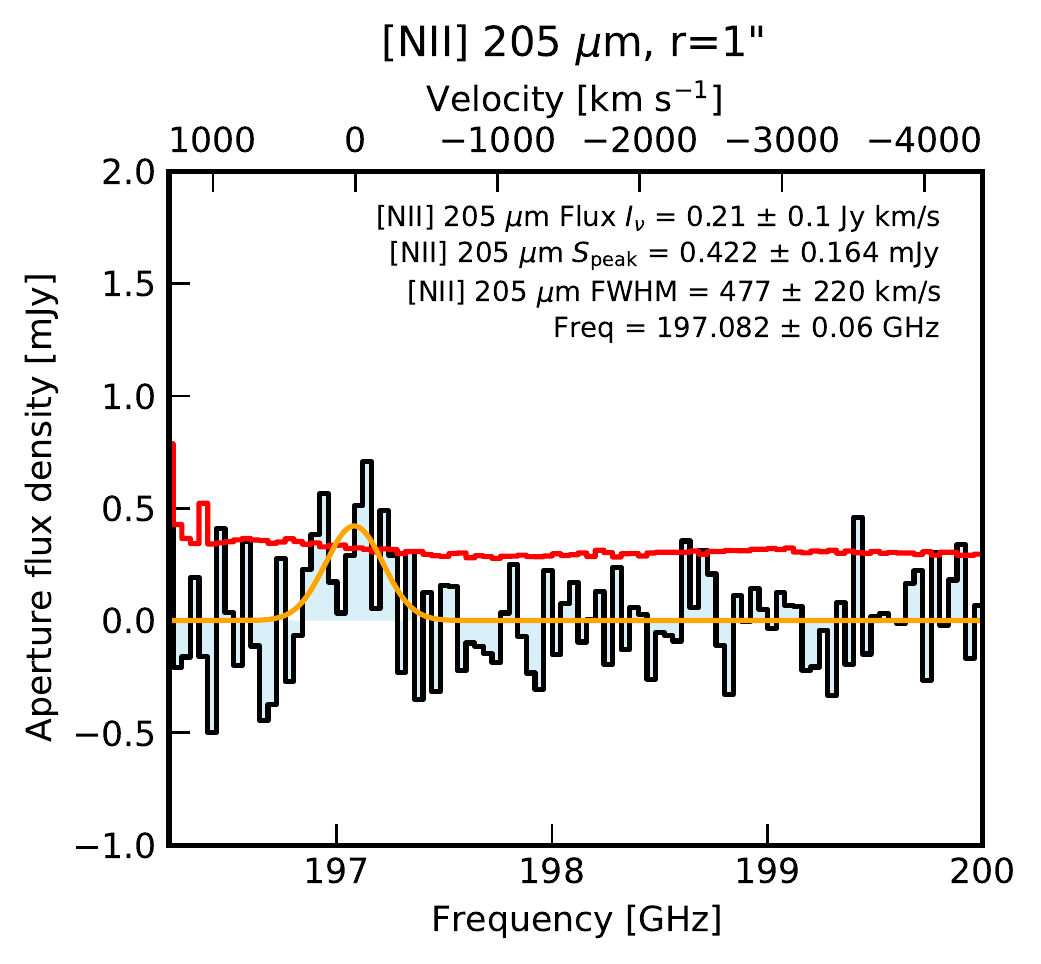}
    \includegraphics[width=0.44\textwidth,trim={0 0 0 0cm},clip]{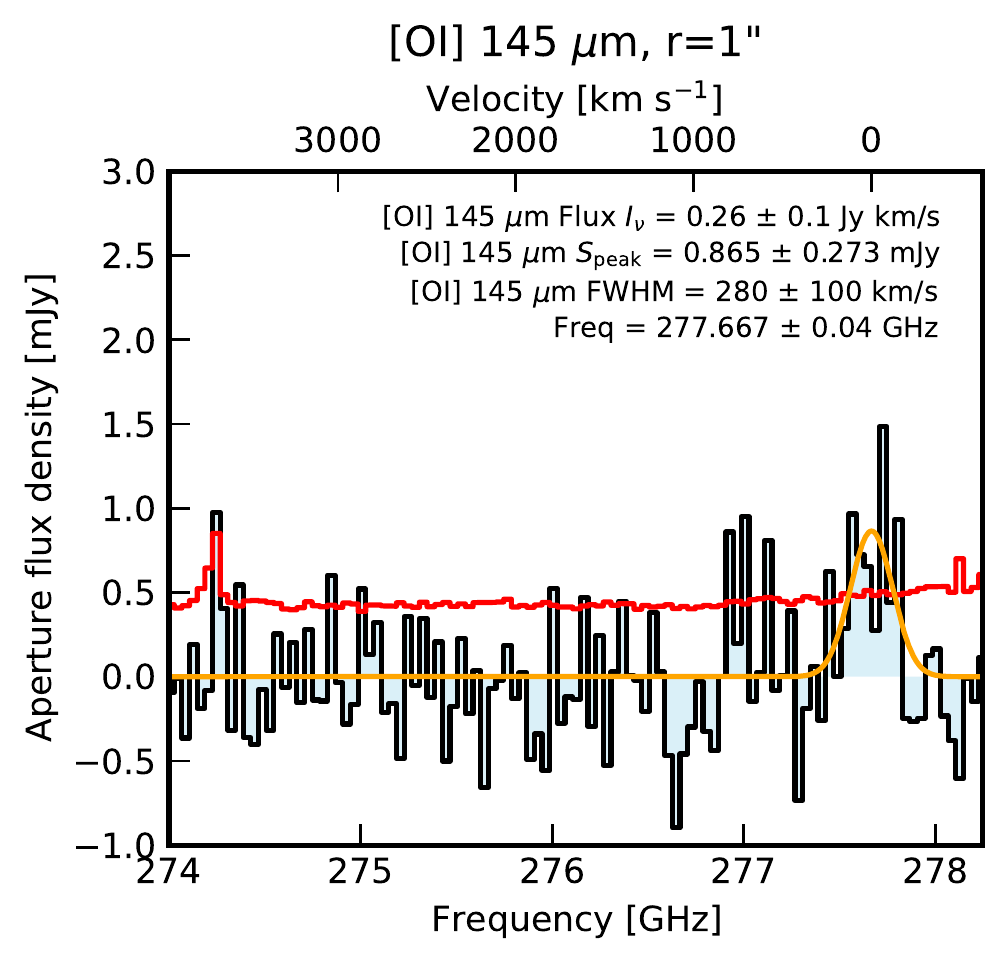}
    \caption{Continuum--subtracted spectrum of the \nii\ (upper panel) and \oi\ (lower panel). The spectra (black lines, shaded blue) are extracted in $r=1"$ apertures centered on the peak of the emission in the velocity integrated maps (Fig. \ref{fig:lines_maps}) using residual scaling (see Sec. \ref{sec:obs}). The $1 \sigma$ noise level per channel for a $r=1"$ aperture is shown in red, and a single Gaussian fit is shown in orange (the best--fit parameters and uncertainties in the upper right corner). }
    \label{fig:spectra_OI_NII}
\end{figure}

We present the line maps for the \cii, \oi\ and \nii\ emission in Figure \ref{fig:lines_maps}. For comparison, we also show maps of the same channel width ($482\,\kms$) as the main line map but offset by $\pm 482\,\kms$ from the line emission to visually assess the robustness of our detections. The spectra of the \cii\ line are presented in Fig.~\ref{fig:cii_fit} (see also Appendix \ref{app:CII_fluxes}) whilst the \oi\ and \nii\ spectra are shown in Fig. \ref{fig:spectra_OI_NII}. 

We detect \cii\ and \oi\ at $42\sigma$ and $5.3\sigma$ (where the SNR is calculated using the peak surface brightness in the line maps and the pixel rms level). \nii\ is only marginally detected ($3.7\sigma$) and its peak emission is potentially offset from the rest--frame UV, dust, \cii, and \oi\ emission. In the central pixel of the \cii\ emission, \nii\ is formally undetected. We defer further discussion of the \nii\ emission line to section \ref{sec:nii_emission}.  In our subsequent  analysis of the ISM of J1148+5251, we focus on the extended aperture--integrated fluxes. All the fluxes and derived line luminosities are tabulated in Table \ref{tab:lines_maps}. For comparison, we have also measured the flux density of the \cii\ emission by fitting the aperture--integrated spectrum with a Gaussian, and show a comparison between our work and earlier studies for various aperture sizes in Appendix \ref{app:CII_fluxes}. In summary, we find a good agreement between the line map flux estimates and the spectrum best-fits values, as well as good agreement between previous studies and the data presented here, as long as the “core" ($<3"$ emission, no velocity offsets) \cii\ emission is considered. 

\begin{table}
\hspace{-1.5cm}
    \begin{tabular}{l|c c c}
    \hline\hline
 3" aperture & \nii & \cii & \oi \\
 \hline 
 S/N & 3.7 & 42.0 & 5.4 \\ 
 $S_{line}\Delta v$ [Jy km s$^{-1}$]  &$0.5\pm0.3$ &$10.2\pm0.5$ &$1.0\pm0.6$  \\
 $L_{\rm{line}}$ [$10^{9}L_\odot$] &  $0.4\pm0.2$ &  $10.6\pm0.5$&  $1.1\pm0.7$ \\
 $L'_{\rm{line}} [10^{9}\,\rm{K \,km\, s}^{-1} \rm{pc}^{-2}$] &$4.4\pm2.1$  & $48.1\pm2.5$  & $4.1\pm2.4$ \\ 
\hline \hline
 Peak [CII] pixel  & \nii & \cii & \oi \\
 \hline
 S/N & 1.1 &$42.0$ &5.4 \\
 $S_{line}\Delta v$ [Jy km s$^{-1}$]  & $<0.12$ &  $6.4\pm0.2$   &  $1.0\pm0.2$  \\
 $L_{\rm{line}}$ [$10^{9}L_\odot$] & $<0.10$ & $6.7\pm0.2$ & $1.2\pm0.2$   \\
  $L'_{\rm{line}} [10^{9}\,\rm{K \,km\, s}^{-1} \rm{pc}^{-2}$] & $<1.0$ & $30.4\pm0.7$ & $4.1\pm0.8$ \\ 
  \hline
    \end{tabular}
    \caption{Atomic fine structure line flux measurements from the line maps ($1.2\times\rm{FWHM}_{\rm{CII}} = 482\,\kms$), flux--corrected by $1/0.84$. Upper limits are given at the 3$\sigma$ level. The first $3$ lines give luminosity for the total integrated flux and the last $3$ for the peak \cii\ surface brightness. All integrated fluxes are derived applying residual--scaling correction, and luminosities are computed following the definitions of \citet[][]{Solomon1997}}
    \label{tab:lines_maps}
\end{table}

\section{The spatial and velocity structure of the \cii\ emission in J1148+5251}
\label{sec:cii}
\subsection{Spectral analysis}
\label{sec:cii_outflow}

\begin{figure*}
    \centering
    \includegraphics[width=0.51\textwidth]{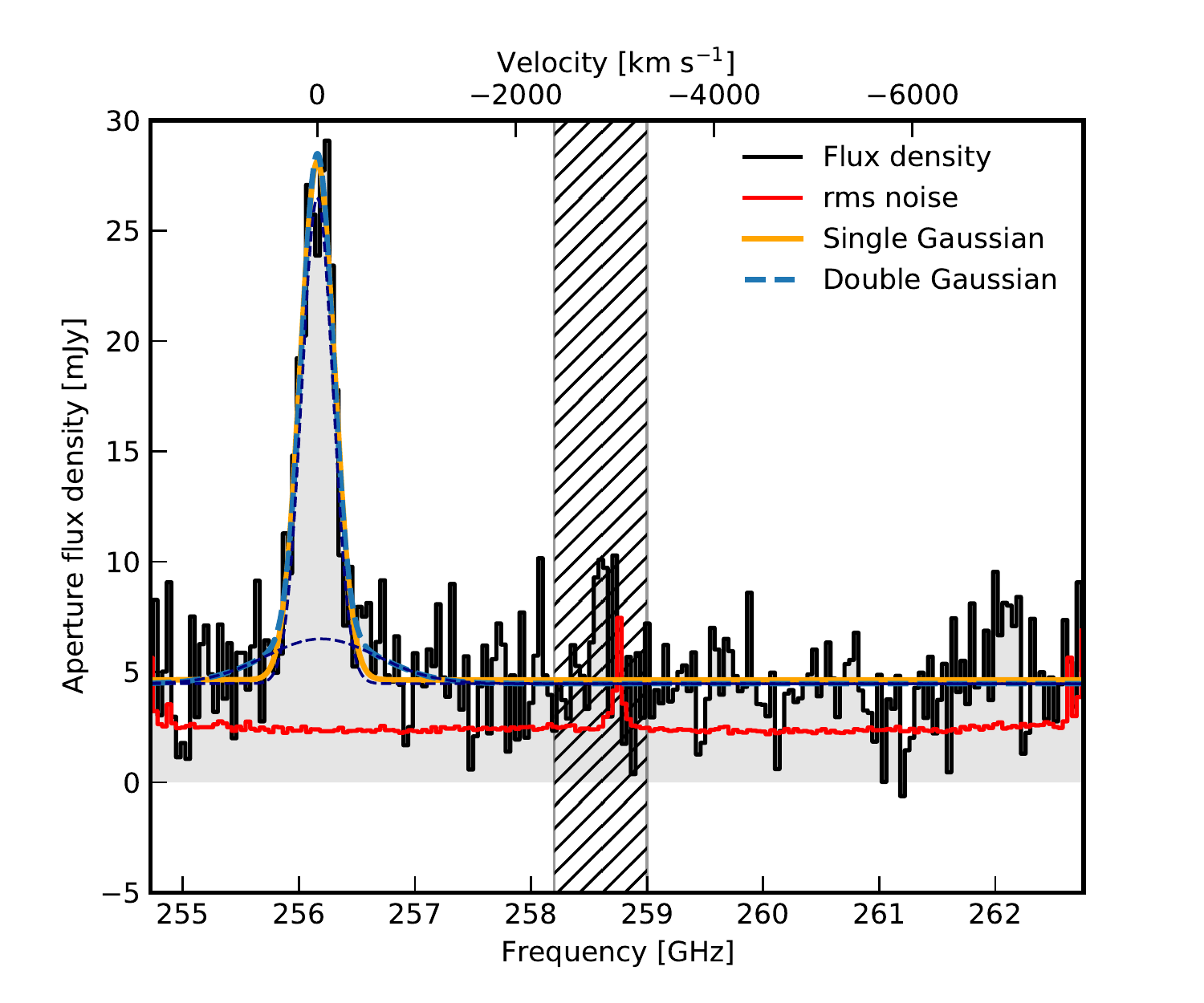}
    \includegraphics[width=0.48\textwidth]{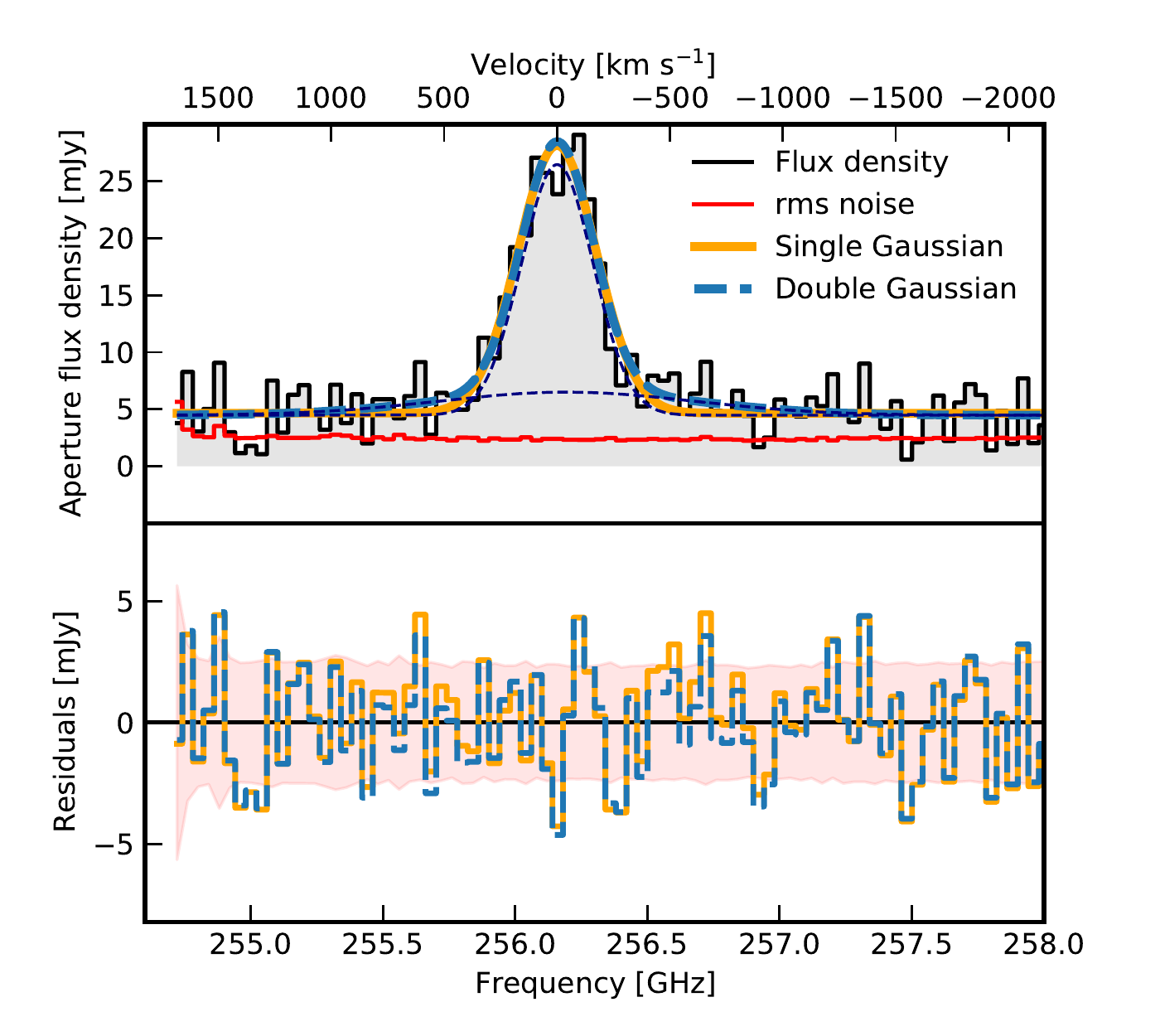}
    \includegraphics[width=0.49\textwidth]{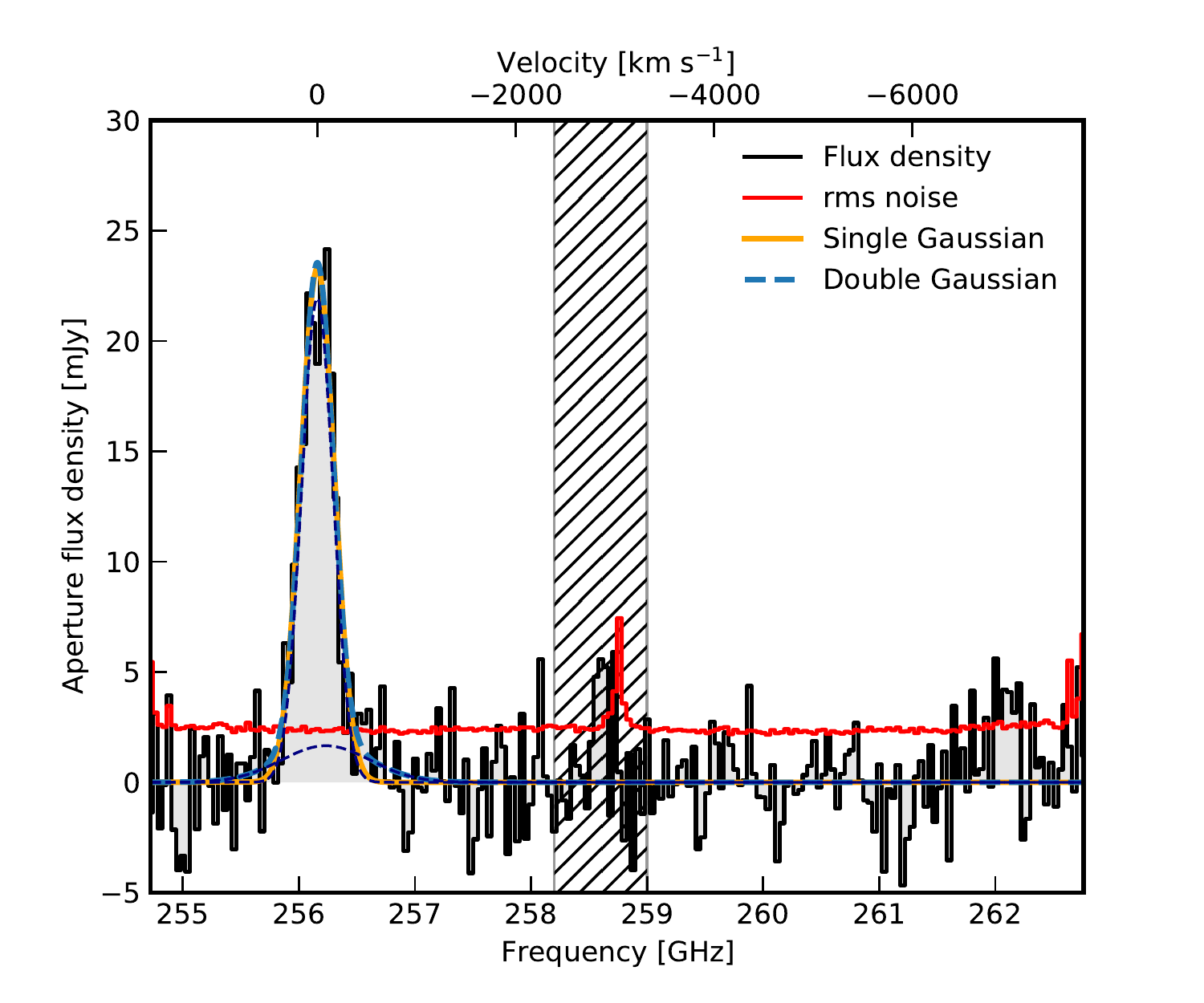}
    \includegraphics[width=0.5\textwidth]{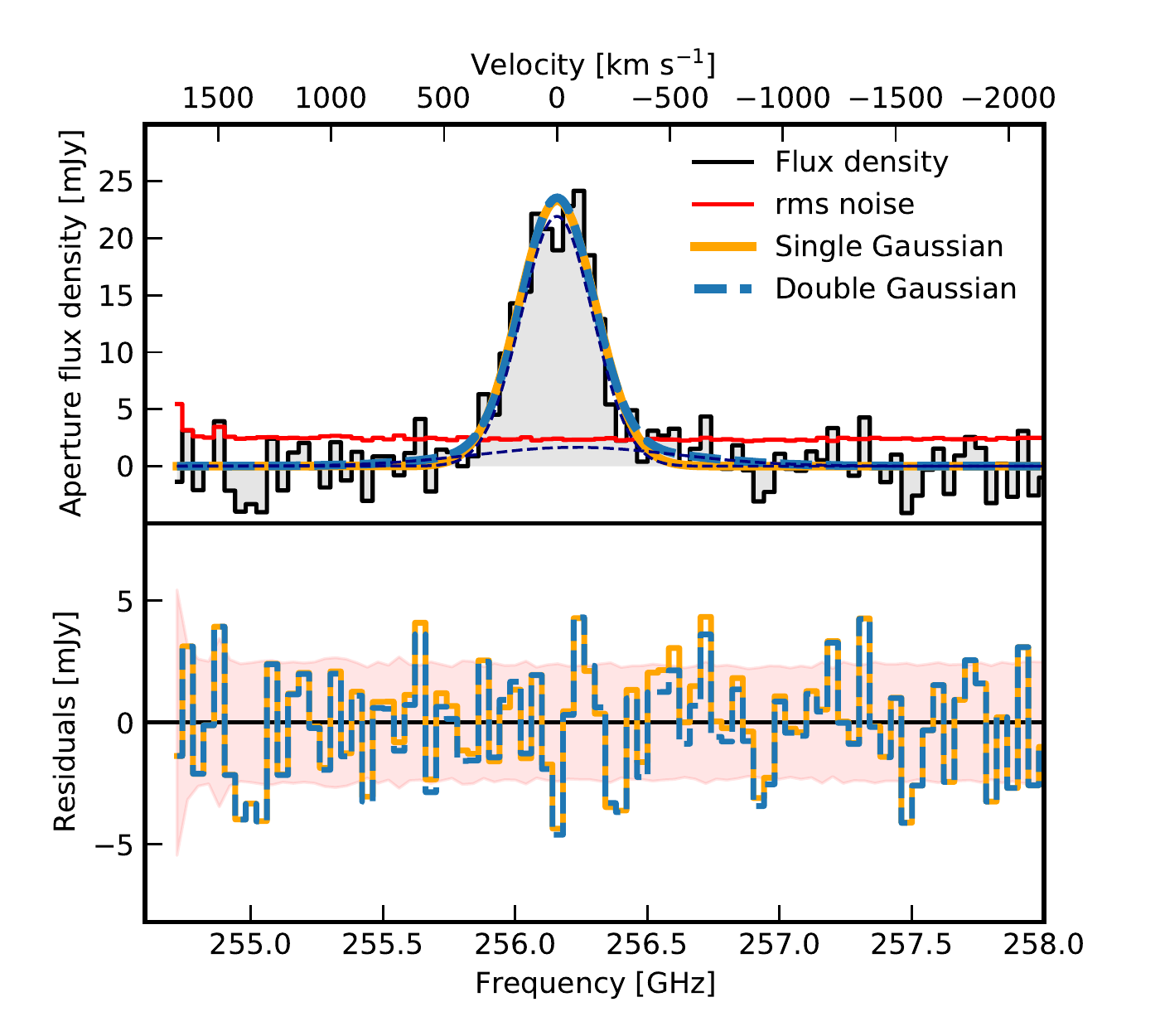} \\
    \caption{Total (top panels) and continuum--subtracted (bottom panels) spectrum of the \cii\ line (black) extracted in a $r=3"$ aperture using residual scaling and 1-- and 2--components Gaussian fits (orange line and dashed--dotted blue, respectively). Thin dashed dark blue lines show the two components of the double Gaussian model. Both models are fitted to the entire frequency range to the exception of the overlap between the two sidebands that is coincident with an atmospheric absorption feature (hatched in grey) and the 262 GHz feature (see text). The right panels show a zoomed--in version of the left plots and the residuals of the two models. The shaded red area in the bottom right panels show the $1\sigma$ noise level, computed using the rms noise in each channel in mJy beam$^{-1}$ with sigma-clipping scaled accordingly for the number of beams in the $r=3"$ aperture. The median noise level at $\nu_{\rm{obs}}>263\ \rm{GHz}$ is $2.36\pm0.14\ \rm{mJy}$, whereas a direct rms estimate from the continuum-subtracted spectrum at the same frequencies gives $2.00\pm0.15\ \rm{mJy}$. We find no evidence for a broad \cii\ emission line in the new NOEMA data (see text for details).}
    \label{fig:cii_fit}
\end{figure*}

Figure \ref{fig:cii_fit} shows the total and continuum--subtracted \cii\ spectra (extracted in a $r=3"$ aperture), as well as single and double Gaussian fits to the data. We found an emission feature at $262.2$ GHz which was subsequently masked to determine and subtract the continuum. An emission line map of the feature does not reveal any convincing emission, and we are not aware of any strong molecular or ionised line which could correspond to this redshifted frequency. We checked that masking this feature or not does not impact any of the results that follow. We find that a single Gaussian describes the \cii\ emission well both in the total and continuum-subtracted spectra, and that the single Gaussian residuals are consistent with the rms noise (Fig.~\ref{fig:cii_fit}). 
For the total spectrum, the best-fit single Gaussian and continuum model has a monochromatic continuum flux density of $S_\nu = 4.6\pm 0.2\ \rm{mJy}$, \cii\ observed frequency $\nu_{\rm{[CII],obs}}= 256.158 \pm 0.008\ \rm{GHz}$, \cii\ FWHM$=408\pm 23\ \kms$ and integrated flux $I_\nu=10.0\pm0.8\ \rm{Jy}\ \kms$. The best-fit double Gaussian and continuum model has a continuum flux density of $S_\nu = 4.5\pm 0.2\ \rm{mJy}$, $\nu_{\rm{[CII],obs}}^{\rm{narrow}}= 256.159 \pm 0.009\ \rm{GHz}$, \cii\ FWHM$^{\rm{narrow}}=373\pm 36\ \kms$ and narrow-component integrated flux $I_\nu^{\rm{narrow}}=8.7\pm1.0\ \rm{Jy}\ \kms$. The broad component has a central frequency $\nu_{\rm{[CII],obs}}^{\rm{broad}}= 256.19 \pm 0.20\ \rm{GHz}$, \cii\ FWHM$^{\rm{broad}}=1330\pm 820\ \kms$ and integrated flux $I_\nu^{\rm{broad}}=2.9\pm3.0\ \rm{Jy}\ \kms$. The $\chi^2$ difference between the single- and double-Gaussian models is only $\Delta \chi^2= 3.66$ for three additional degrees of freedom, which corresponds to a p--value of $p=0.7$ or evidence at the $0.52\sigma$ level for a broad component. Equivalently, the Bayesian Information Criterion (BIC) difference is $\rm{BIC}_1 - \rm{BIC}_2 = -12.0$, where a lower BIC implies a better model, and we adopt a threshold $\Delta \rm{BIC}_{12}>10$ for strong significance to prefer the more complex model \citep[e.g.][]{Kass1995}. We also perform an Anderson-Darling test on the residuals to find any deviations from normally distributed residuals (with a variance equal to the measured rms squared). We find no significant deviation, for either fit, at the $p>0.15$ level. We thus do not find evidence for a broad spectral \cii\ component in J1148+5251 based on the new NOEMA data alone. 

For the continuum-subtracted spectrum, the best--fit single and double Gaussian models yield consistent results within the uncertainties (for more details, including the impact of different aperture sizes, see appendix \ref{app:CII_fluxes}), and the residuals are nearly identical (Fig. \ref{fig:cii_fit}, second row, lower right). The $\chi^2$ improvement obtained with a double Gaussian is $\Delta \chi^{2}=2.24$, which for an increase of $3$ free parameters gives a $p$--value of $0.46$ of rejecting the simple model in favor of a double Gaussian, in agreement with the analysis of the total spectrum. The BIC difference is $\rm{BIC}_1 - \rm{BIC}_2 = -14.0$, e.g. the single-Gaussian model is strongly favoured, and an Anderson-Darling test on the residuals also concludes that they are consistent with the measured noise at the $p>0.15$ level.

We note that, regardless of the aperture chosen or the use of residual scaling, the addition of a broad \cii\ component to the spectral fit is not preferred by the NOEMA data (see Appendix \ref{app:CII_fluxes}). We have checked that the continuum fluxes in the lower and higher frequency sidebands are consistent within $\sim 2\%$. This ensures that the continuum of J1148+5251, determined from the full $\sim 7.6$ GHz sideband, is consistent with that determined solely from the single $\sim 3.8$ GHz baseband containing the \cii\ line. We also show in Appendix \ref{app:continuum} that the choice of the line masking width and the flux calibration differences between the two $\sim 3.8$ GHz basebands do not impact the recovered \cii\ line. 

The presence of a broad \cii\ wing reported in previous works \citep[][]{Maiolino2012,Cicone2015} and its absence in the NOEMA data seems puzzling at first. Indeed \citet[][]{Cicone2015} reach a sensitivity in the \cii\ line comparable to ours ($0.46\, \rm{mJy\, beam}^{-1}$ vs.\ $0.45 \, \rm{mJy\, beam}^{-1}$ per $100\,\kms$ channel) and have similar angular resolution ($1.3" \times 1.2"$ vs.\ $1.97" \times 1.59"$). However, their observations were performed with a smaller number of antennas (6, corresponding to 15 baselines, where the new observations were done with 8--9, corresponding to 28--36 baselines), using a correlator that provided a spectral coverage ($\sim3.6$ GHz) twice as narrow than the new observations ($\sim 7.7$ GHz) around the \cii\ line. Hence the discrepancy cannot stem from sensitivity or resolution issues and we expect that due to our slightly larger beam, larger number of antennas and higher imaging fidelity we would be more sensitive, if anything, to extended \cii\ structures in J1148+5251 compared to earlier data. The wider spectral coverage of the new \textit{Polyfix} also enables us to better constrain the dust continuum emission and the exact shape of a potential broad \cii\ emission line.

In Appendix \ref{app:pdbi_data} we discuss in detail the origin of these discrepancies which can be ascribed to a combination of (i) continuum subtraction and (ii) residual-scaling effects. We show that when taking these into account, the previous PdBI and new NOEMA data are consistent. We present the combined data in Appendix \ref{app:pdbi_data}, and find that, using the same approach as used here, broad \cii line wings are only tentatively recovered in the merged PdBI and NOEMA dataset.
We finally note that the statistical methodology used here to assess the presence of an outflow is arguably quite conservative. Indeed, with the sensitivity and spectral coverage of the data at hand, only extremely powerful outflows would produce a broad line component whose significance is such that a $\xi^2$ or BIC criterion would prefer a double component fit over a single-component one. In low-redshift studies, it is common to use multiple diagnostics (PV diagrams, other outflow tracers) to assume the presence of an outflow first and then fit the spectral profile with two components \citep[see, e.g., the discussion in ][]{Cicone2014}. In any case, these new NOEMA data, thanks to the improved continuum subtraction and analysis methodology, indicate a much less prominent broad component than previously reported, and so rule out the presence of an extremely strong high--velocity outflow. We have however provided above the 2-Gaussian best-fit results in case future observations provide ancillary evidence in favour of an outflow.

\subsection{Spatially extended \cii\ emission}
\label{sec:cii_extension}
\begin{figure*}
    \centering
    \includegraphics[width =\textwidth]{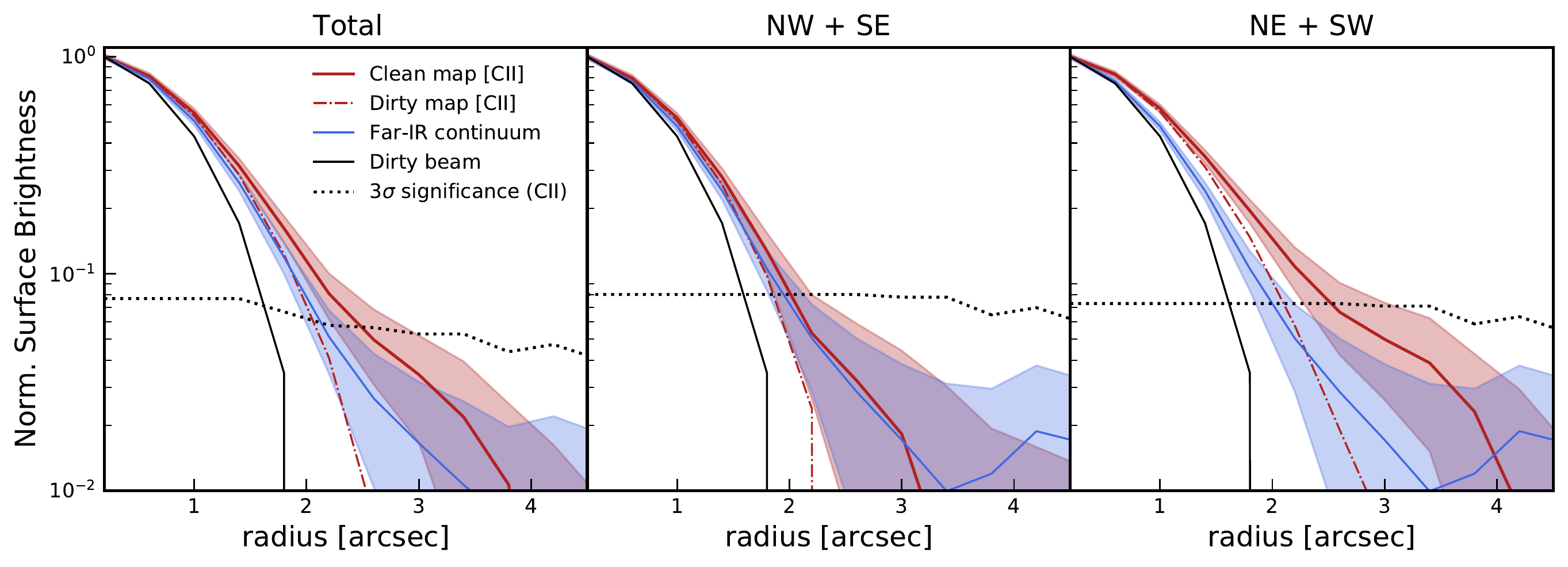}
    \caption{\cii\ (red) and dust continuum emission (blue) profiles for J1148+5251, which both extend up to $\theta = 2.51^{+0.46}_{-0.25}\ \rm{arcsec}$ at the $3\sigma$ level, corresponding to $r= 9.8^{+3.3}_{-2.1}$ (accounting for beam-convolution). The shaded areas show the corresponding $1$ sigma noise level. A point source profile is shown in black and the dirty (before cleaning and residual--scaling) profile is shown in dashed--dotted red. The three panels show, from left to right the radial profile integrated, over the full $2\pi$, only in the North--West and South--East quadrant, and finally only in the North--East and South--East quadrant (e.g. along the extended emission seen in Figure \ref{fig:lines_maps}). In the latter case, the \cii\ emission is more extended than the FIR continuum ($5.8 \sigma$ significance).}
    \label{fig:profile_r}
\end{figure*}

The \cii\ emission is spatially extended (Fig. \ref{fig:lines_maps}). In Figure \ref{fig:profile_r} we show the dust and \cii\ radial profiles to investigate their spatial extension. The dust continuum emission (at $\sim 259$ GHz) is as extended (within $1\sigma$ errors) as the \cii\ emission (up to $\theta = 2.51^{+0.46}_{-0.25}\ \rm{arcsec}$, at the $3\sigma$ level, see Figure \ref{fig:profile_r}, leftmost panel), suggesting that both trace star-forming material with physical conditions similar to those that we normally ascribe to the ISM of galaxies rather than an extended \cii\ halo \footnote{We consider here and thereafter a \cii\ halo to be significant emission beyond that of the dust continuum emission \citep[e.g.,][]{Fujimoto2019,Fujimoto2020,Herrera-Camus2021}} or outflow. This corresponds to a radius of $r= 9.8^{+3.3}_{-2.1} \rm{kpc}$ once accounting for beam--convolution. This is twice larger than any of the 27 quasars observed in \cii\ by \citet[][]{Venemans2020}, confirming earlier reports \citep[][]{Maiolino2012,Cicone2015} that J1148+5251 is an outlier in terms of \cii\ emission extension. This result holds as well if the dust extension is measured from other spectral setups (e.g., Fig. \ref{fig:cont_maps}).

\begin{figure*}
    \centering
    \includegraphics[height=0.25\textheight]{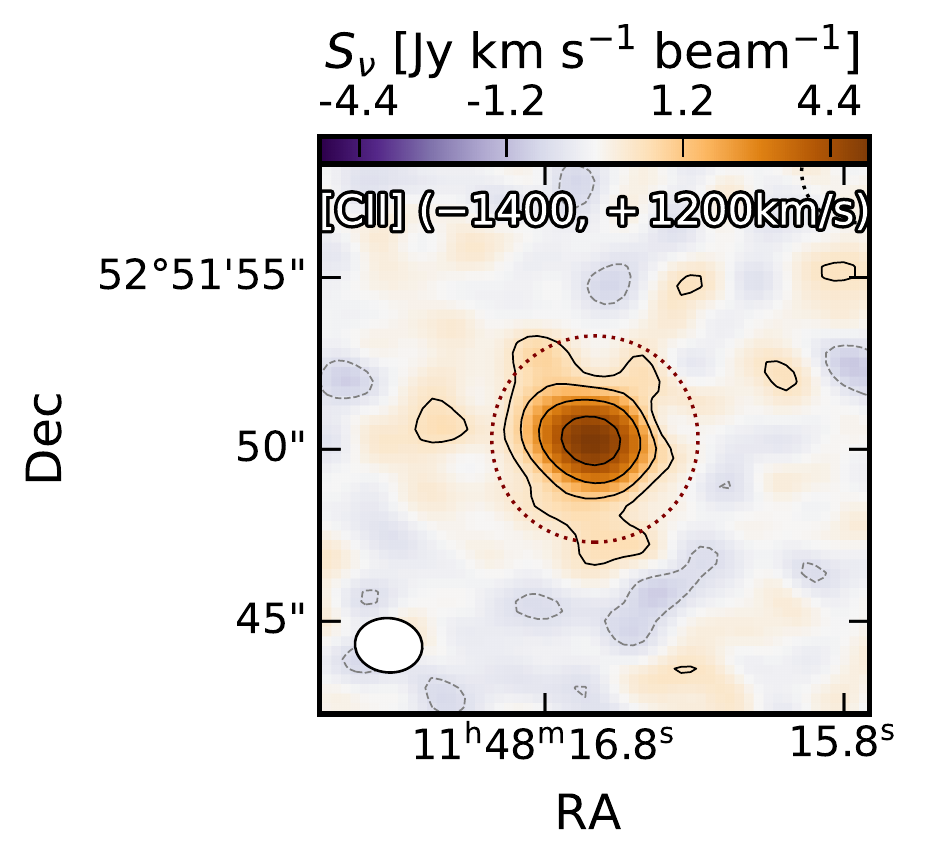}
    \includegraphics[height=0.25\textheight,trim={3.2cm 0 0 0 },clip]{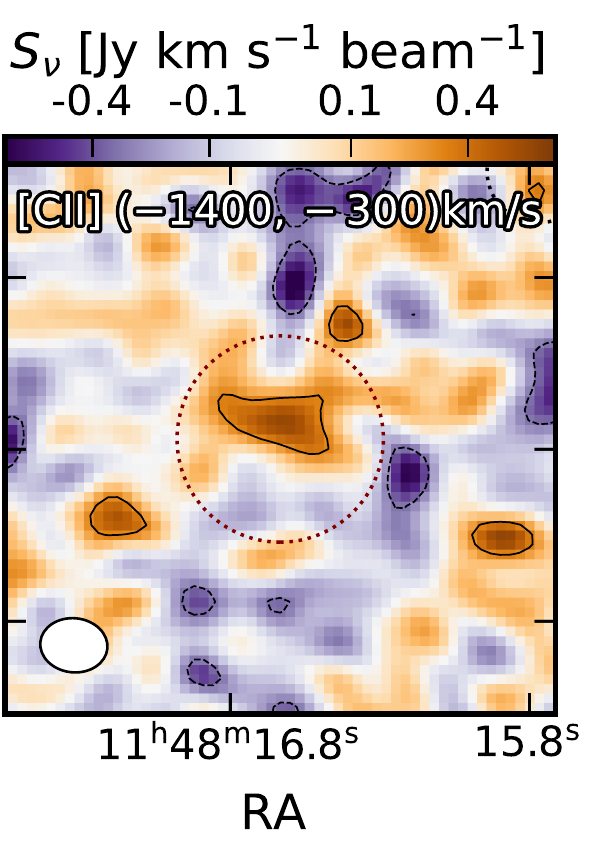}
    \includegraphics[height=0.25\textheight,trim={3.2cm 0 0 0 },clip]{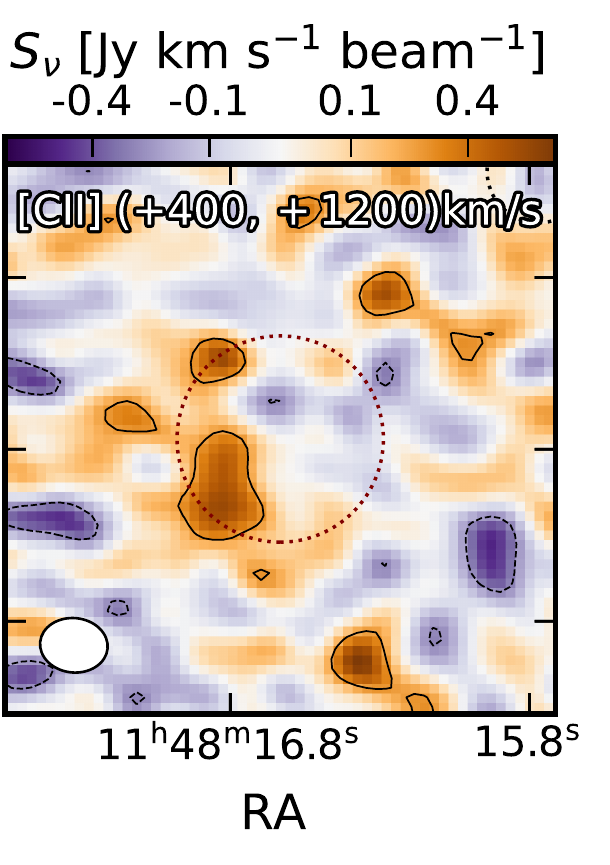}
    \caption{From left to right: map of the \cii\ emission, averaged over the velocity ranges $(-1400,1200) \, \kms$, $(-1400,-300) \, \kms$ and $(+400,+1200) \, \kms$. These velocity ranges are similar to that presented in \citet[][]{Cicone2015} who reported $6\sigma$ \cii\  clumps at $2-4$ arcsec from the source center in the blue/red--shifted wings of the \cii\ emission. Such spatially and spectrally offset emission is not recovered in the NOEMA data. The $1\sigma$ rms values in the above three maps are $0.27,0.14,0.15\, \rm{mJy\,beam}^{-1}$. This figure shows contours in logarithmic increments (-4,-2,2,4,8,16,32)$\sigma$.  }
    \label{fig:map_CII_1500kms}
\end{figure*}

The extension of \cii\ is however asymmetric, with a prominent NE--SW axis (Fig. \ref{fig:lines_maps}, although this is less pronounced with a larger channel width, see Fig. \ref{fig:map_CII_1500kms}, first panel). In the second and third panel of Figure \ref{fig:profile_r} we compare the dust continuum and the \cii\ radial surface brightness profile for the radially averaged case, the NE--SW axis and the NW--SE axis (see Fig. \ref{fig:lines_maps}). Whilst we find no evidence for an extended \cii\ halo when averaging radially (or along the NW--SE axis), along the NE--SW axis the \cii\ emission is significantly more extended than the dust continuum (5.8$\sigma$).

\begin{figure}
    \centering
    \includegraphics[width=0.45\textwidth]{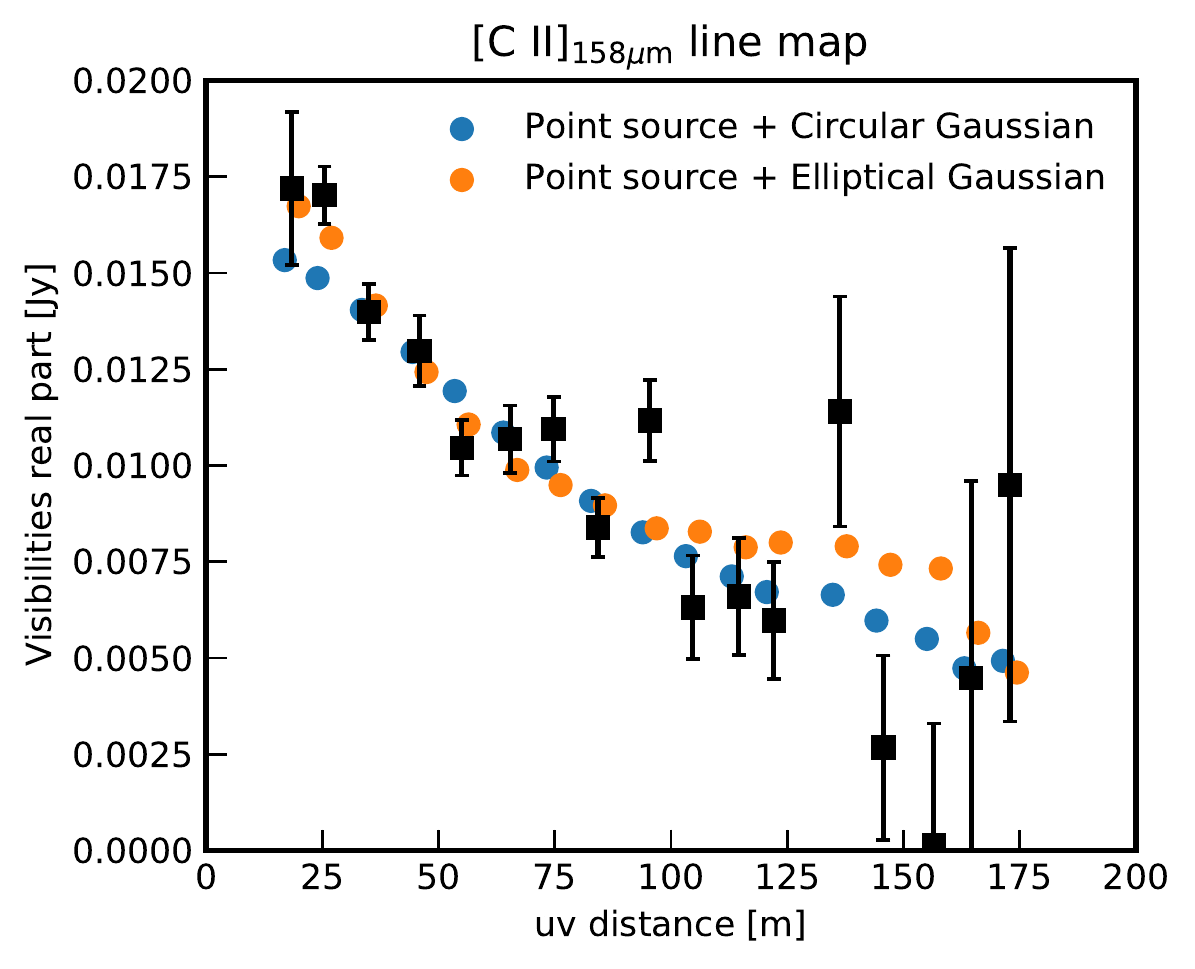}
    \includegraphics[width=0.45\textwidth]{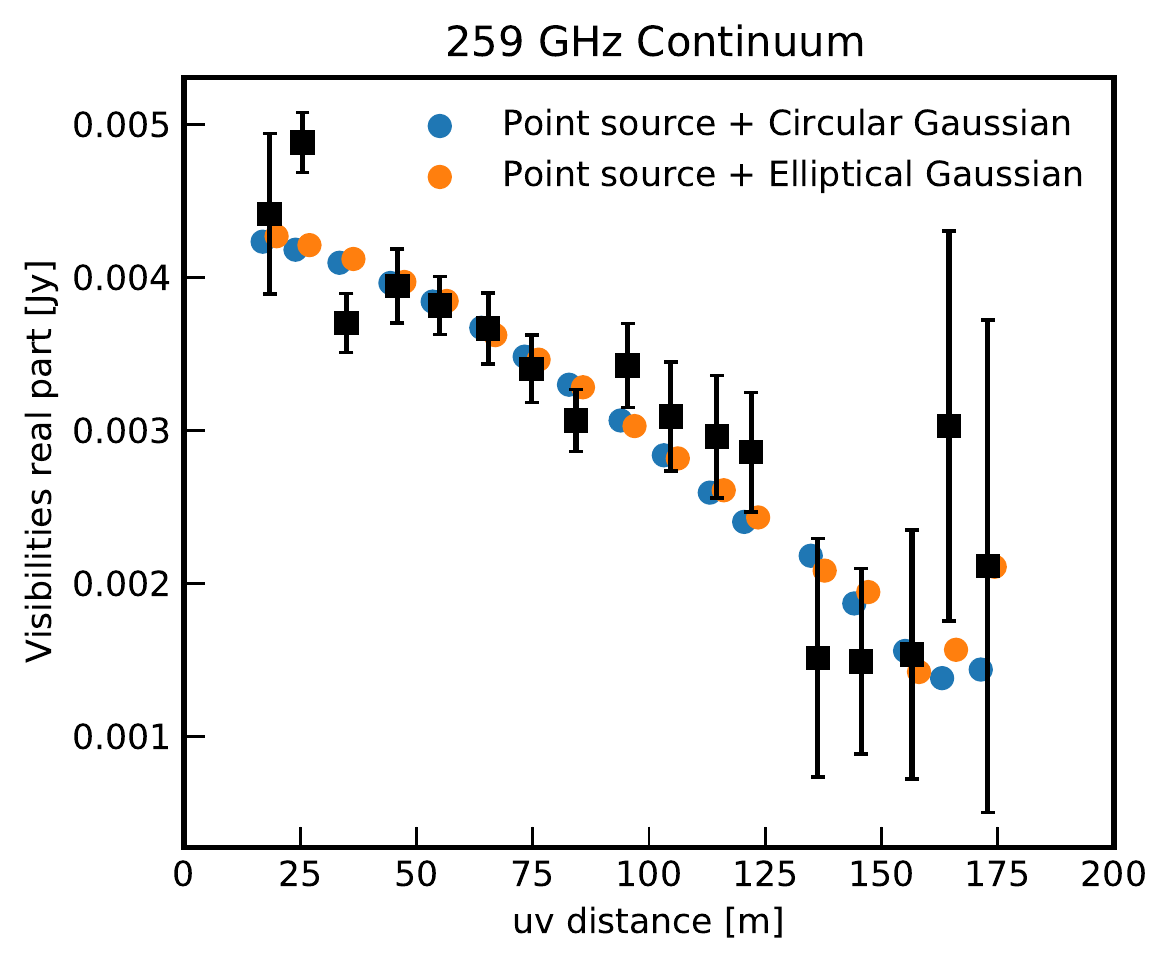}
    \caption{Observed (black) and modelled (blue,orange) real part of the visibilities for the \cii\ emission (integrated over $\Delta v= 482 \kms$) and dust continuum in J1148+5251. The model points are slightly offset for presentation purposes. The difference between the two models (blue: point source + circular Gaussian, orange: point source + elliptical Gaussian) is not statistically significant (p--value $= 0.83$, e.g. $0.96\sigma$ significance). }
    \label{fig:uv_plane}
\end{figure}

We finally explore the extension of the \cii\ emission and the dust continuum directly in the \textit{uv} plane. To that end, we use the UV\_FIT and UV\_CIRCLE routines in GILDAS/MAPPING to fit the visibilities with a point source and 2D Gaussian emission model, and then bin the modelled and observed visibilities radially. We plot the real part of the visibilities against the \textit{uv} radius in Figure \ref{fig:uv_plane}. In such plots, a point source gives a constant flux density at all \textit{uv} radii, while a Gaussian emission model yields a Gaussian profile centered at $r=0$. We find good agreement between the observed visibilities and a composite emission model comprising a point source and a 2D Gaussian. We fit both a circular Gaussian and an elliptical Gaussian to the dust and \cii\ continuum visibilities. For the \cii\ emission, the best--fit model gives corrected velocity--integrated flux of $4.2\pm 0.5 \,\rm{Jy}\, \kms $ in the point source component, and $5.0\pm0.5 \,\rm{Jy}\,\kms $ in the extended Gaussian, in good agreement with the flux derived from the line map (Table \ref{tab:lines_maps}) and the fitted spectrum (see Appendix \ref{app:CII_fluxes}). The elliptical Gaussian models yields velocity--integrated fluxes of $5.0\pm 0.3 \,\rm{Jy} \,\kms $ and $5.5\pm 0.4 \,\rm{Jy}\, \kms $ for the point source and resolved components, respectively. In either case, we are consistent with the large fraction of emission coming from the unresolved component reported in \citet[][]{Cicone2015} but not with the overall flux, which is possibly due to continuum subtraction differences as discussed in Appendix \ref{app:pdbi_data}.
Additionally, the unresolved flux is consistent with that measured at higher resolution data \citep[][]{Walter2009}, suggesting the difference in total flux is due to the extended component being resolved out in high--resolution configurations.

The FWHM of the \cii\ Gaussian component is $\rm{FWHM}=1.6"\pm0.2" \,(9\pm1 \rm{\, kpc})$, or minor/major axes $a=3.6"\pm0.3"\, (20\pm2\, \rm{ kpc}), b =  1.8"\pm0.2" \,(10\pm 1 \,\rm{kpc})$ for the elliptical model. This is in agreement with the scale of the $3\sigma$ extension directly measured on the cleaned image. The elliptical Gaussian has a major/minor axis difference and angle (PA=$53\pm4$ deg) in agreement with the asymmetric \cii\ emission discussed above. However, a likelihood ratio test does not prefer the elliptical Gaussian model over the circular one (p--value $= 0.83$, e.g. $0.96\sigma$ significance). We do not find any difference between the elliptical Gaussian and circular Gaussian model for the dust emission (see Fig. \ref{fig:uv_plane}). The best--fit model for the dust emission gives a flux density of $1.0\pm0.7 \, \rm{mJy}$ and $3.3\pm 0.7 \rm{\, mJy}$ for the unresolved/resolved components, with the sum in agreement with the measurement aperture--integrated flux density (Table \ref{tab:continuum}). The dust \textit{uv} profile (Fig. \ref{fig:uv_plane}, bottom panel) does not show clear evidence for a flattening at large \textit{uv} distances that is characteristic of a point source. The Gaussian component has a FWHM of $r=0.7"\pm0.1" \,(3.9\pm0.6\ \rm{kpc})$.

\citet[][]{Cicone2015} reported a complex velocity structure of the \cii\ outflow, with emission clumps detected up to $r\sim 30$ kpc and $\sim 1000\,\kms$ offsets from the central emission, particularly visible in an emission line map integrated over a $(-1400,1200)\,\kms$ interval (we show in Appendix \ref{app:pdbi_data} how we recover these structures in the previous PdBI data, although at a lower significance level). Such features are not recovered in the new NOEMA data when averaging over a similar velocity interval (Fig. \ref{fig:map_CII_1500kms}, left, see also Appendix \ref{app:pdbi_data}). The \cii\ emission is concentrated mostly within $3"$ ($\sim 11 \rm{kpc}$) no significant emission is recovered beyond that radius \footnote{Following the analysis of \citet{Cicone2015}, we also plot the blue-- and red--shifted component ($(-1400,-300) \, \kms$ and $(+400,+1200) \, \kms$) of the \cii\ emission, and find only marginal detections ($3.0$ and $3.4\sigma$) which are not spatially offset (Fig. \ref{fig:map_CII_1500kms}).}. We further show in Appendix \ref{app:velocity} that no significant velocity structure is detected in J1148+5251 in the NOEMA data, neither by inspecting the channel maps (with channel width $\Delta v = 120\, \kms$) nor by means of a kinematical analysis. 

\section{ISM properties}
\label{sec:ism}
\subsection{\cii\ emission}
In the following analysis, we adopt luminosities measured from the line maps. We measure a \cii\  line--luminosity $L_{\rm{[C~II]}} = (10.6\pm0.5)\times10^{8}\rm{L}_\odot \, (L_{\rm{[C~II]}}' = (48.1\pm2.5)\times10^{9} \,\rm{K} \, \kms \rm{pc}^{-2} )$. Following the [CII]--SFR relation of \citet[][]{DeLooze2014} for high--redshift ($z>0.5$) galaxies, we find $\rm{SFR}= 2041\pm111 \, \rm{M}_\odot\,\rm{yr}^{-1}$, in good agreement with the SFR inferred from the continuum luminosities. Using the \citet[][]{DeLooze2014} calibration for AGNs does lower the result ($\rm{SFR}= 857\pm35 \, \rm{M}_\odot\,\rm{yr}^{-1}$), and that for ULIRGS increases the inferred SFR ($\rm{SFR}= 5562\pm257 \, \rm{M}_\odot\,\rm{yr}^{-1}$). This agreement between the \cii\ and the infrared-luminosity based SFR is also observed in other high--redshift quasars \citep[e.g.,][]{Venemans2020,Novak2019} where the FIR luminosity is smaller. We caution that the SFR-\cii\ relations used above are dependent on the FIR luminosity and the FIR surface brightness of the galaxy populations used to calibrate them \citep[e.g.,][]{Diaz-Santos2017,Herrera-Camus2018a}. These SFRs values should only be used to compare with other high--redshift sources where the FIR is not well constrained. For J1148+5251, the [CII]/FIR ratio is  $\simeq (7.8\pm 1.3)\times 10^{-4}$. We thus use the appropriate “high $\Sigma_{\rm{FIR}}$" \cii-SFR calibration from \citet[][]{Herrera-Camus2018a} and find $\rm{SFR}=3100\pm1600\ \rm{M}_\odot\,\rm{yr}^{-1}$, in agreement with the other estimates presented above.

Following \citet[][]{Weiss2003,Weiss2005}, in the optically thin limit, the total mass of ions $X$ emitting photons with rest--frame frequency $\nu_{ij}$ stemming from a transition between the $i$ and $j$ levels can be derived from the observed luminosity as
\begin{equation}
    M_{X} = m_{X} \frac{8\pi k\nu_{ij}^2 }{hc^3 A_{ij}} Q(T_{\rm{ex}}) \frac{1}{g_i}e^{-T_{\rm{ex}}/T_i} L_{[X_{i\rightarrow j}]}'
    \label{eq:mass_ion}
\end{equation}
where $k$ is the Boltzmann constant, $c$ the speed of light, $h$ the Plank constant, $m_{X}$ is the mass of a single ion $X$, $Q(T_{\rm{ex}})= \sum_i g_i e^{-T_i/T_{\rm{ex}}}$ is the partition function of the species, with $g_i$ the statistical weight of level $i$, $T_i$ the energy of (above--ground) level $i$, $T_{\rm{ex}}$ is the excitation temperature of the $i\rightarrow j$ transition, and $L_{[X_{ij}]}'$ is the observed integrated source brightness temperature of the line in $K \,\kms pc^{-2}$. Note that Eq. \ref{eq:mass_ion} neglects heating from the CMB which is negligible for the \oi, \cii\ and \nii\ transitions, but not \ci\ at $z\sim 6.4$. For $\rm{C}^{+}$,  Eq. \ref{eq:mass_ion}  reduces to 
\begin{equation}
    M_{C^{+}} / M_\odot = 2.92\times10^{-4} Q(T_{\rm{ex}})\frac{1}{4}e^{-91.2/T_{\rm{ex}}} L_{\rm{[C~II]}}'
\end{equation}
where $Q(T_{\rm{ex}})= 2+4e^{91.2/T_{\rm{ex}}}$ is the \cii\ partition function. The optically thin limit assumption, while widespread in the literature, is uncertain at higher--redshift since optical depths measurements, although pointing to moderate values, are scarce \citep[e.g.,][]{Neri2014,Gullberg2015}. \citet{Lagache2018} and \citet{Vallini2015} note that in the optically thick limit, only emission from PDRs would reach the observer. As we will show in Section \ref{sec:cloudy}, most of the \cii\ emission in J1148+5251 comes indeed from PDRs, and hence we do no expect the optically thin limit assumption to affect our results. Assuming an excitation temperature $T_{\rm{ex}}=50\rm{K}$ \citep[from the CO modelling, ][]{Riechers2009}, we find $M_{C^{+}} = (5.8\pm 0.3) \times10^{7}M_\odot$. This is five times the neutral carbon mass ($M_{C} = 1.1\times10^{7} M_\odot$) measured by \citet[][]{Riechers2009}. 

\subsection{\nii\ emission}
\label{sec:nii_emission}
We report a \nii\ marginal detection in J1148+5251 at SNR$=3.7$ and an integrated ($r=3"$) luminosity $L_{[NII]} = (0.4\pm0.2)\times10^{9} L_\odot$. This is in slight tension with the $3 \sigma$ limit ($<0.4\times10^{9} L_\odot$) reported by \citet[][]{Walter2009a} using the IRAM 30m telescope. Nonetheless, we would expect better image fidelity with the new NOEMA facility. We consider the \nii\ detection marginal ($3.7\sigma$) and urge caution in interpretations that rely upon it.

The marginal ($1.59" / 8.9 \rm{kpc}$ $\sim$ 1 beam) offset between \nii\ and \cii\ in J1148+5251 could be of interest. On the one hand, spatial offsets between low and high--ionization lines have been reported in  other high--redshift quasars, most often between \cii\ and [O III]$_{88\,\mu\rm{m}}$ \citep[e.g.,][]{Novak2019}. Spatial offsets have also been predicted in theoretical simulations \citep[][]{Katz2017,Katz2019} where they arise from different gas phases with different temperatures and densities. Although nitrogen and carbon have a similar ionization level, \citet[][]{Katz2019} show that the \nii\ and \cii\ can arise from different gas phases, with \cii\ originating in lower temperature and higher density regions\footnote{Although a minor fraction of \cii\ traces the same H{~\small II} regions as \nii, see Sec \ref{sec:cloudy}.}, which could explain the offset of \nii\ towards the outskirts of J1148+5251. On the other hand, offsets often seen at low--resolution in high--redshift galaxies can disappear in higher--resolution data once fainter emission components are detected \citep[e.g. see HZ10 in ][]{Pavesi2016,Pavesi2019}. 
Following \citet[][]{Ferkinhoff2010,Ferkinhoff2011}, we derive the minimum $H^{+}$ for the \nii\ luminosity observed. To do so, we assume high densities and a high temperature (as found around O and B stars), such that all nitrogen in the HII regions is ionized. We use Eq. \ref{eq:mass_ion} to derive the mass of $N^{+}$ in J1148+5251 from the observed luminosity, with A$_{10}=2.1\times10^{-6}$ the Einstein coefficient of the $^{3}P_1 \rightarrow ^{3}P_0$ transition, $g_1=3$ the statistical weight of the $^{3}P_1$ emitting level, $\nu_{10} = 1461.1 \,\rm{GHz}$ the rest--frame frequency, and $g_t\simeq9$ the partition function. We can then derive the minimum H$^{+}$ mass by assuming the upper limit on the ionized nitrogen to ionized hydrogen ratio (i.e. $\chi(N^{+}) = N^{+} / H^{+}$) to be the total nitrogen abundance ratio $\chi(N)$, such that
\begin{equation}
 M(H^{+}) \geq M_{N^{+}} \frac{m_N}{m_H \chi(N^{+})} = \frac{L_{[NII] 205\mu\rm{m}} }{\frac{g_1}{g_t} A_{10} h \nu_{10} }\frac{m_H}{\chi(N^{+})} \text{\,\,\, ,} 
\end{equation}
We adopt the abundance value for HII regions $\xi(N)=9.3\times10^{-5}$ from \citet[][]{Savage1996}. Hence we estimate an ionized hydrogen mass $M(H^{+}) \geq (2.1\pm1.0) \times 10^{9} M_\odot (2\sigma \,\rm{level})$. Using the H$_2$ gas mass from the CO luminosity \citep[][]{Riechers2009}, the ionized--to--molecular gas ratio is $M(H^{+})/ M(H_2) > 0.1$. This is significantly higher than what is found by  \citet{Ferkinhoff2011} (based on the compilation by \citet[][]{Brauher2008}) for local galaxies  using the [NII] 122 $\mu \rm{m}$ line ($<0.01$). We caution here again that the \nii\ line is marginal.

\begin{figure}
    \includegraphics[width=0.5\textwidth]{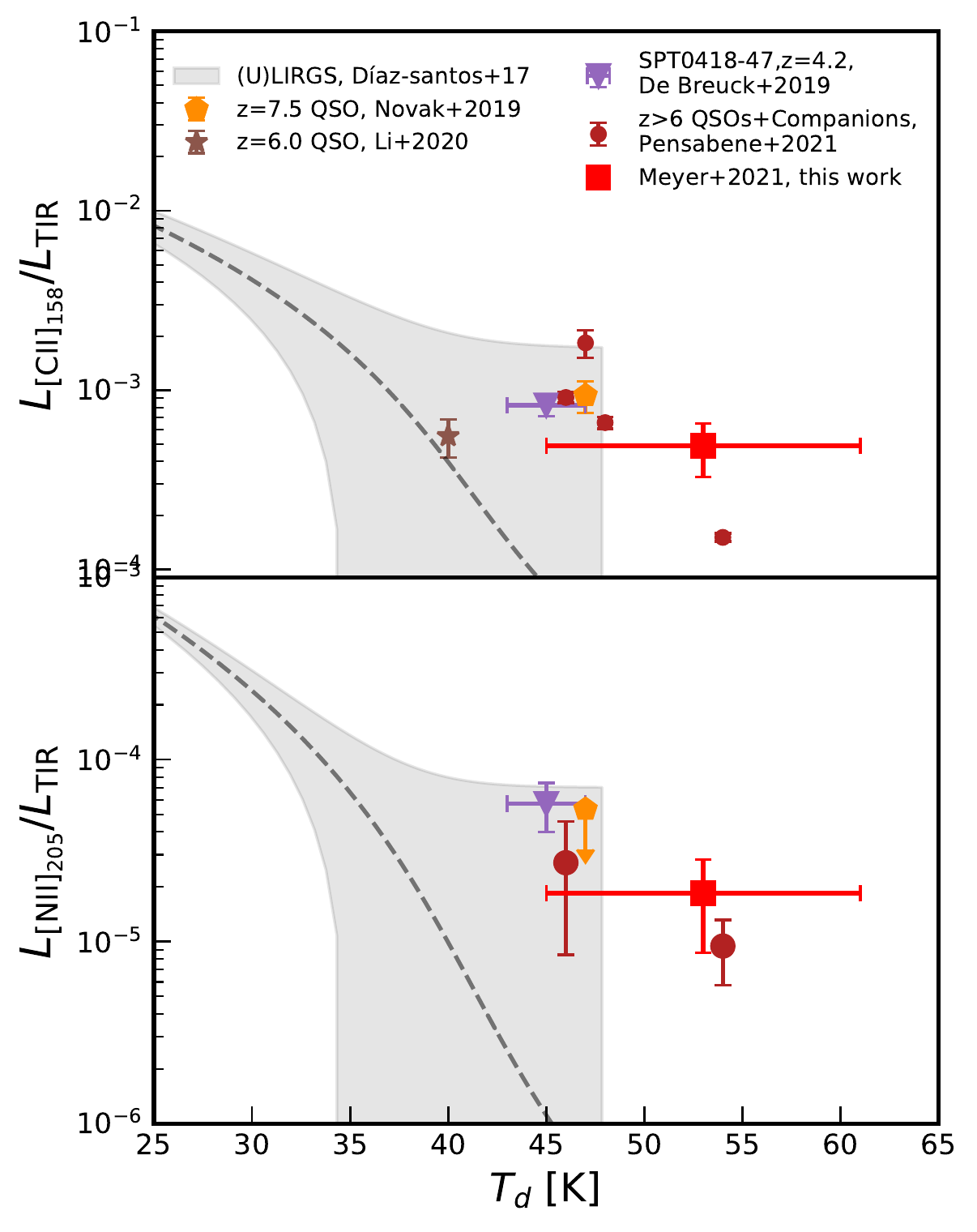}
    \caption{\cii\ and \nii\ line deficits of local (U)LIRGS \citep[shaded grey][]{Diaz-Santos2017}, high--redshift galaxy SPT0418--47 \citep[purple triangle][]{DeBreuck2019}, and high--redshift quasars \citep[orange pentagon, dark orange star, dark red circles][ respectively]{Novak2019,Li2020b,Pensabene2021} and companion galaxies \citep[dark red circles][]{Pensabene2021}. Our measurements for J1148+5251 are shown in red squares.}
    \label{fig:deficits}
\end{figure}

The [CII]/FIR ratio ($\simeq (7.8\pm 1.3)\times 10^{-4}$), [NII]/FIR ratio ($\sim (2.9\pm 1.5)\times 10^{-5}$) and dust temperature ($53\pm 8$ K) are all in agreement with the ratio trends (extrapolated to high temperature observed in J1148+5251) observed in local ULIRGS \citep[][]{Diaz-Santos2017} and high--redshift galaxies or quasars \citep[][see Figure \ref{fig:deficits}]{Pavesi2019, DeBreuck2019,Novak2019,Li2020b, Pensabene2021}. We note that the [CII]/[NII]  luminosity ratio ($10.6\pm 3.7$) is slightly low for dense PDRs/XDRs regions expected around high--redshift quasars \citep[e.g.,][]{Decarli2014,Pavesi2019}, though the uncertainties are large given the marginal detection in the \nii\ line. Finally, we also test \citet[][]{Zhao2016} \nii--SFR scaling relation derived from local ULIRGS which gives $\rm{SFR}= 794\pm274\  \rm{M}_\odot\,\rm{yr}^{-1}$, lower than the \cii\ and FIR-derived values. We note however that \nii\ is a poor tracer of the SFR in intense star-forming regions such as J1148+5251 due to its low critical density \citep[e.g., ][]{Zhao2016,Herrera-Camus2016}.

\subsection{\oi\ emission}
The peak of the \oi\ emission is located at the same position as the \cii\ emission. The \oi/FIR ratio is $ 1.3\times 10^{-5}$  in good agreement with that of local ULIRGS with comparable FIR surface flux density \citep{Herrera-Camus2018}. The \oi/\cii\ line luminosity ratio $L_{\rm{[OI]}} / L_{\rm{[CII]}} = 0.10 \pm 0.07 $ is typical of that observed in high-redshift SMGs and quasars \citep[e.g.,][]{DeBreuck2019,Yang2019,Li2020b,Lee2021}, and is at the higher-end of the range spanned by local AGNs \citep[e.g.,][]{Fernandez-Ontiveros2016}. 

Using again Eq. \ref{eq:mass_ion} for the \oi\ $^{3}P_0\rightarrow ^{3}P_1$ transition ($T_0=329 $ K, $g_0=1$, $Q(T_{\rm{ex}}) = 5+ 3e^{-329/T_{\rm{ex}}} + e^{-228/T_{\rm{ex}}}$), we derive a neutral oxygen mass 
\begin{equation}
    M_{O} / M_\odot = 6.19\times10^{-5} Q(T_{\rm{ex}})e^{329/T_{\rm{ex}}} L_{[OI]}' \text{\,\,\,\, .}
\end{equation} 
Assuming an excitation temperature $T_{\rm{ex}}=50\,\rm{K}$ \citep[from the CO modelling, ][]{Riechers2009}, we find $M_{O} = (9 \pm 5) \times10^{8}M_\odot$. This estimate is extremely sensitive to the excitation temperature, e.g. ranging from $0.7\times 10^7 M_\odot$ at $200$\,K to $8\times 10^8 M_\odot$ at $T_{\rm{ex}}=50 K$. Note that the above expression only applies in the optically thin limit which is probably not the case of \oi, therefore underestimating the neutral oxygen mass.
\newpage
\subsection{CLOUDY modelling of the fine--structure line ratios}
\label{sec:cloudy}
We now use the FSL ratios to determine the physical properties of the ISM of J1148+5251. In order to do so, we make use of CLOUDY \citep{Ferland2017}, a spectral synthesis code designed to simulate the spectra of astrophysical plasmas used to study both low and high redshift galaxies sub--mm lines. The grid of models used in this work were generated to study both high--redshift quasar hosts and their companions in \citet[][]{Pensabene2021}, which we refer to for further details. To summarise briefly, the model grids includes both PDR and XDR predictions for total hydrogen column densities $ N_{\rm{H}} / [\rm{cm}^{-2}]=10^{23}-10^{24}$, total hydrogen number density $1 < \log n / [\rm{cm}^{-3}] < 6$, and local FUV radiation field $2< \log G/[G_0]< 6$ in units of Habing flux (for the PDRs) or X-ray flux $-2 < \log F_X/[\rm{erg s}^{-1}\,\rm{cm}^{-2}<2]<2$ (for the XDRs). For each combination of parameters, the flux of various lines of interest is predicted and the ISM conditions can be determined by comparing to the observed line ratios.

Whereas \cii\ can originate from both the neutral and ionized gas phase, \oi\ and \ci\ originate solely in the neutral gas phase/PDRs. It is therefore crucial to determine which fraction of the \cii\ emission in J1148+5251 comes from the neutral gas. \nii\ and \cii\ have very close critical density in ionized media, therefore the \cii--to--\nii\ ratio is primarily a function of the N$^{+}$/C$^{+}$ abundance ratio \citep[e.g.,][]{Oberst2006}. Photoionization models \citep[e.g.,][]{Oberst2006,Pavesi2016,Croxall2017} predict a relatively constant ratio ($2.5-3$) for \nii/[CII]$^{\rm{ion}}_{158 \mu \rm{m}}$ for a large range of electron density. For consistency with the existing literature, we adopt \nii/[CII]$^{\rm{ion}}_{158 \mu \rm{m}} = 3$. Any deviation from that ratio can be interpreted as \cii\ emission from the neutral phase. The fraction of \cii\ emission originating in the neutral phase is therefore
\begin{equation}
    f(\rm{[CII],neutral}) \simeq 1-3\frac{L_{[\rm{NII} ]_{205 \mu \rm{m}}}}{L_{[\rm{CII}]_{158 \, \mu \rm{m}}} } \text{\,\,\,\,  .}
\end{equation}
The ratio of the luminosities measured in this work for J1148+5251 is $27\pm13$ for the aperture--integrated ($r=3"$) flux and $>100 (2\sigma \,\rm{level})$ at the peak of the \cii\ emission. In either case, most of the \cii\ emission ($88\%-97\%$) comes from the neutral phase.

\begin{figure}
    \centering
    \includegraphics[width=0.5\textwidth]{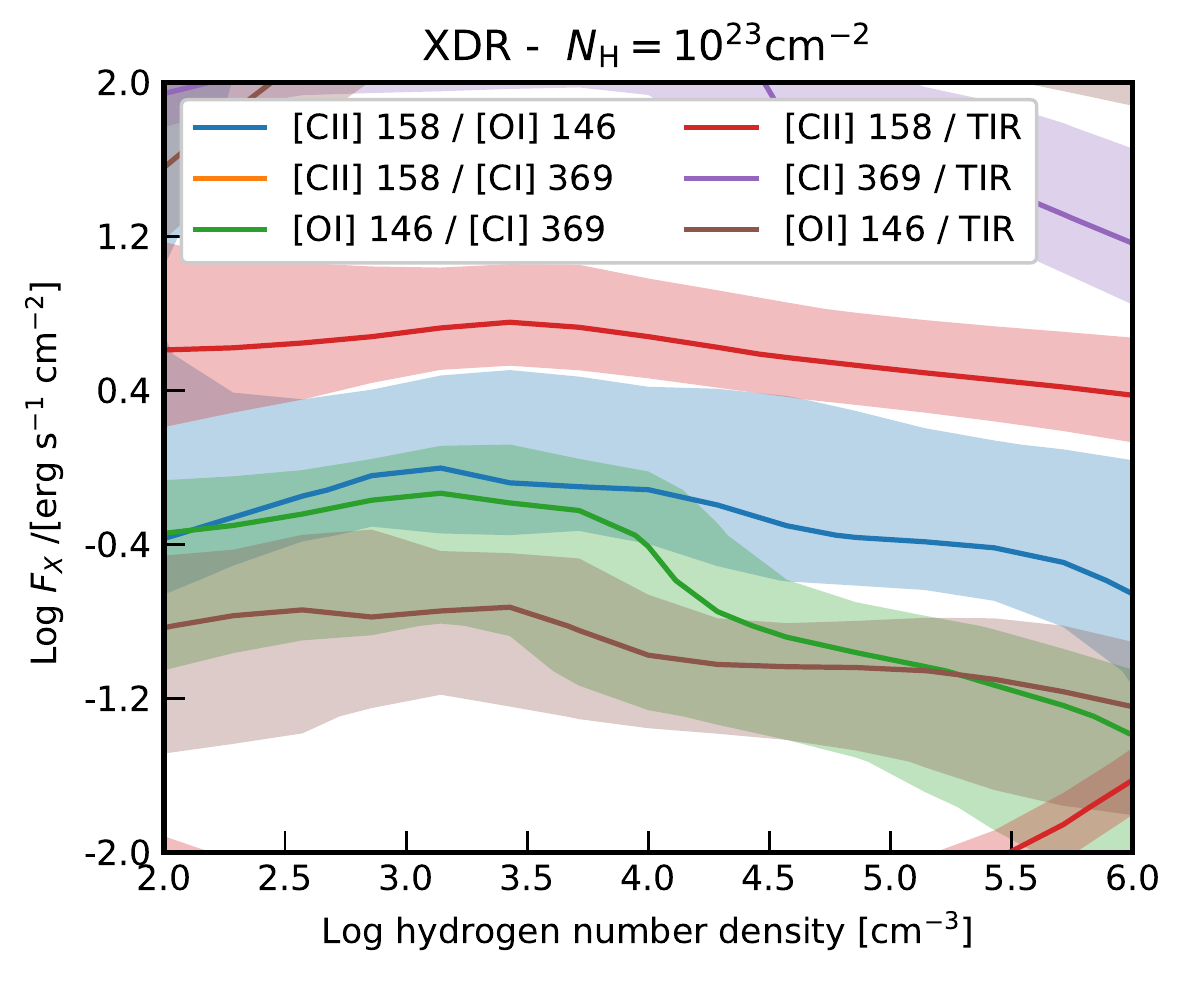} \\
    \includegraphics[width=0.5\textwidth]{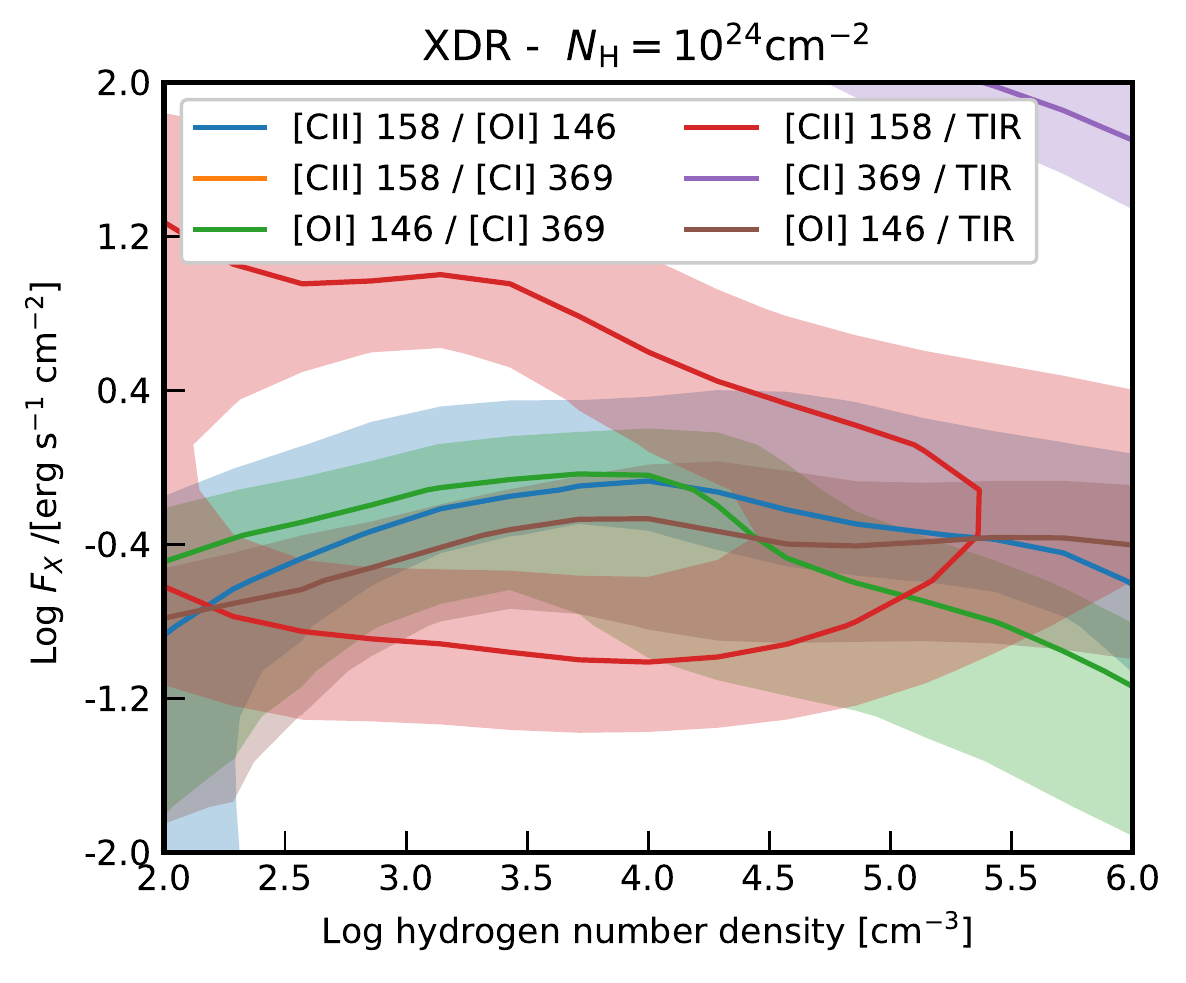}    \caption{ISM constraints for J1148+5251 from the FSL ratios predicted by CLOUDY for XDRs with hydrogen columns densities $N_{\rm{H}} / [\rm{cm}^{-2}]=10^{23}$ (top panel) and $N_{\rm{H}} / [\rm{cm}^{-2}]=10^{24}$ (lower panel). The observed ratio constraints from the aperture integrated fluxes are plotted in solid lines and the $\pm1\sigma$ values in shaded area of the same color. Ratios absent from the plots (such as \cii\ ratios) are not reproduced by the XDR grid of models. }
    \label{fig:CLOUDY_1}
\end{figure}

\begin{figure}
    \centering
    \includegraphics[width=0.5\textwidth]{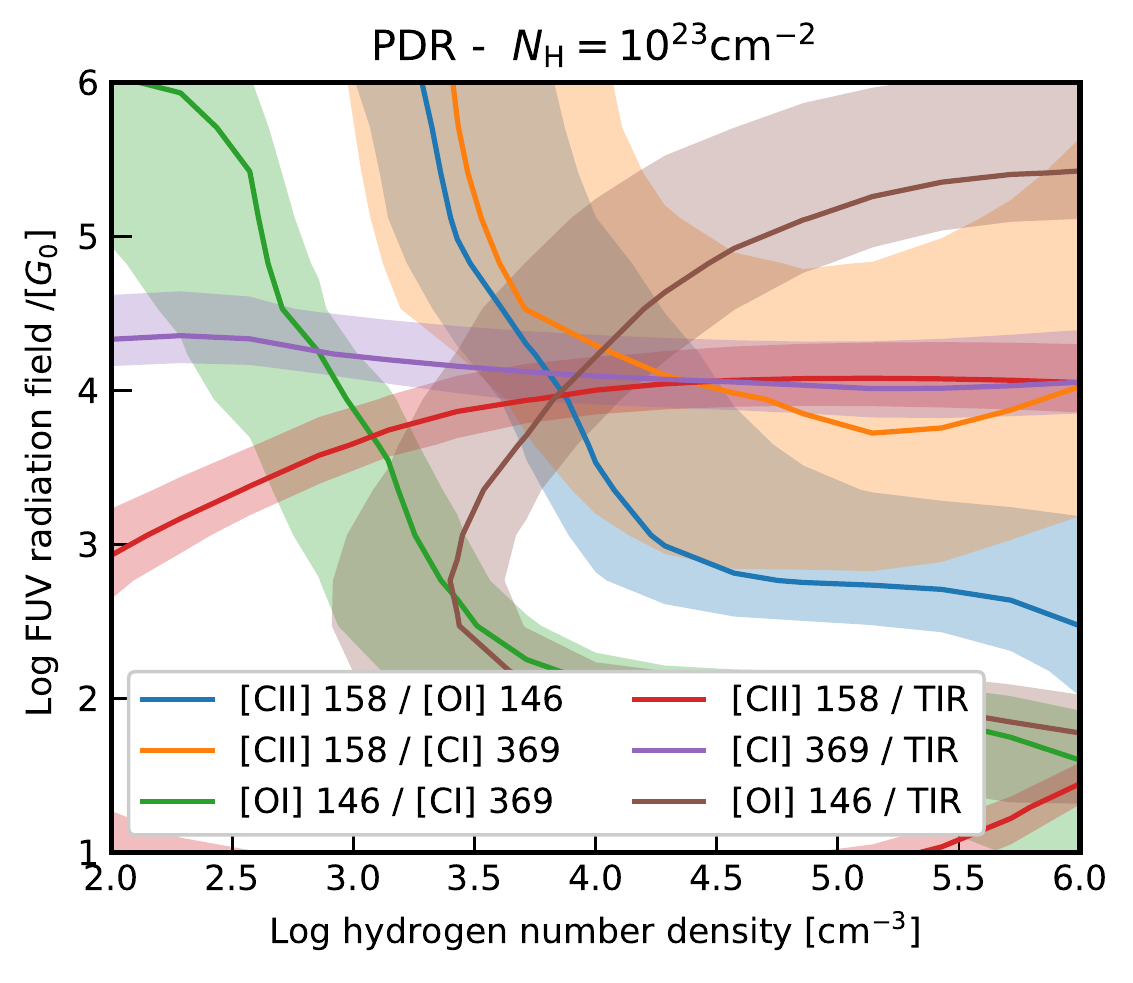} \\
    \includegraphics[width=0.48\textwidth]{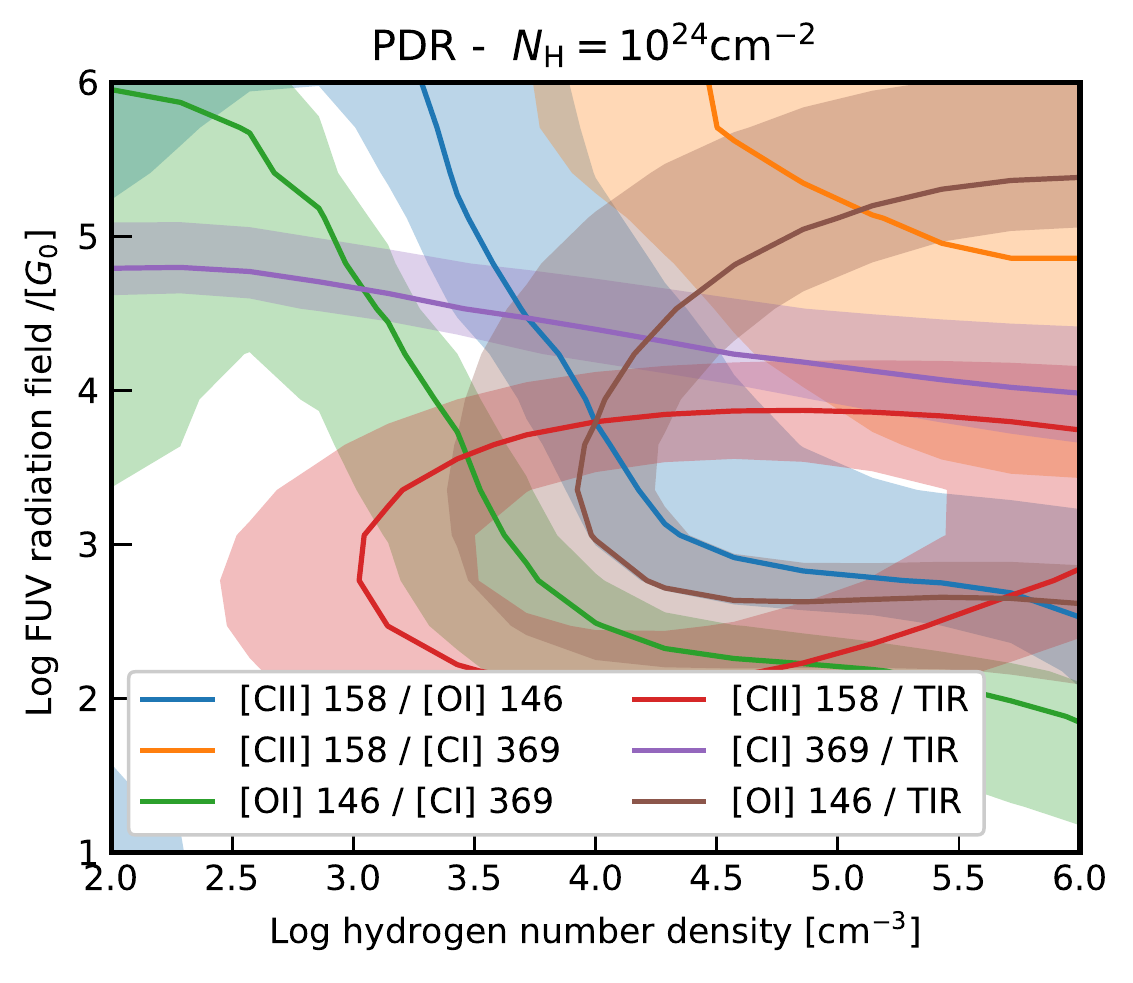} \caption{ISM constraints  from the FSL ratios predicted by CLOUDY for PDRs with hydrogen columns densities $N_{\rm{H}} / [\rm{cm}^{-2}]=10^{23}$ (top panel) and $N_{\rm{H}} / [\rm{cm}^{-2}]=10^{24}$ (lower panel). The observed ratio constraints from the aperture integrated fluxes are plotted in solid lines and the $\pm1\sigma$ values in shaded area of the same color.}
    \label{fig:CLOUDY_2}
\end{figure}

Given the resolution of the NOEMA data, we use the $r=3"$ aperture--integrated line luminosities to compute line luminosity ratios and compare to the grid of CLOUDY models. In order to include properly the \ci\ measurements from \citet[][]{Riechers2009}, we have repeated the line map procedure detailed in Section \ref{sec:obs} on the original \ci\ data. We produce a line map with width $\Delta v = 482 \, \kms$ and integrate the flux up to $r=3"$, finding $S_{\rm{[CI] 369}\mu m}\Delta v = 0.45\pm0.16\, \rm{Jy}\, \kms\,$ and a line luminosity $L_{\rm{[CI] 369}\mu m}\ = (0.20 \pm 0.07) \times 10^{9} L_\odot$. The new luminosity value is agreement, within $1.4\sigma$, with the previously published of $(0.10 \pm 0.02) \times 10^{9} L_\odot$ \citep[][]{Riechers2009}. The difference between the two measurements can be explained by the larger aperture, and is comparable to that seen for \cii\ (see Appendix \ref{sec:curve_growth}).

We show the radiation field and density predictions of CLOUDY for the observed FSL ratios in Figs.\ \ref{fig:CLOUDY_1} and \ref{fig:CLOUDY_2}. We find that the different line ratios are not perfectly reproduced for a single density and radiation field in the chosen grid of models. However, the \cii$^{\rm{neutral}}$ / \ci\ luminosity ratio ($47 \pm 3$) excludes XDR models which cannot reproduce such high ratios (the maximum being $\sim 15$) \citep[Fig. \ref{fig:CLOUDY_1}, e.g.,][]{Venemans2017,Venemans2017a,Novak2019,Pensabene2021}. Besides, our XDR model grid cannot reproduce all the FSL--to--FIR luminosity ratios with a single hydrogen number density and X-ray flux. We note that this result, based on the analysis of the fine structure lines, is in tension with \citet[][]{Gallerani2014} who concluded that an XDR component is needed in J1148+5251 to explain their reported CO(17-16) line. However, as noted by the authors, this line is potentially contaminated by a nearby OH$^{+}$ emission line, and thus additional observations of high--J CO lines are needed to clarify the situation.

The PDR models in agreement with the observed ratios (except \oi / \ci) have an HI column density of $N_{HI}=10^{23} \rm{cm}^{-2}$, high radiation fields ($10^{3.5-4.5}\,G_0$) and moderate hydrogen number densities ($n \simeq 10^{3.5-4.5}\, \rm{cm}^{-3}$), commensurable with other studies of high--redshift quasars \citep[e.g.,][]{Novak2019,Pensabene2021}. The model grids do not reproduce the observed \nii/\cii\ ratio since it was not created to model the ionized phase (H{\small\ II} regions) traced by \nii\ emission. The \oi / \ci\ discrepancy could stem from the different quality and resolution of the data, as well as the low SNR of both lines. Another possibility is that a fraction of the \cii\ emission could come from collisional excitation in the cold neutral medium, rather than solely from the PDRs as is the case for \oi\ and \ci, leading to a small discrepancy between ratios involving the \cii\ line and others in our PDR-only models. This hypothesis is supported by the spatial extent of the \cii\ line emission (Section \ref{sec:cii_extension}), which is beyond what is normally observed for the disk component of $z\sim6$ massive star-forming galaxies and other quasar hosts \citep[e.g.][]{Venemans2020}. Deeper and higher resolution of the inner and outer \cii\ emitting regions, combining the presence or absence of \oi, \nii\ and \ci, would be required to further investigate the ISM of J1148+5251. 

\section{Conclusions}
\label{sec:conclusion}
We report new NOEMA observations of the z=6.42 quasar J1148+5251 in the atomic fine structure lines of \cii, \oi\ and \nii\ and the underlying dust continuum emission. The high--fidelity data, together with the large instantaneous bandwidths that NOEMA provides, enabled us to revisit the properties of the quasar's host galaxy and derive the physical conditions of its interstellar medium. The main conclusions of this paper are as follows:    

\begin{itemize}
    \item  A \textit{uv} plane analysis confirms the presence of an extended \cii\ and dust emission component ($\rm{FWHM}=1.6"\pm0.2" \,(9\pm1 \rm{\, kpc})$) accounting for $\sim 50-60\%$ of the total \cii\ and most of the dust emission, in agreement with earlier studies \citep[][]{Maiolino2012,Cicone2015}. J1148+5251 thus remains an outlier with very extended \cii\ and dust emission compared to other $z>6$ quasars \citep[e.g.,][]{Venemans2020}.
    \item The \cii\ emission shows a similar spatial extent as the FIR dust when averaged radially (up to $\theta = 2.51^{+0.46}_{-0.25}\ \rm{arcsec}$, corresponding to $r= 9.8^{+3.3}_{-2.1}$ physical kpc,  accounting for the beam convolution). However, if the \cii\ emission is examined only along its NE--SW axis, a significant ($5.8\sigma$) excess \cii\ emission (w.r.t. to the dust) is detected.
    \item The \cii\ line profile can be fitted with a single Gaussian with a FWHM of $408\pm23 \, \kms$. The new NOEMA data has enabled us to rassess the significance of the broad \cii\ line wings reported in previous studies. We find that an additional broad \cii\ component is not statistically required to describe the NOEMA data (or the merged NOEMA and previous PdBI data).
    \item We report the detection of \oi\ and \nii\ (tentatively) in J1148+5251. Using various empirical relations, we report a $C^{+}$ mass of $M_{C^{+}} = (2.7\pm0.1)\times 10^{7}M_\odot$, an oxygen mass of $M_{O} = (12 \pm 4) \times10^{8}M_\odot$, and a lower limit on the ionized hydrogen mass $M_{H^{+}} > (4.6\pm0.16) \times 10^{8} M_{\odot} (2\sigma \,\rm{level})$.
    \item The FSL line ratios are consistent with the trends observed in local (U)LIRGs. We find that a large fraction ($\sim 90\%$) of the \cii\ emission originates in the neutral phase (PDR) of the gas. 
    \item We have compared CLOUDY models to the observed FSL ratios in J1148+5251. The \cii /\ci\ ratio and the multiple FSL--to--FIR ratios exclude XDR models in favour of PDRs. We find good agreement for models that have a high radiation field ($10^{3.5-4.5}\,G_0$), a moderate hydrogen number densities ($n \simeq 10^{3.5-4.5}\, \rm{cm}^{-3}$) and HI column density $N_{HI} = 10^{23} \rm{cm}^{-2}$.
 \end{itemize}

The results described here highlight the importance of large instantaneous bandwidths when observing high--redshift quasars (or galaxies) to search for weak extended emission of atomic or molecular lines. Our findings enabled a renewed view on the host galaxy of the J1148+5251 quasar shedding light on the feedback activity and providing new constraints on the excitation conditions of its interstellar medium that appear similar to what is found in local ULIRGs. Higher angular resolution and sensitivity data, in particular in the \ci\ emission line, would be required to put additional constraints on the gas density and the ionisation source of J1148+5251 and further explore its properties.  

\acknowledgments
The authors thank the anonymous referee for useful suggestions which improved the manuscript. R.A.M., F.W., M.N. acknowledge support from the ERC Advanced Grant 740246 (Cosmic\_Gas). D.R. acknowledges support from the National Science Foundation under grant Nos. AST--1614213 and AST--1910107. D.R. also acknowledges support
from the Alexander von Humboldt Foundation through a Humboldt Research
Fellowship for Experienced Researchers. 

This work is based on observations carried out with the IRAM NOrthern Extended Millimeter Array (NOEMA). IRAM is supported by INSU/CNRS (France), MPG(Germany), and IGN (Spain). The authors would like to thank wholeheartedly the staff at IRAM for their help with the reduction and analysis software GILDAS; and in particular M. Krips, V. de Souza, C. Herrera and J.-M. Winters.

\facilities{ NOEMA (IRAM)} 

\software{astropy \citep{TheAstropyCollaboration2018},  
          CLOUDY \citep{Ferland2017}, 
          Numpy \citep{Numpy2020}, Scipy \citep[][]{Virtanen2020}, Matplotlib \citep[][]{Hunter2007},
          Interferopy \citep*[][]{interferopy}
          }

\vspace{5cm}

\appendix
\section{Curve of growth analysis}
\label{sec:curve_growth}
\begin{figure}[h]
    \centering
    \includegraphics[width= 0.7\textwidth]{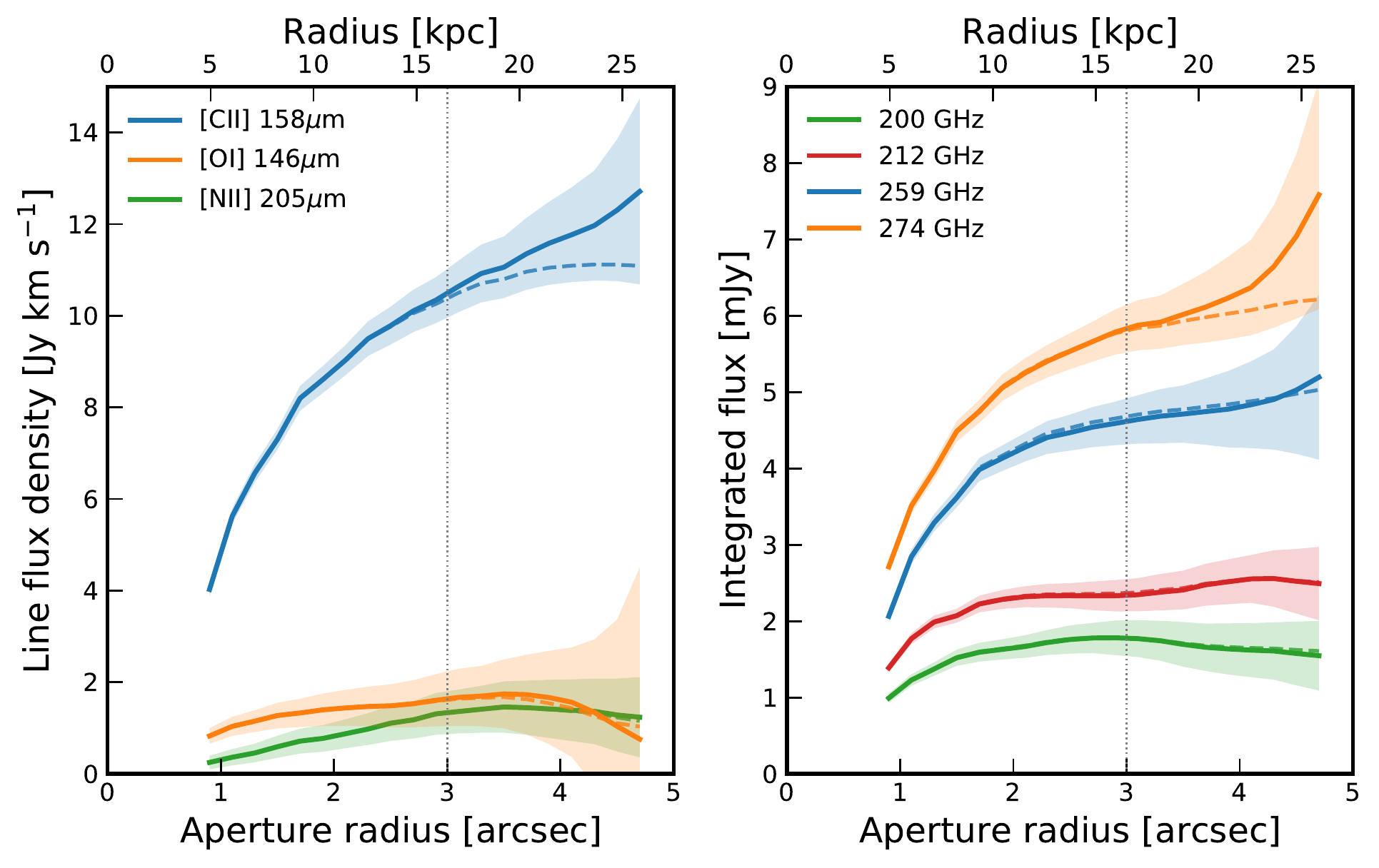}
    \caption{Fine--structure line flux densities (left panel) and integrated continuum flux (right panel) as a function of aperture radius. The solid and dashed lines show the fluxes with and without residual scaling correction (see Sec. \ref{sec:obs}). The final aperture radius of $3"$ was chosen to encompass all of the \cii\ emission and the continuum at the higher frequencies.}
    \label{fig:growing_apertures}
\end{figure}

In this appendix, we investigate what is the size of the aperture that is necessary to recover most of the continuum or \cii\ line fluxes. Figure \ref{fig:growing_apertures} shows the line flux density of \cii, \oi\ and \nii\   as well as the continuum flux density as a function of increasing aperture radius. We find that all line/continuum fluxes reach a maximum or plateau at an aperture radius $r=3"$, which corresponds to $16.9$ kpc at $z=6.42$. The \cii\ flux presents tentative evidence for additional flux up to $4"$ ($\sim 1\, \rm{Jy\,} \kms$), but this is within the $1\sigma$ errors. Note that at large radii where there is no more cleaned flux, residual-scaling can become numerically unstable. Throughout this paper, an aperture of $r=3"$ is therefore adopted for measurements unless specified otherwise.

\section{Multiscale and H\"ogbom cleaning methods}
\label{sec:cleaning}
In this appendix, we briefly detail our experiments with different cleaning methods for our interferometric data. H\"ogbom cleaning \citep[][]{Hogbom1974} is one of the standard method for cleaning interferometric data. It relies on iteratively finding peaks in the data and subtracting the dirty beam at that location until the residuals reach a desired level. It is particularly efficient for point sources, but struggles with large--scale emission which it tries to reconstruct using a multitude of point sources. In that case, so--called multi--scale algorithms which convolve the beam with various Gaussians to subtract larger scales are preferable \citep[e.g.,][]{Wakker1988,Cornwell2008}. 

\begin{figure*}
    \centering
    \includegraphics[width = 0.8\textwidth]{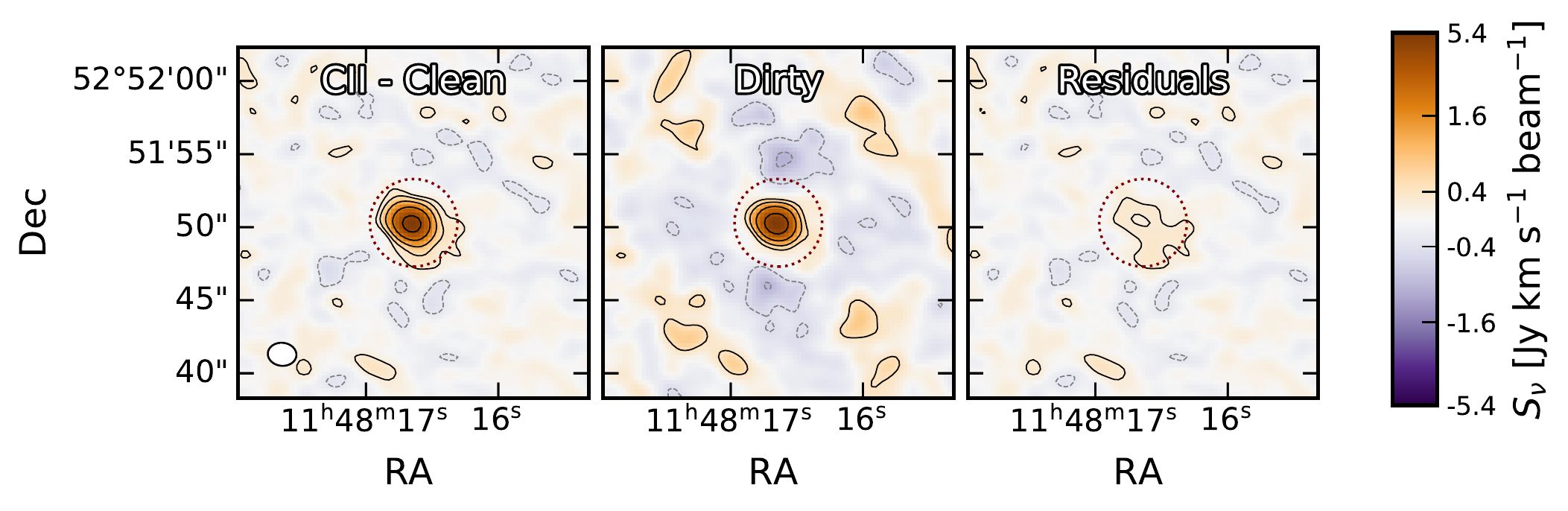}
    \includegraphics[width=0.8\textwidth]{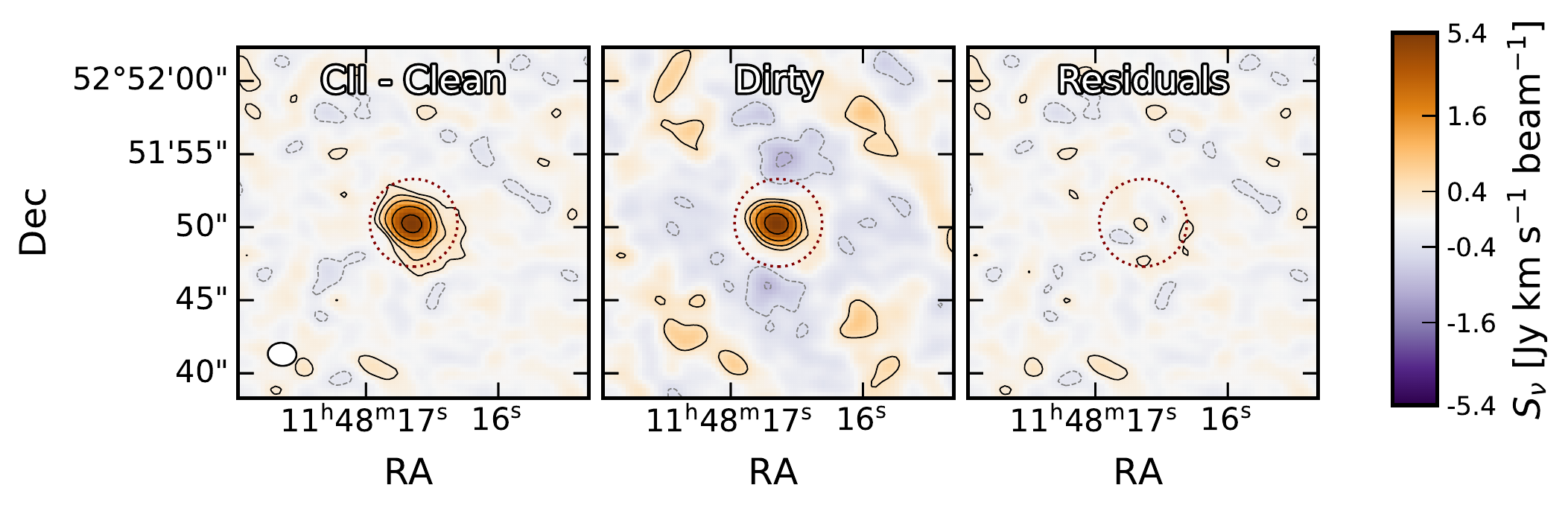}
    \caption{Clean, dirty and residual maps of the \cii\ emission for two different cleaning algorithms: H\"ogbom (first row) and Multiscale (second row). Both are cleaned down to $2\sigma$, where $\sigma$ is the RMS noise of the dirty map.}
    \label{fig:cii_technical_clean}
\end{figure*}

Figure \ref{fig:cii_technical_clean} shows the clean map, dirty map and residuals for H\"ogbom and Multiscale clean on the \cii\ map (integrated over $482\,\kms$). Clearly, the H\"ogbom clean residuals show a flat excess of 2$\sigma$ flux filling the $3"$ aperture which would not be expected if the source was a single (or a limited number of) point source(s). On the contrary, the residuals of the Multiscale algorithm are closer to zero on average. Therefore, we chose the Multiscale algorithm for all \cii--derived quantities and images in this paper.

\begin{figure*}
    \centering
    \includegraphics[width=0.49\textwidth]{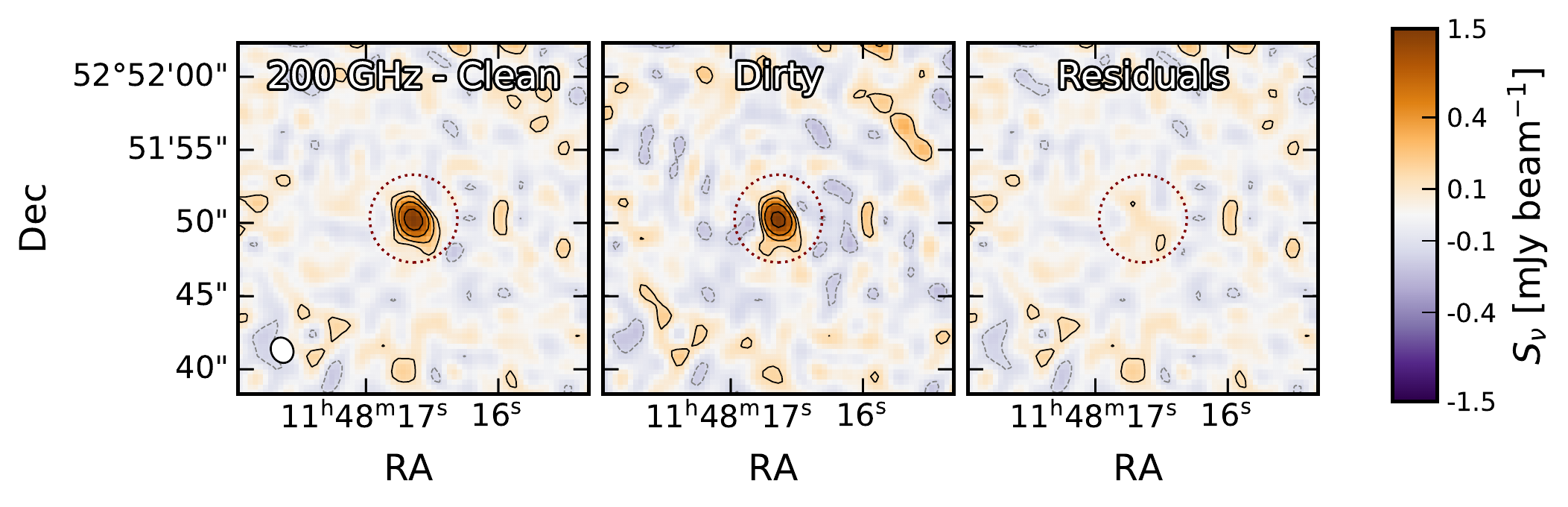}
    \includegraphics[width=0.49\textwidth]{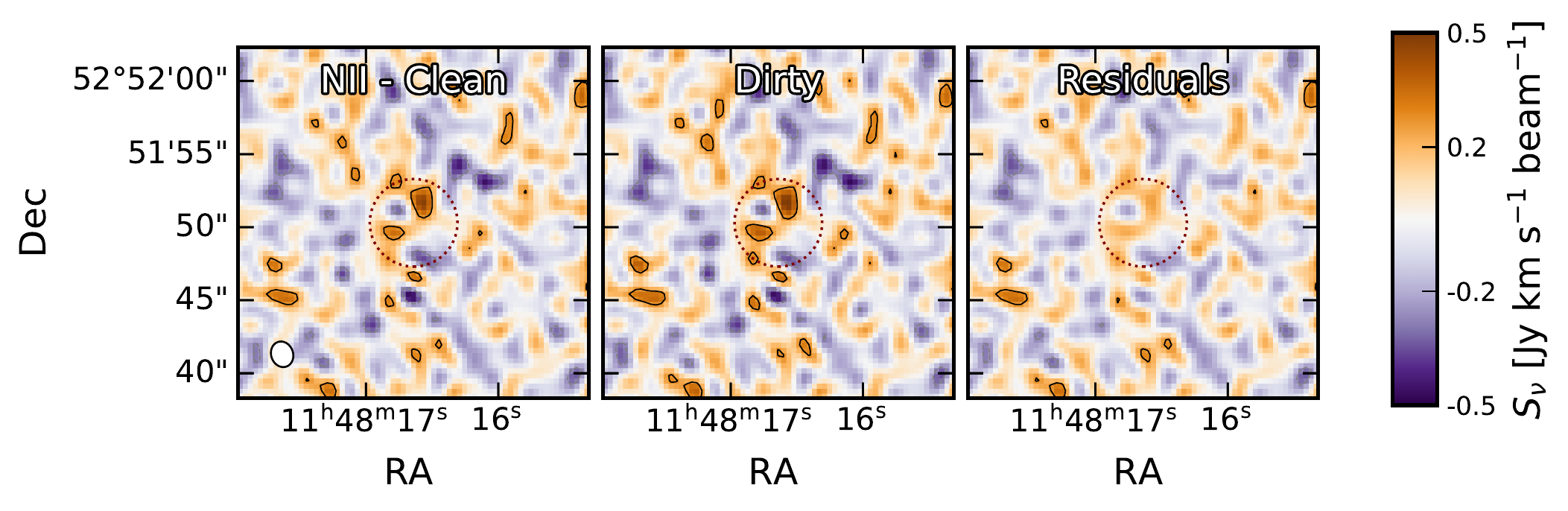}\\
    \includegraphics[width=0.49\textwidth]{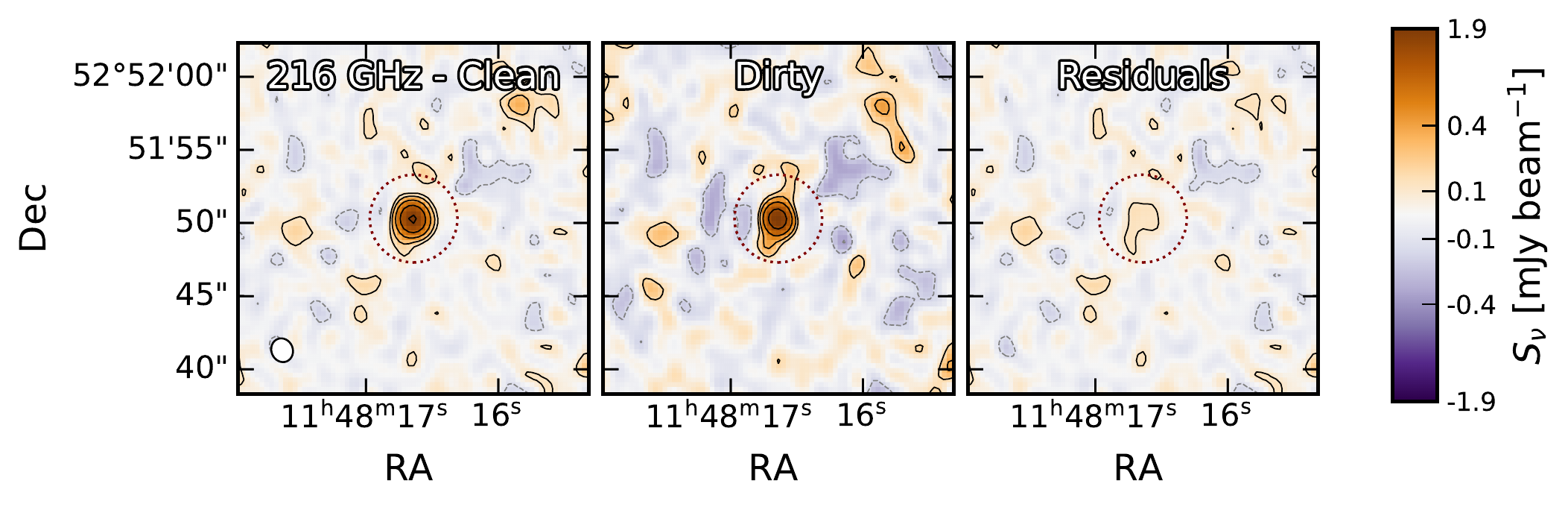}
    \includegraphics[width=0.49\textwidth]{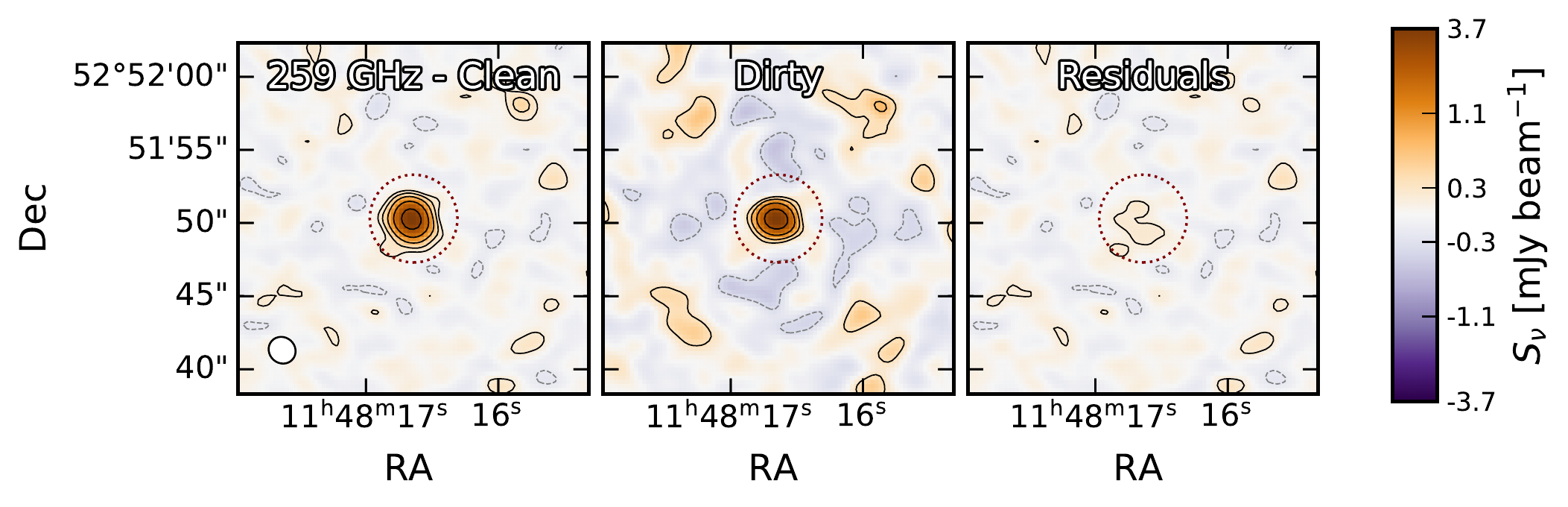} \\
    \includegraphics[width=0.49\textwidth]{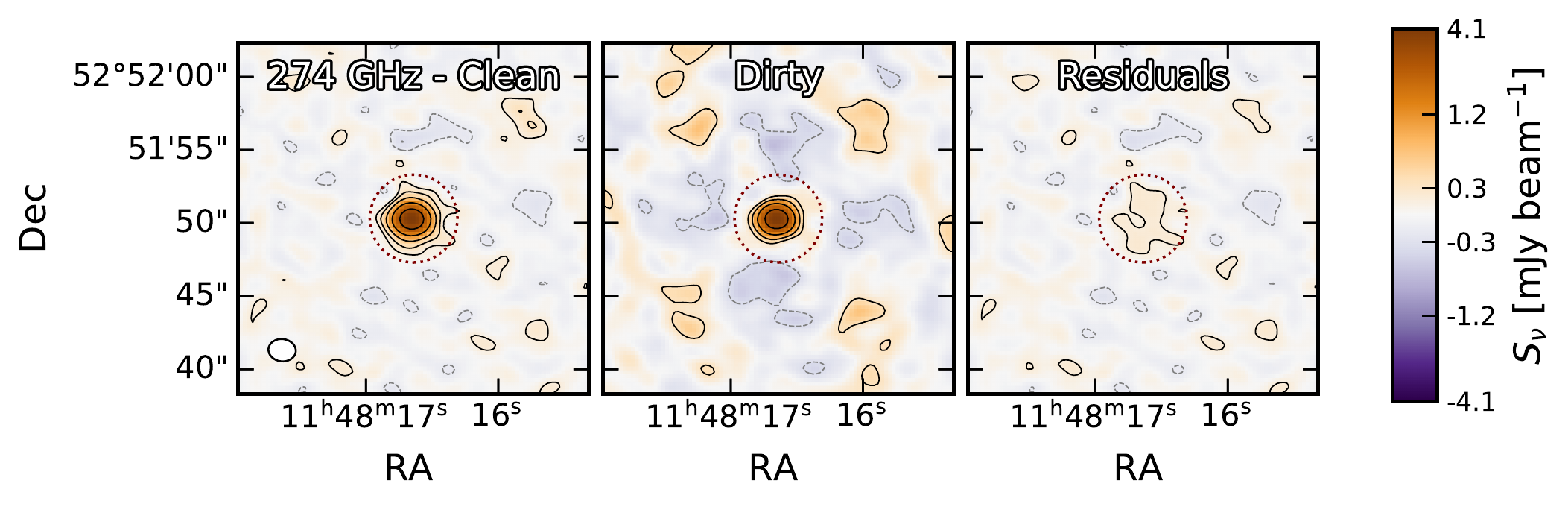}
    \includegraphics[width=0.49\textwidth]{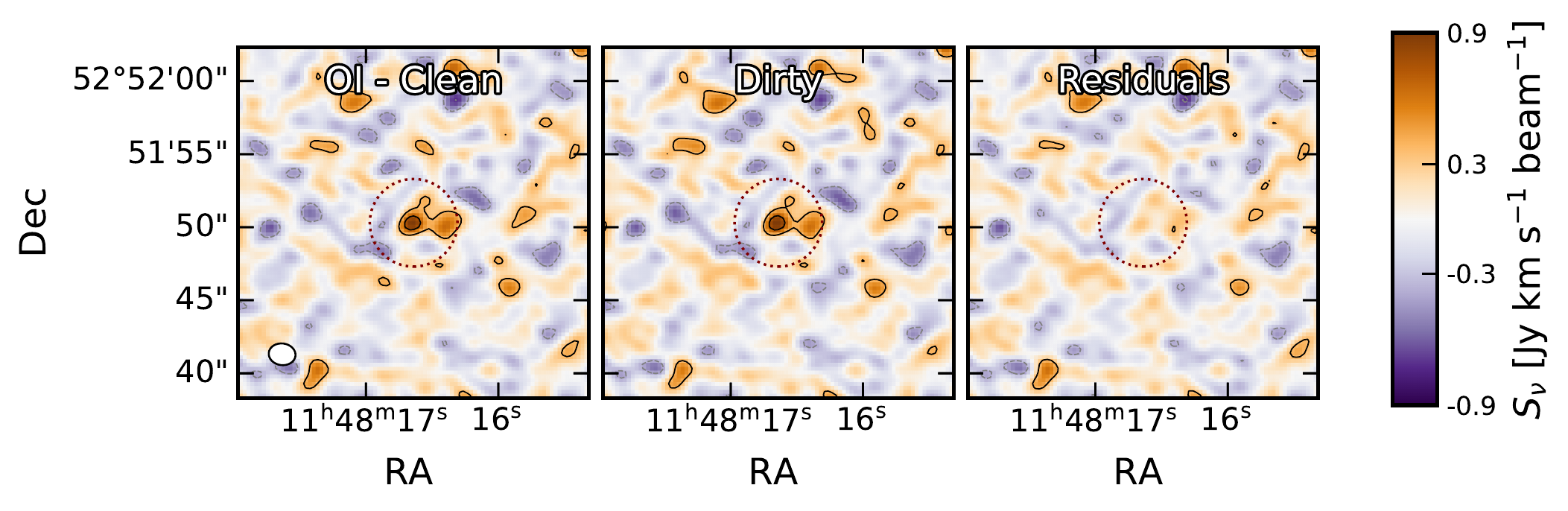}
    \caption{Clean,dirty and residual maps for a H\"ogbom clean to $2\sigma$ of the 200 GHz continuum, \nii\ emission, $216,259,274$ GHz continuum and finally \oi\ emission.}
    \label{fig:residual_technical}
\end{figure*}

Figure \ref{fig:residual_technical} shows the clean, dirty and residual maps for the continuum maps and the other FSL maps for a H\"ogbom clean. For the  lower frequency continuum maps, \oi, and \nii\   maps expect the residuals are well--behaved and do not require the use of multiscale cleaning. For the $259,274$ GHz continuum, some residuals are seen and the multiscale algorithm is adopted for those in the paper, albeit changing only the final continuum flux by $\lesssim 2\%$.

\section{Dust SED parameter posterior distribution}
\label{app:dust_sed_parameters}
\begin{figure}
    \centering
    \includegraphics[width=0.5\textwidth]{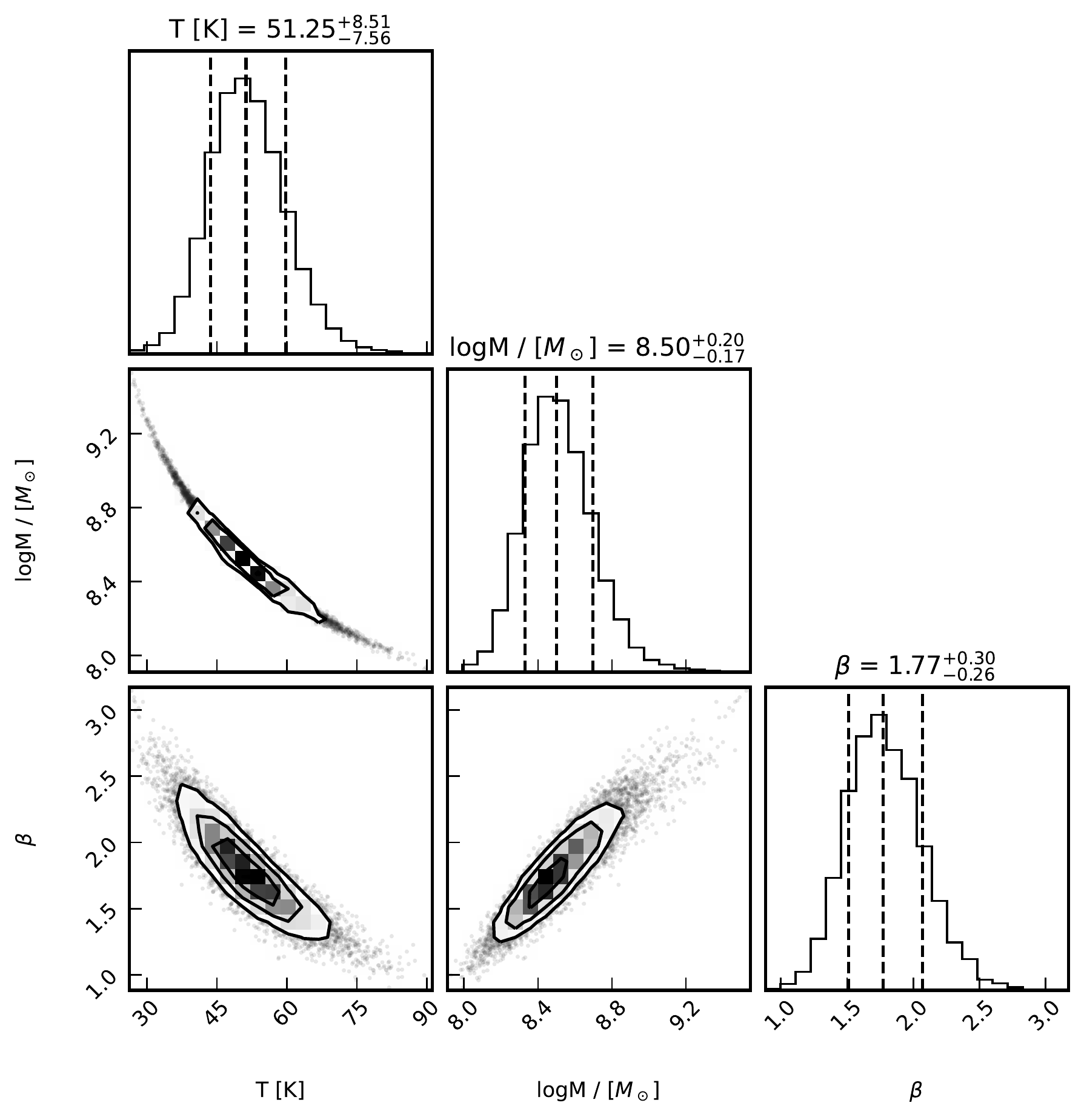}
    \caption{Dust SED fitting posterior distribution of the dust temperature $T$, dust mass $M$ and the dust spectral emissivity index $\beta$.}
    \label{fig:dust_sed_posterior}
\end{figure}

We present in Figure \ref{fig:dust_sed_posterior} the posterior distribution of the dust SED parameters fitted in Section \ref{sec:obs}. The median dust properties derived are consistent with the existing literature on high--redshift quasars \citep[e.g.,][]{Venemans2020}.

\section{\cii\ flux density}
\label{app:CII_fluxes}

\begin{figure*}
    \centering
    \includegraphics[width=0.25\textheight]{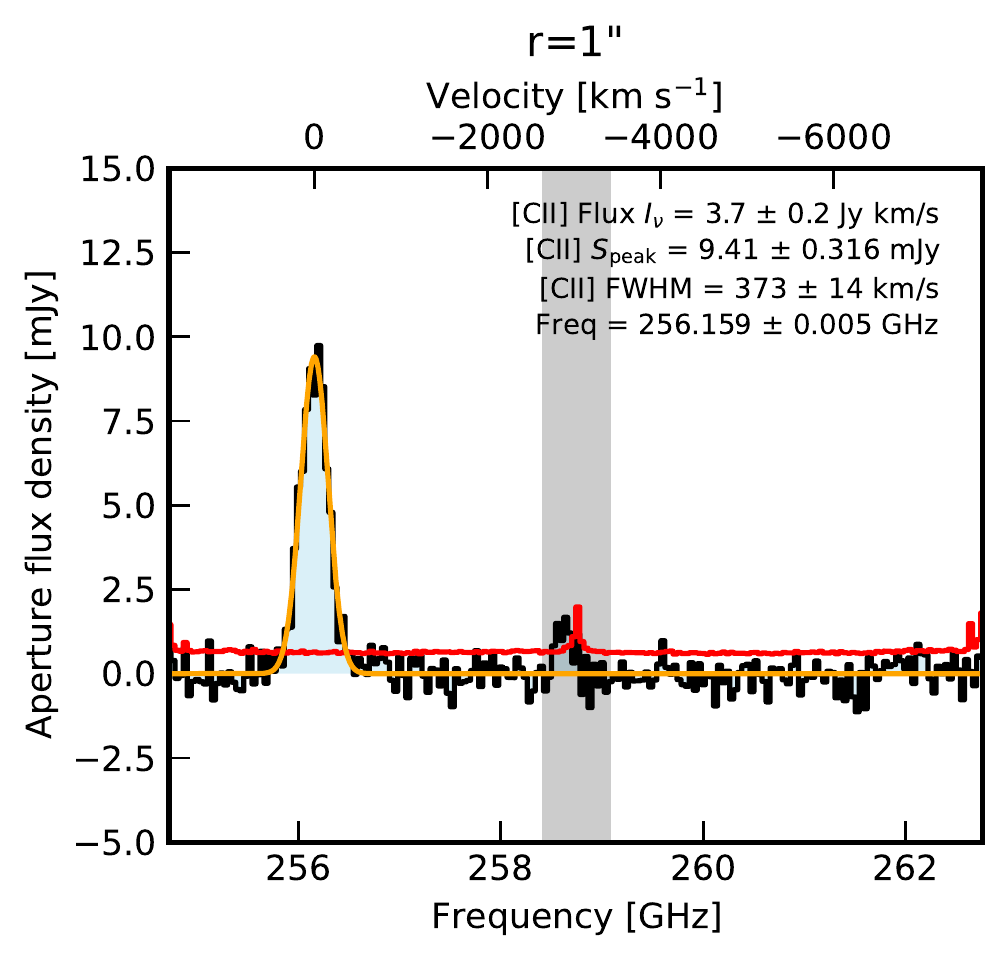}
    \includegraphics[width=0.25\textheight]{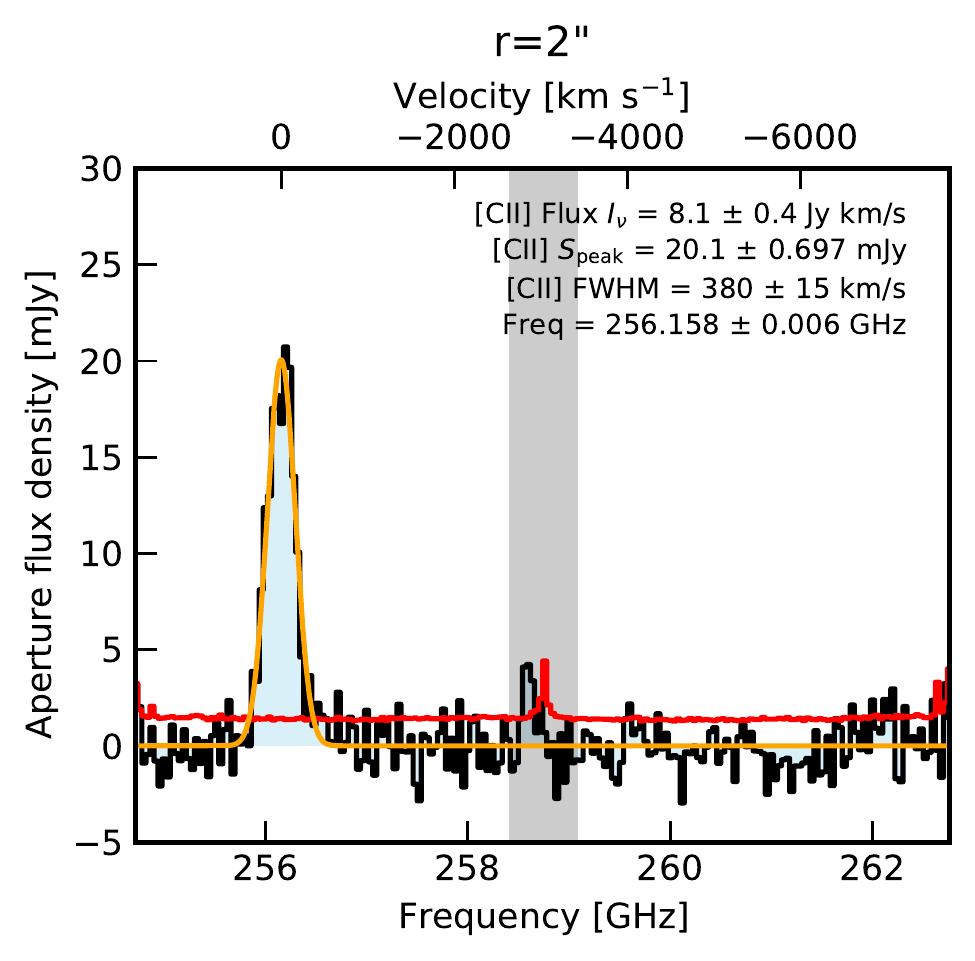}
    \includegraphics[width=0.25\textheight]{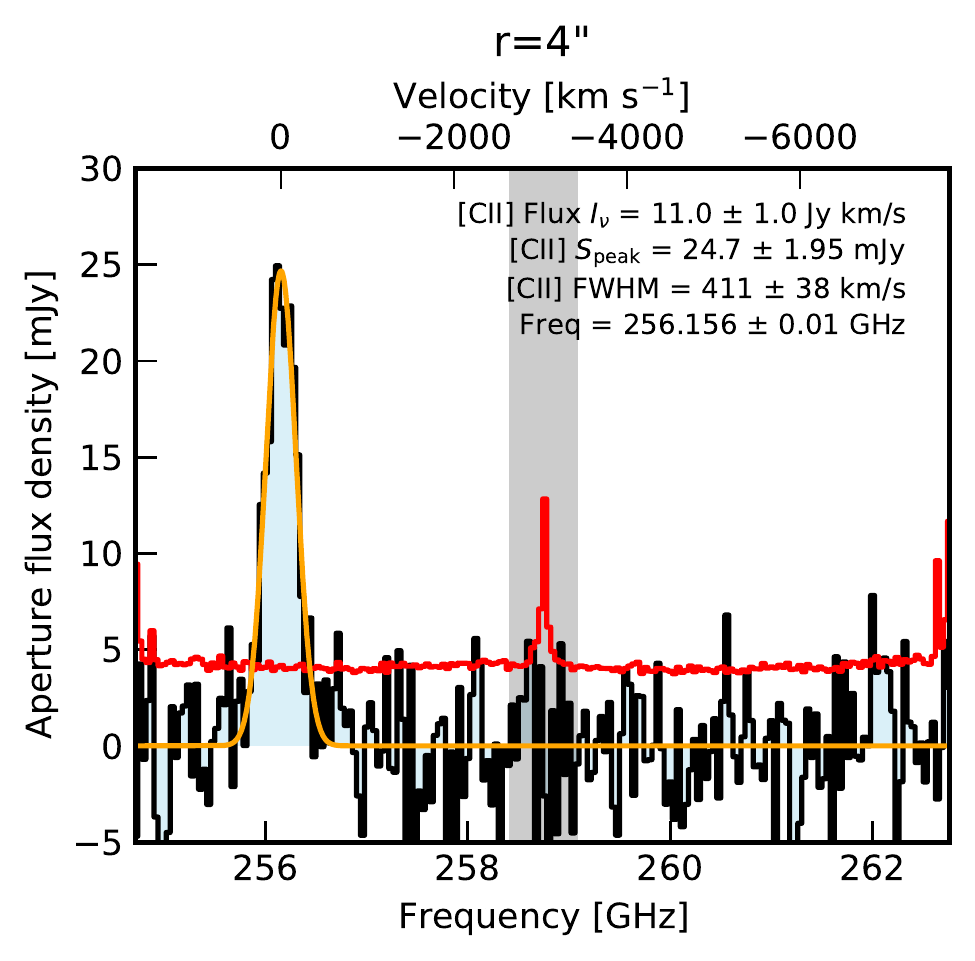}
    \caption{Continuum--subtracted aperture--integrated \cii\ spectra (black) for circular apertures with radii $r=1",2",4"$. The best fit Gaussian is shown in orange and the best--fit parameters are displayed in the upper--right corner of each plot.}
    \label{fig:cii_fits}
\end{figure*}

In table \ref{tab:CII_fluxes}, we compare the \cii\ fluxes measured from the extracted aperture--integrated \cii\ spectra for various aperture sizes between previous studies and this work. This work's spectra and best--fit Gaussians are displayed in Figure \ref{fig:cii_fits} for circular apertures with radii $r=1",2",4"$ and a central--pixel--only spectrum. The $r=3"$ case is already presented in the main text (Fig. \ref{fig:cii_fit}). For \citet[][]{Cicone2015}, we report their best--fit parameters and fluxes for the "narrow" $(-200,200)\,\kms$ component defined in their study. We find good agreement for all apertures between this work and previous studies. This implies that most of the excess flux in larger apertures ($r\gtrsim 3"$) described in \citet{Cicone2015} is  due to the reported presence of blue/red--shifted components with large spatial offsets which are not recovered in this work, as discussed in Section \ref{sec:cii}. 
\begin{table*}
    \centering
    \begin{tabular}{l|c|c|c|c|c|c|c|c|c}
    \hline \hline
    &\multicolumn{2}{c}{$r=1"$} & \multicolumn{2}{|c|}{$r=2"$} & \multicolumn{2}{|c|}{$r=3"$} & \multicolumn{2}{c}{$r=4"$} \\
     & C15 & This work & C15 & This work & M12 & This work & C15 & This work  \\ \hline
    $\sigma_v$ [km s$^{-1}$]  &$146\pm11$ & $158\pm6$ & $148\pm16$ & $162\pm4$ &  $150\pm 20$ &$171\pm10$  & $150\pm20$  & $175\pm16$ \\
    $S_{\rm{peak}}$ [mJy]  &  $14.5\pm0.9$  & $9.41\pm0.32$ & $30\pm3$  & $20.1\pm0.7$ &$23\pm 2$ & $23.5\pm 1.2$& $34\pm4$  & $24.7\pm2.0$\\
    $I_\nu$ [Jy $\kms$]  &  $5.3\pm0.5$ &  $3.7\pm0.2$ & $11.0\pm1.5$ & $8.1\pm0.4$& $14\pm 3$& $10.0\pm0.8$ & $13\pm3$& $11.0\pm1.0$ \\ 
    \hline 
    \end{tabular}
    \caption{\cii\ line fluxes measured from a Gaussian fit the aperture--integrated spectra in this work and previous studies. For aperture radius and study, we give the best--fit velocity width $\sigma_v$, the peak line flux density $S_{peak}$ and the integrated flux $I_\nu$. For the \citet[][C15]{Cicone2015} values, only the narrow component results (\cii\ emission integrated between $(-200,200) \, \kms$) are reported. The \citet{Walter2009} best--fit values are: $\sigma_v = 122 \pm 12  \,\kms$,  $S_{\rm{peak}}=12.7 \pm 1.1  \,\rm{mJy}$, $I_\nu =  3.9 \pm 0.3 \,\rm{Jy}\, \kms$. Similarly, the \citet[][]{Maiolino2005} IRAM 30m measurement gives: $\sigma_v=149 \pm 21 \, \kms$, $S_{\rm{peak}}=11.8\, \rm{mJy}$, $I_\nu = 4.1\pm 0.5 \,\rm{Jy}\, \kms$.}
    \label{tab:CII_fluxes}
\end{table*}

\section{Continuum subtraction and masking of the \cii\ line}
\label{app:continuum}
\begin{figure}
    \centering
    \includegraphics[width=0.44\textwidth]{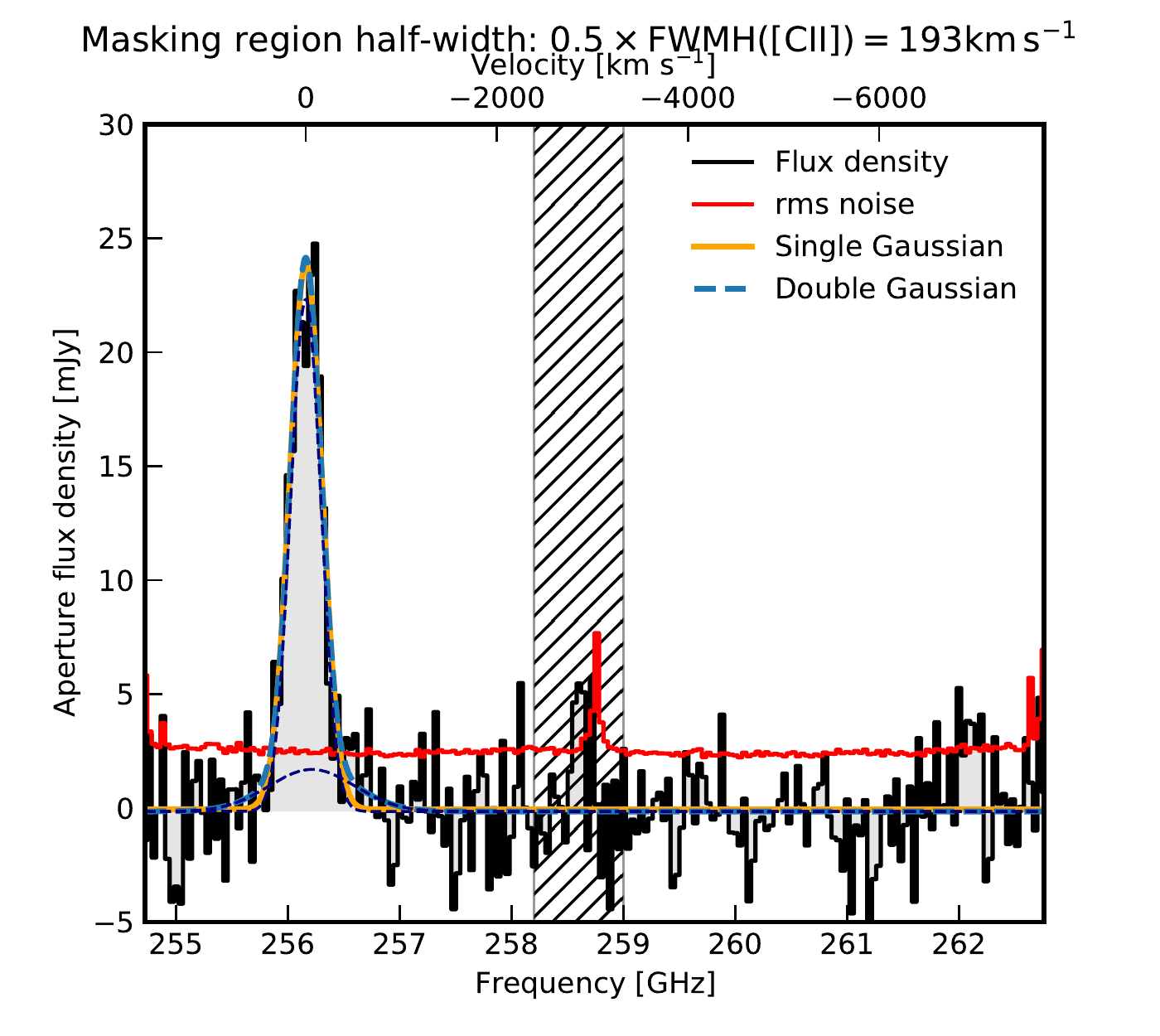}
    \includegraphics[width=0.44\textwidth]{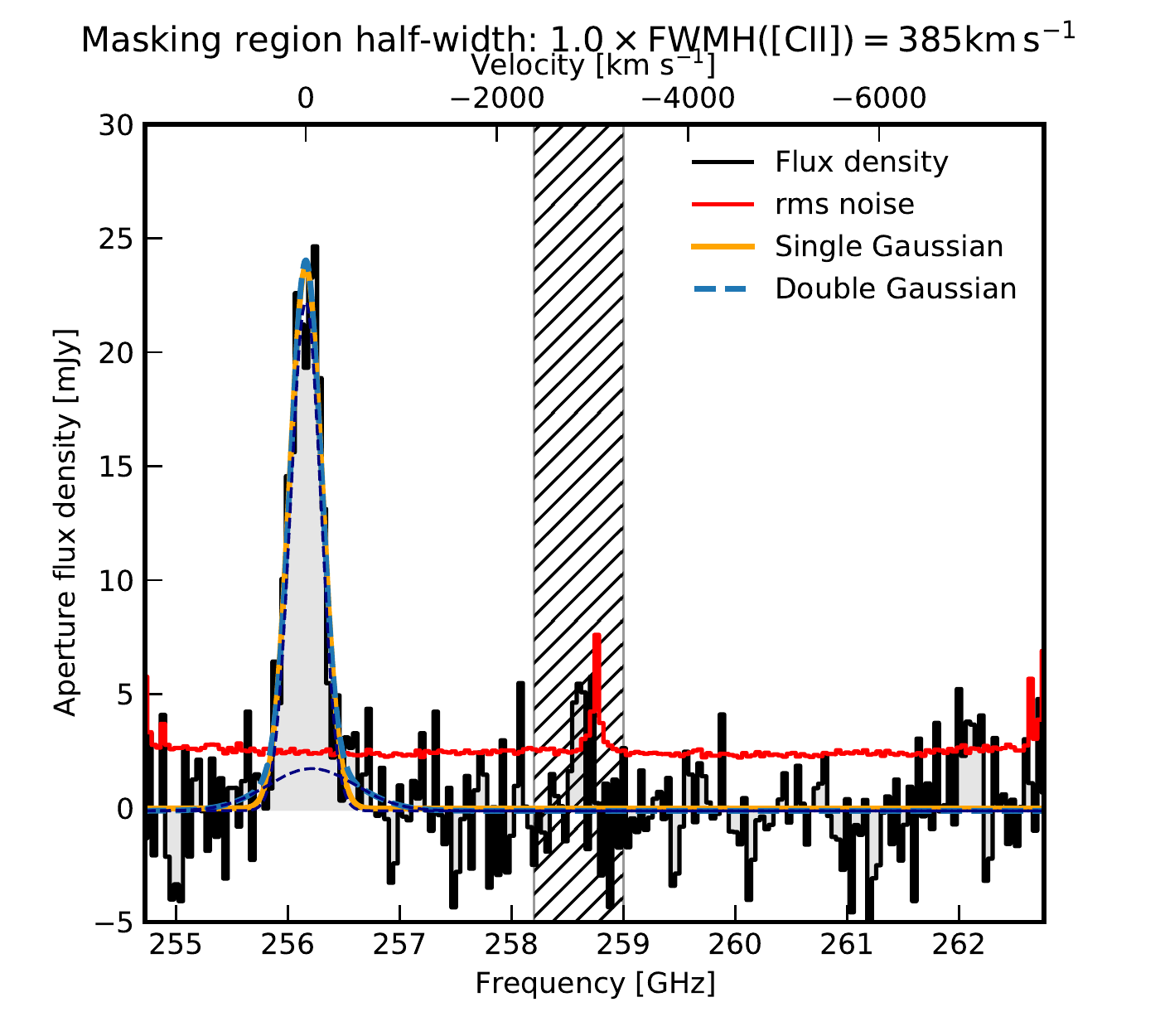}
    \includegraphics[width=0.44\textwidth]{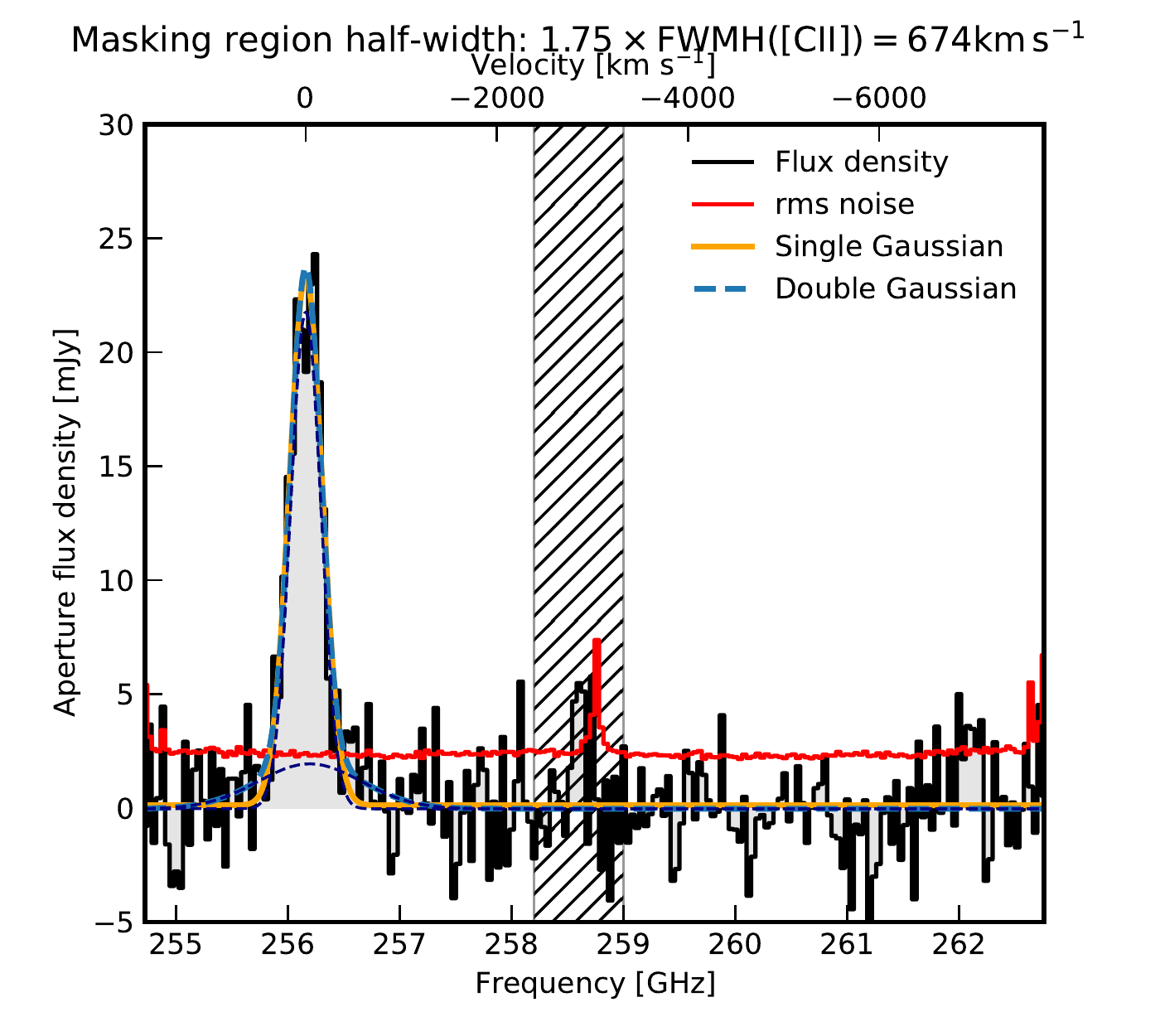}    
    \includegraphics[width=0.44\textwidth]{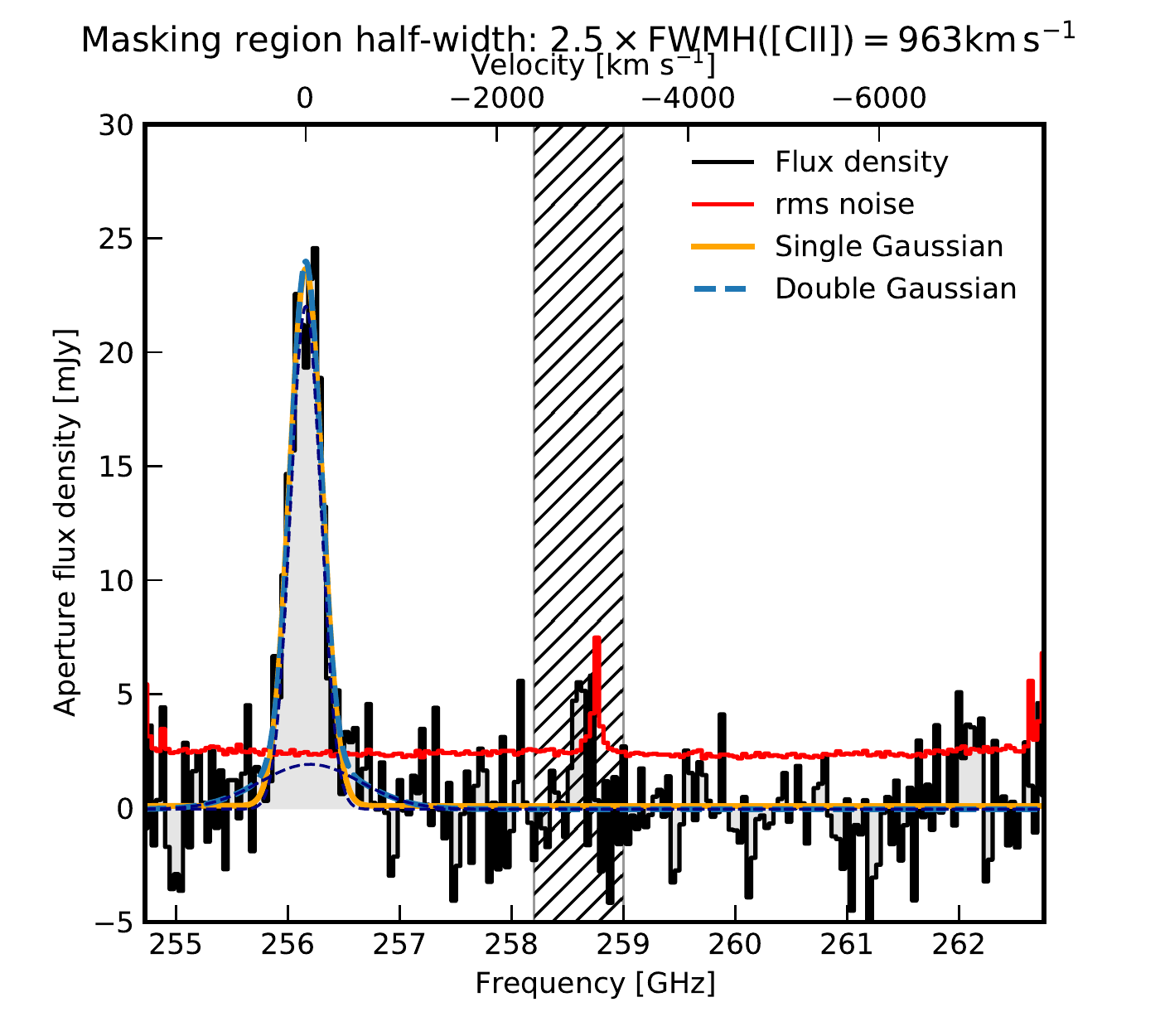}
    \includegraphics[width=0.44\textwidth]{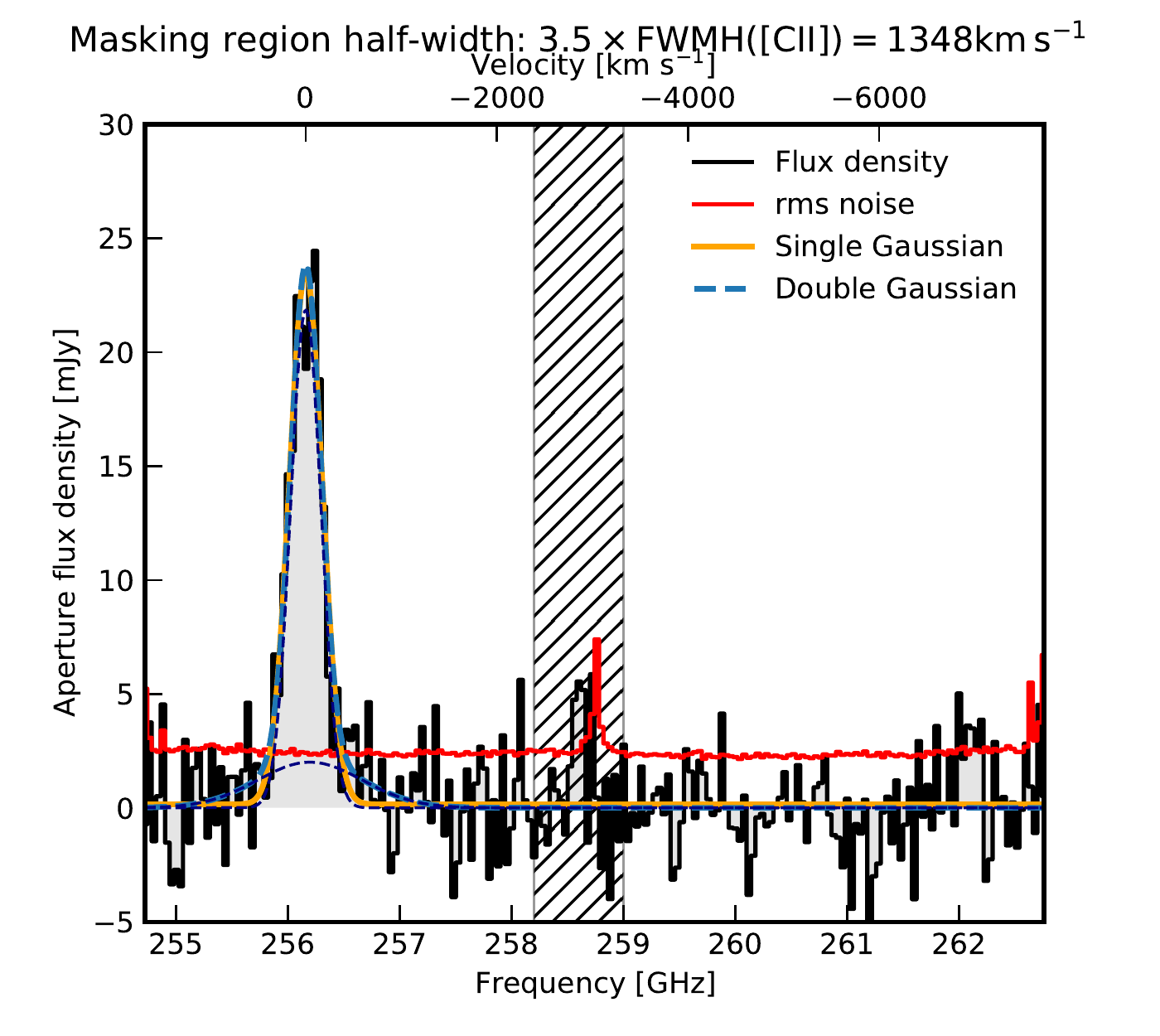}
    \includegraphics[width=0.44\textwidth]{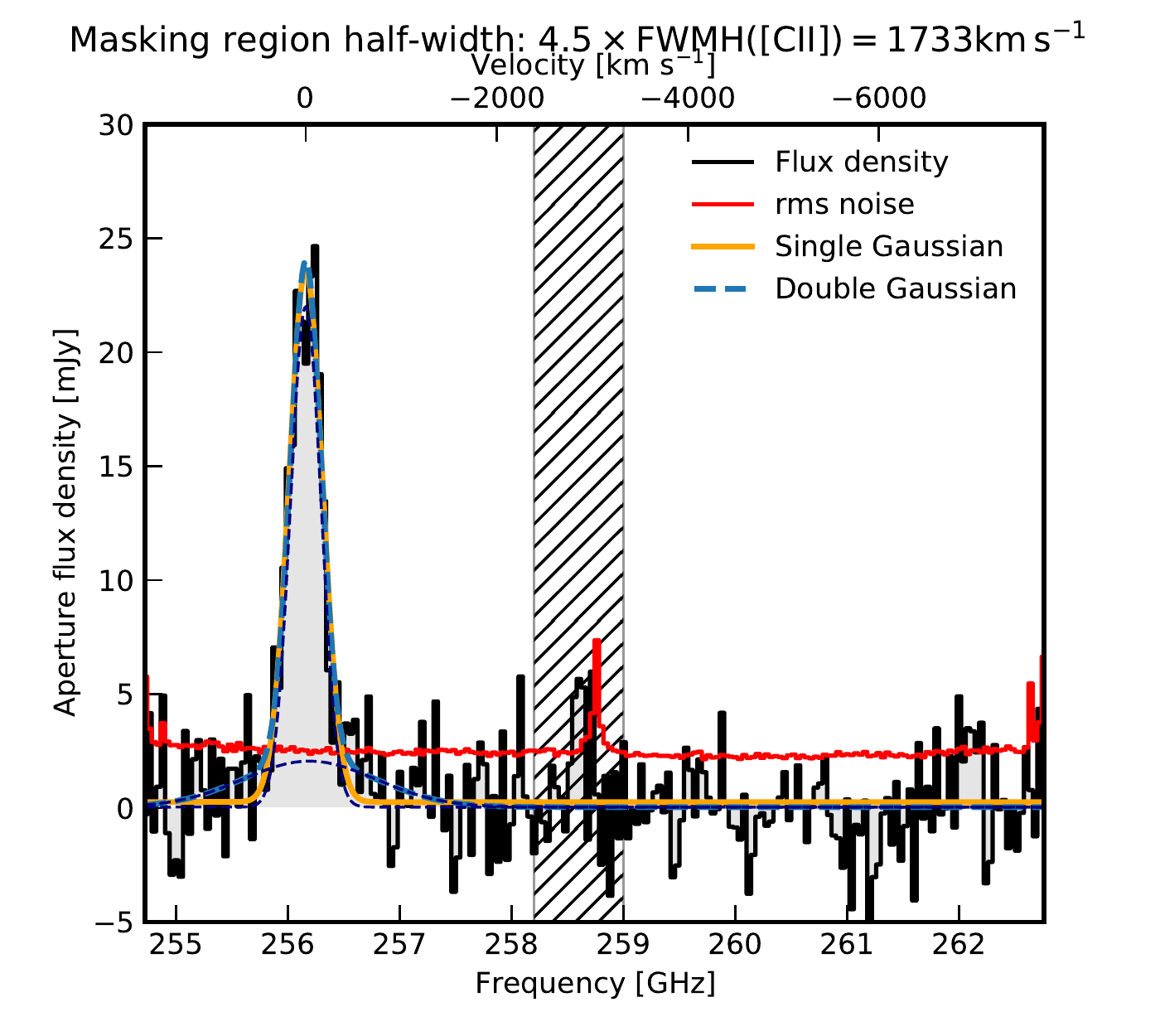}
    \caption{Spectra of the \cii\ line for different continuum subtractions. The line half-width masking region is indicated in the title of each panel. The colors and symbols are the same as in Figure \ref{fig:cii_fit}.}
    \label{fig:spectra_masking}
\end{figure}

\begin{figure}
    \centering
    \includegraphics[width=0.5\textwidth]{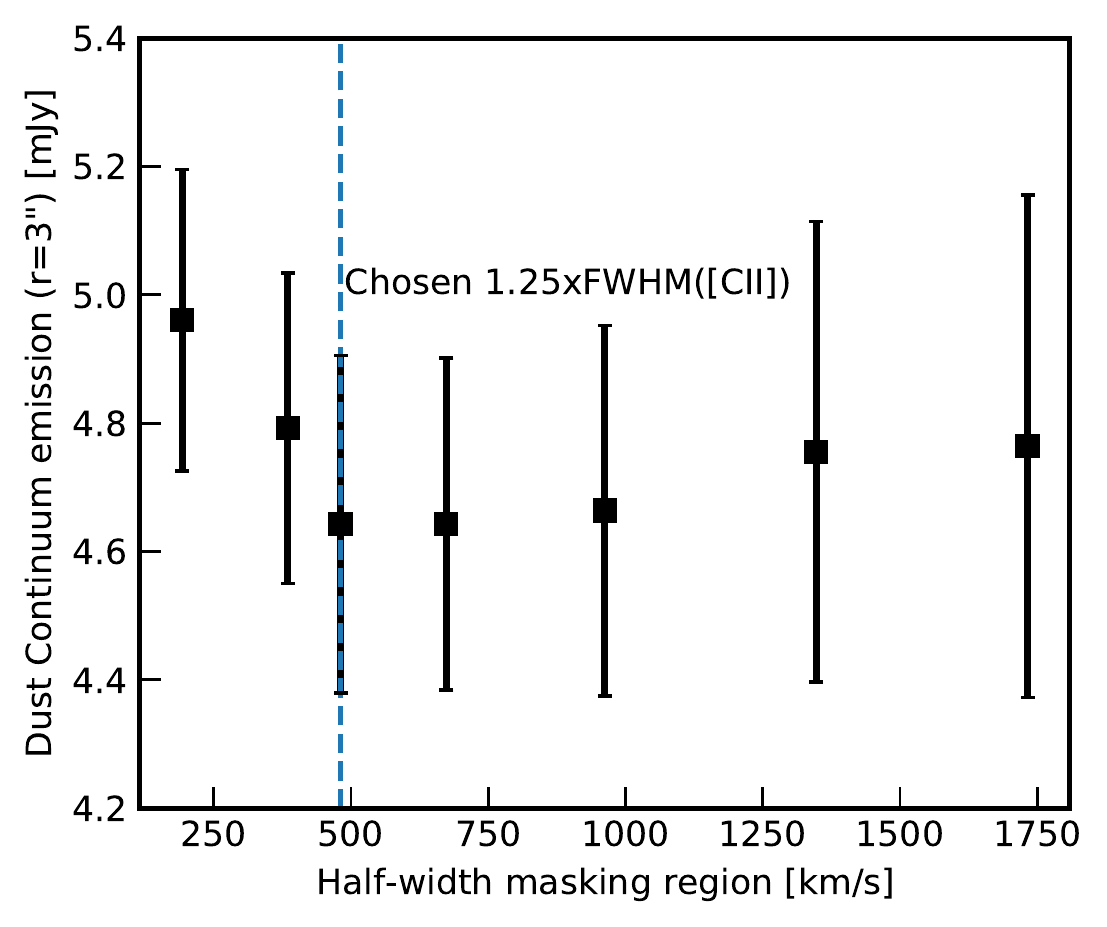}
    \caption{Aperture-integrated continuum flux density at $259$ GHz as a function of the line masking half-width. The measurement has converged at the adopted masking width of $1.25\times \rm{FWHM([CII])}$.}
    \label{fig:test_continuum_masking}
\end{figure}

In this appendix, we provide details on the continuum subtraction procedures, focussing on the \cii\ spectral setup. Figure \ref{fig:spectra_masking} shows the aperture-integrated \cii\ spectrum for various half-width masking regions ranging from $4.5\times \rm{FWHM([CII])]} = 1733 \, \kms$ to $0.5\times \rm{FWHM([CII])]} = 193 \, \kms$. In all cases we fit a single and double Gaussian models to the \cii\ emission (as in Section \ref{sec:cii_outflow}) and find no evidence to reject the single Gaussian model in favor of a double Gaussian emission profile. Finally, Figure \ref{fig:test_continuum_masking} shows the dust continuum measurement as a function of the masking half-width. The continuum is predictably higher for small masking regions, but has converged at the masking half-width adopted in this paper ($1.25 \times \rm{FWHM([CII])}$).

\section{\textbf{Previous PdBI data and merge with the new NOEMA data}}
\label{app:pdbi_data}

\begin{figure*}
    \centering
    \includegraphics[width=0.49\textwidth]{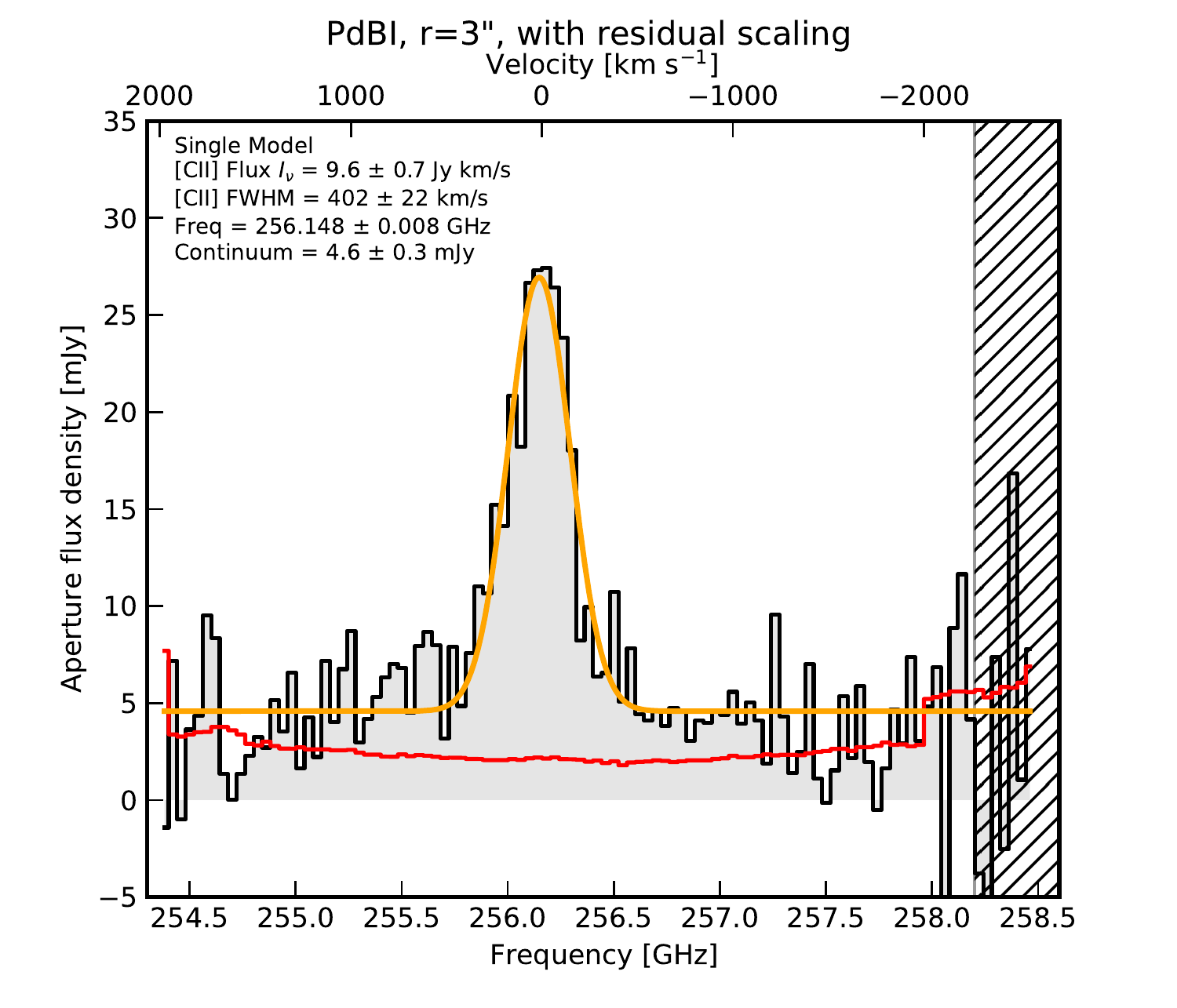}
    \includegraphics[width=0.49\textwidth]{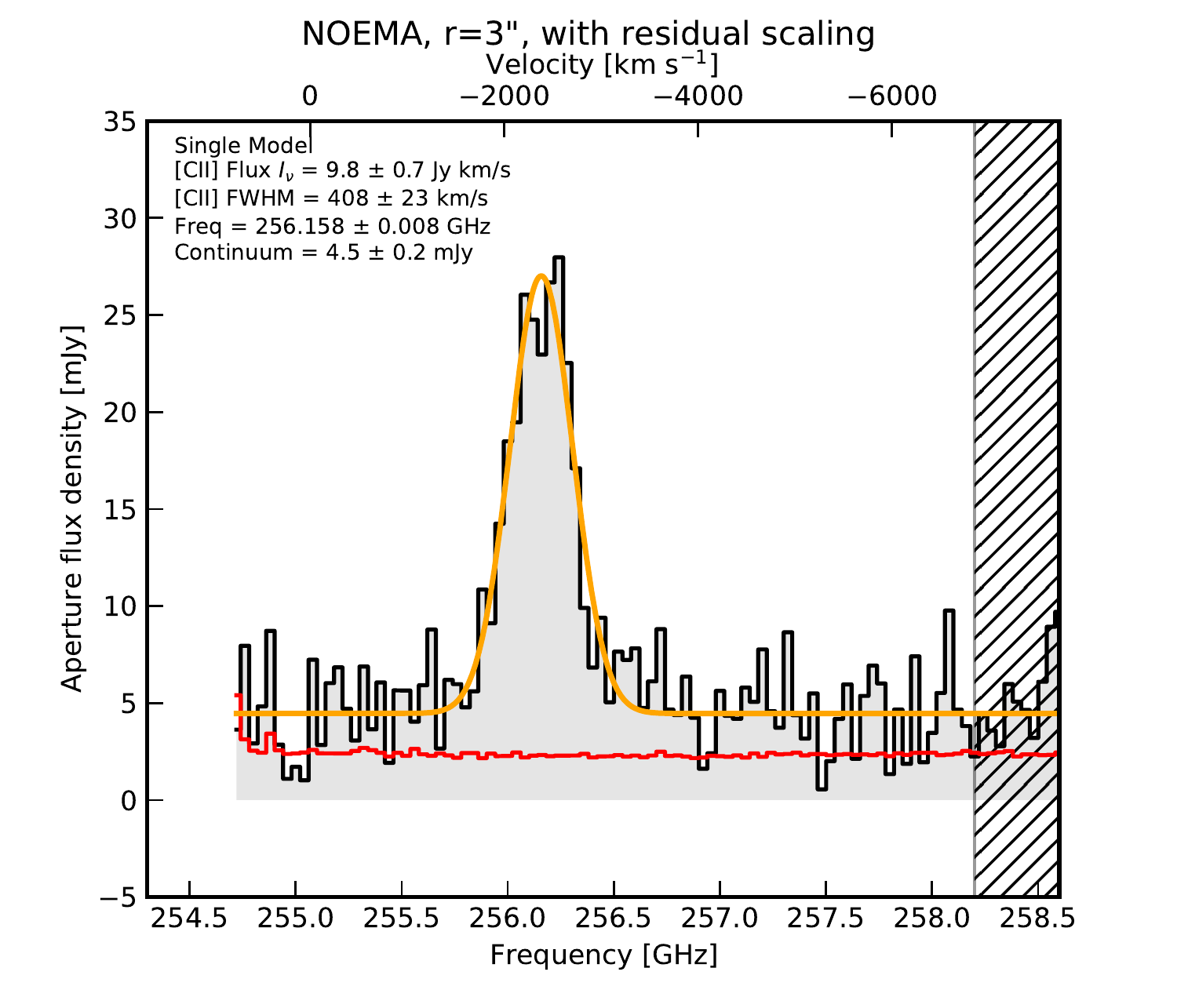}
    \caption{Total spectrum (black) of the \cii\ line  extracted in $r=3"$  apertures  using  residual  scaling  in the previous PdBI data (left) and the new NOEMA data (right). Single Gaussian (+continuum) fits to both dataset are shown in orange and the parameters are given in the upper left corner. The best-fit continuum (constant) and single-Gaussian model parameters are consistent between the two datasets. We note that the increase in noise (shown as red line) in the PdBI data at low ($<254.7$ GHz) and high ($>258$ GHz) frequencies is due to the way the data were frequency-stitched at the time.}
    \label{fig:pdbi_noema_total_spectra}
\end{figure*}

In this appendix, we analyse the previous PdBI observations of J1148+5251 published in \citet[][]{Cicone2015}. We do not recalibrate their data, and follow our analysis method presented in Sections \ref{sec:obs} and \ref{sec:results}. Based on the previous PdBI data, we show the total spectrum extracted in a $r=3"$ aperture applying residual scaling corrections  \citep[e.g.,][]{Jorsater1995,Walter1999,Walter2008, Novak2019} in Figure \ref{fig:pdbi_noema_total_spectra} (where we also show the new NOEMA spectrum for comparison). Clearly the two datasets are compatible, and a single--Gaussian+continuum fit gives similar continuum levels and \cii\ peak fluxes within the $\sim 10-20\%$ amplitude calibration errors. We note that the noise in the previous PdBI spectrum is not uniform, as different datasets with
different frequency coverages and antenna configuration were stiched together \citep[][]{Cicone2015}. This leads to increased noise in the ranges
$\nu\lesssim 254.7\ \rm{GHz}$ and $\nu\gtrsim 258.0\ \rm{GHz}$. The best-fit continuum to the total spectrum in the $r=3"$ ($4.6\pm0.3\ \rm{mJy}$) is perfectly consistent with the new measurement ($4.5\pm0.2\ \rm{mJy}$) and the best-fit dust SED (Section \ref{sec:dust_cont}), and in tension with the $3.3\ \rm{mJy}$ continuum value (at $256\ \rm{GHz}$) published in \citet[][]{Cicone2015} derived from a best-fit model in the UV plane to line-free channels. We therefore infer that this tension is probably due to an unfortunate choice of the continuum channels or an unstable UV-plane fitting routine in the old pipeline.

\begin{figure}
    \centering
    \includegraphics[width=0.49\textwidth]{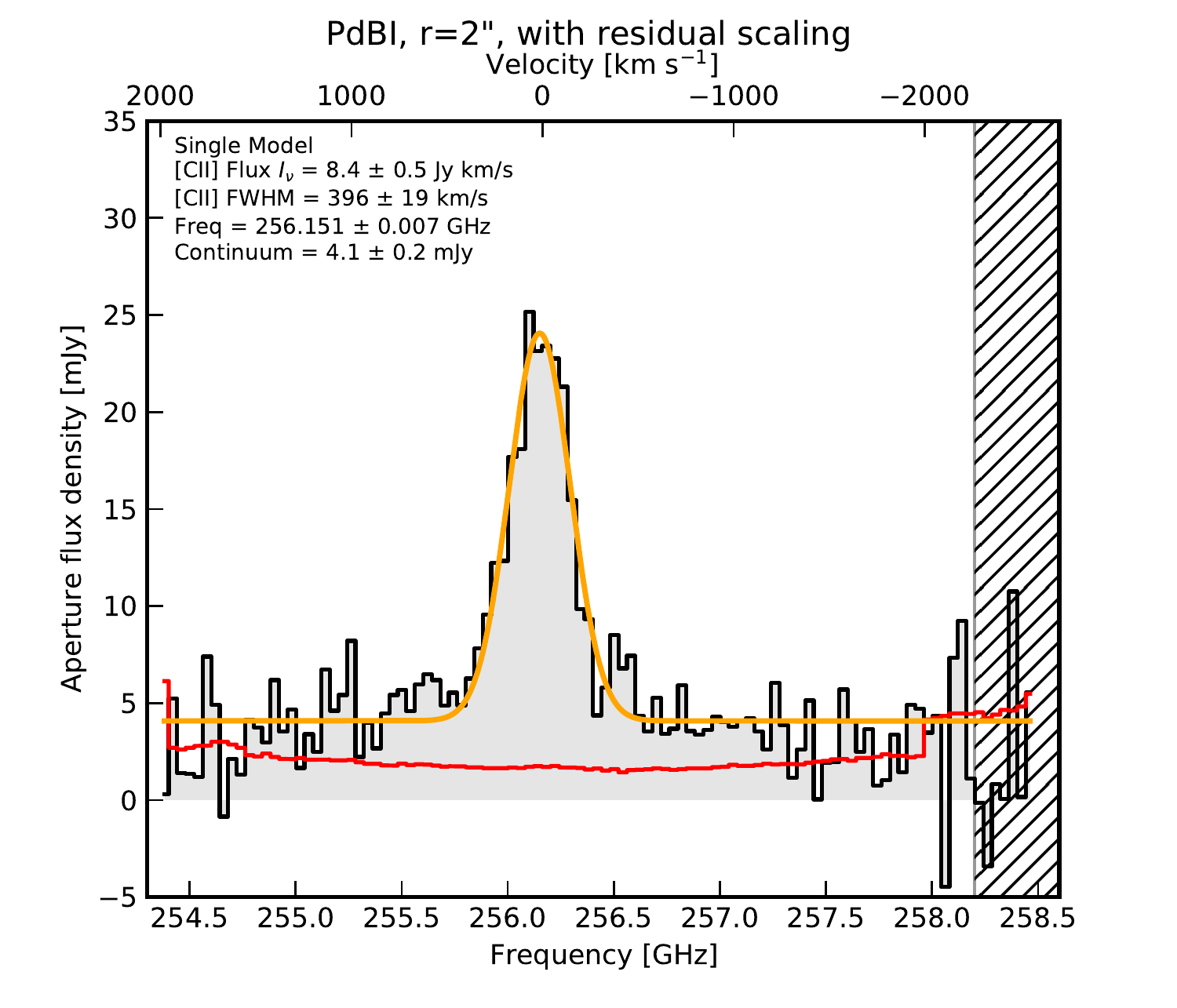}
    \includegraphics[width=0.49\textwidth]{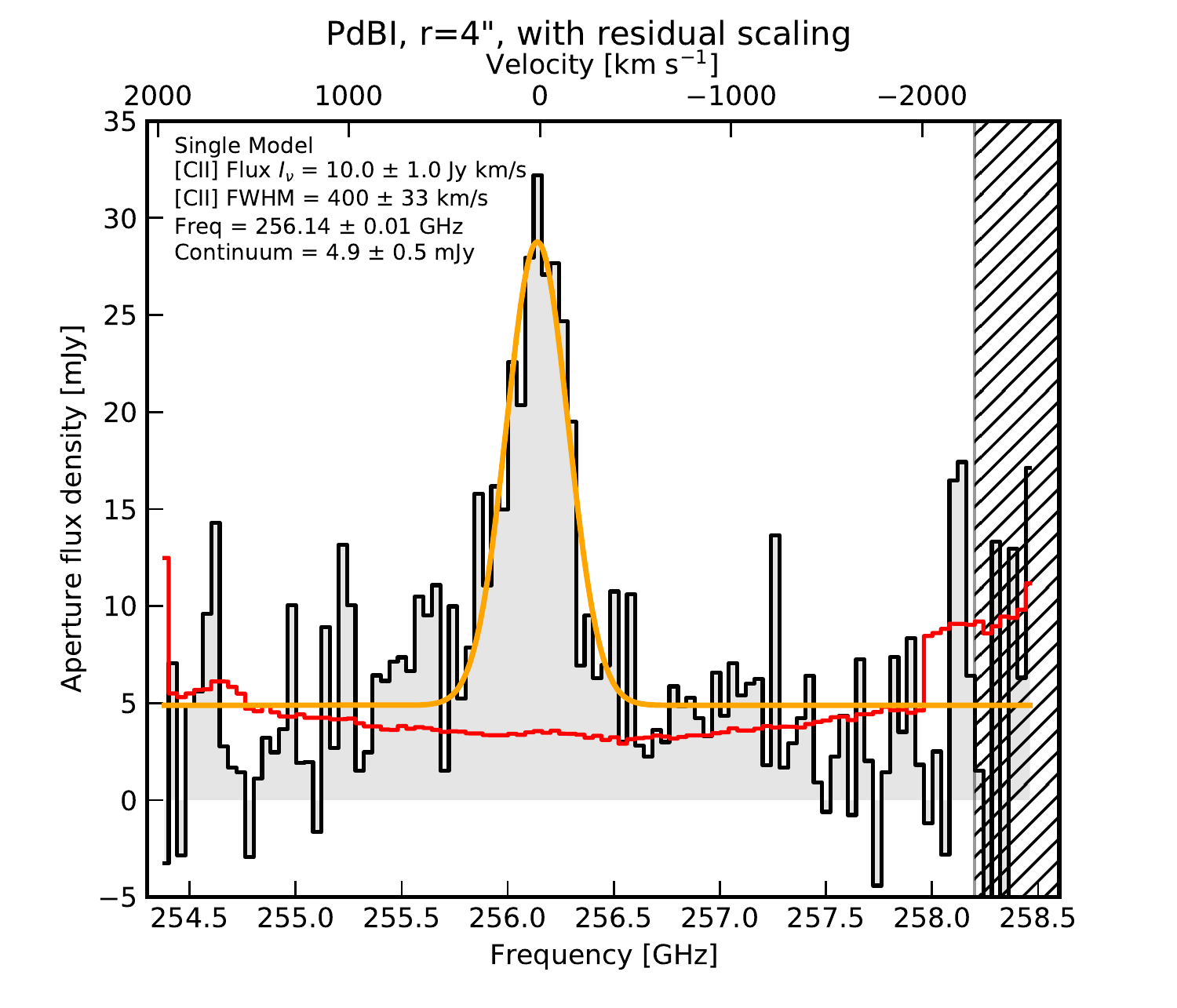}
    \includegraphics[width=0.49\textwidth]{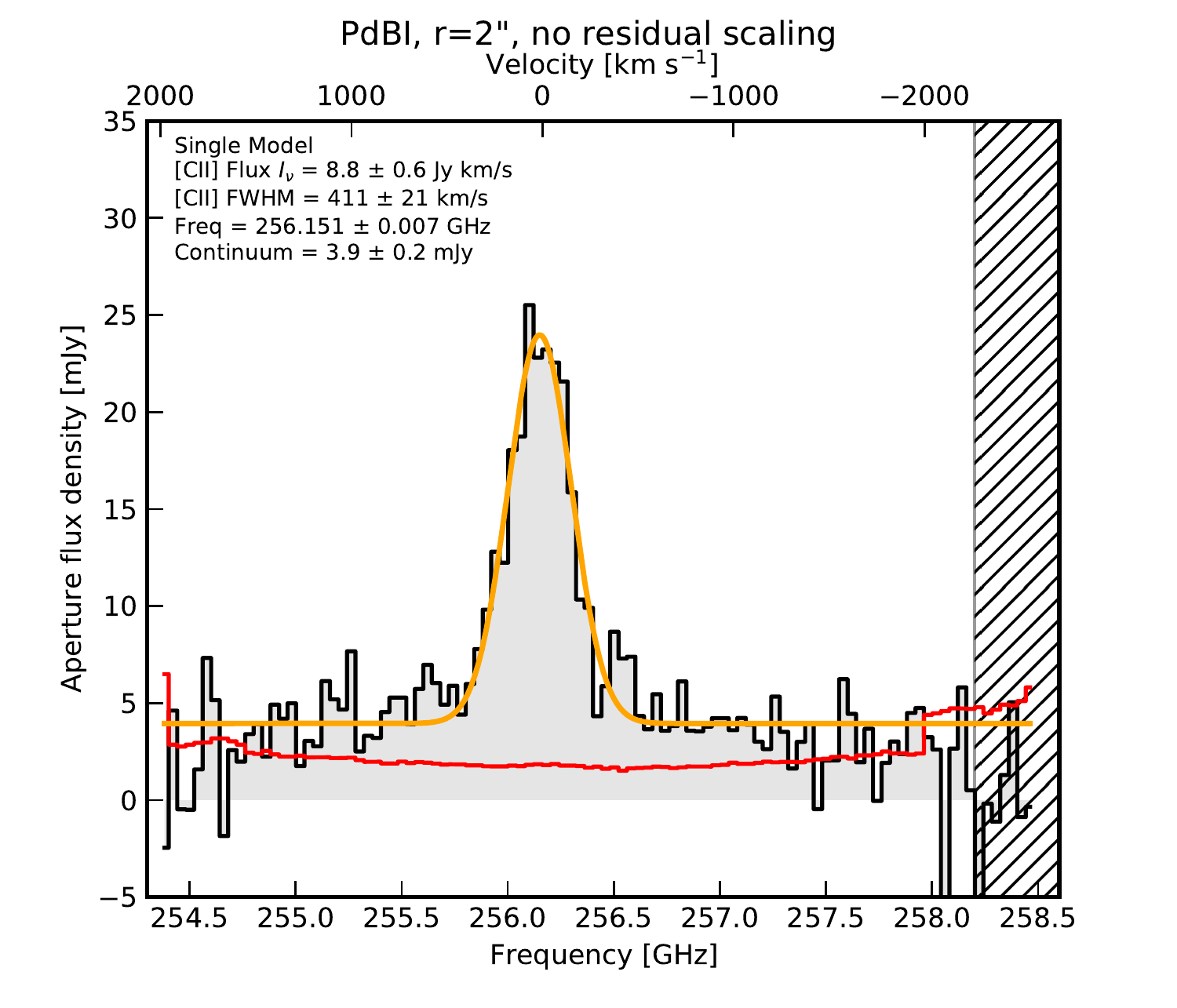}
    \includegraphics[width=0.49\textwidth]{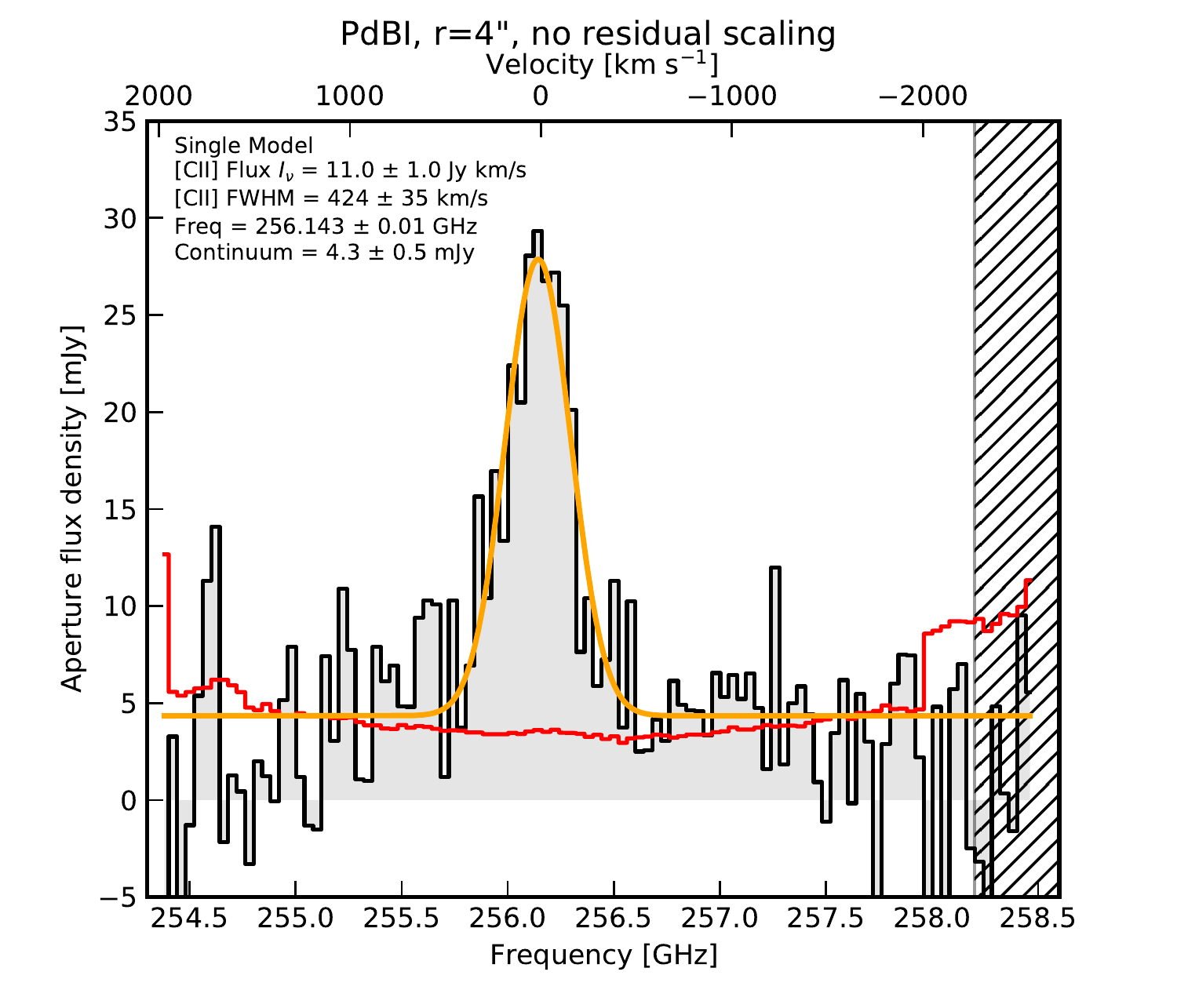}
    \caption{Total spectra of the \cii\ line with (first row) and without (second row) residual scaling
    extracted in $r=2"$ and $r=4"$  apertures (first and second column respectively) in the PdBI data. Single Gaussian (+continuum) fits to both dataset are shown in orange. In the absence of residual scaling, the continuum flux drops away from the \cii line in the larger apertures (see text for details). Even though no continuum was subtracted in the UV-plane, the continuum flux approaches $\sim 0\ \rm{mJy}$ at the band edges in the $r=4"$ aperture (without residual scaling), creating the illusion of a broad emission line.}
    \label{fig:pdbi_residual_scaling}
\end{figure}

\begin{figure}
    \centering
    \includegraphics[width=0.51\textwidth]{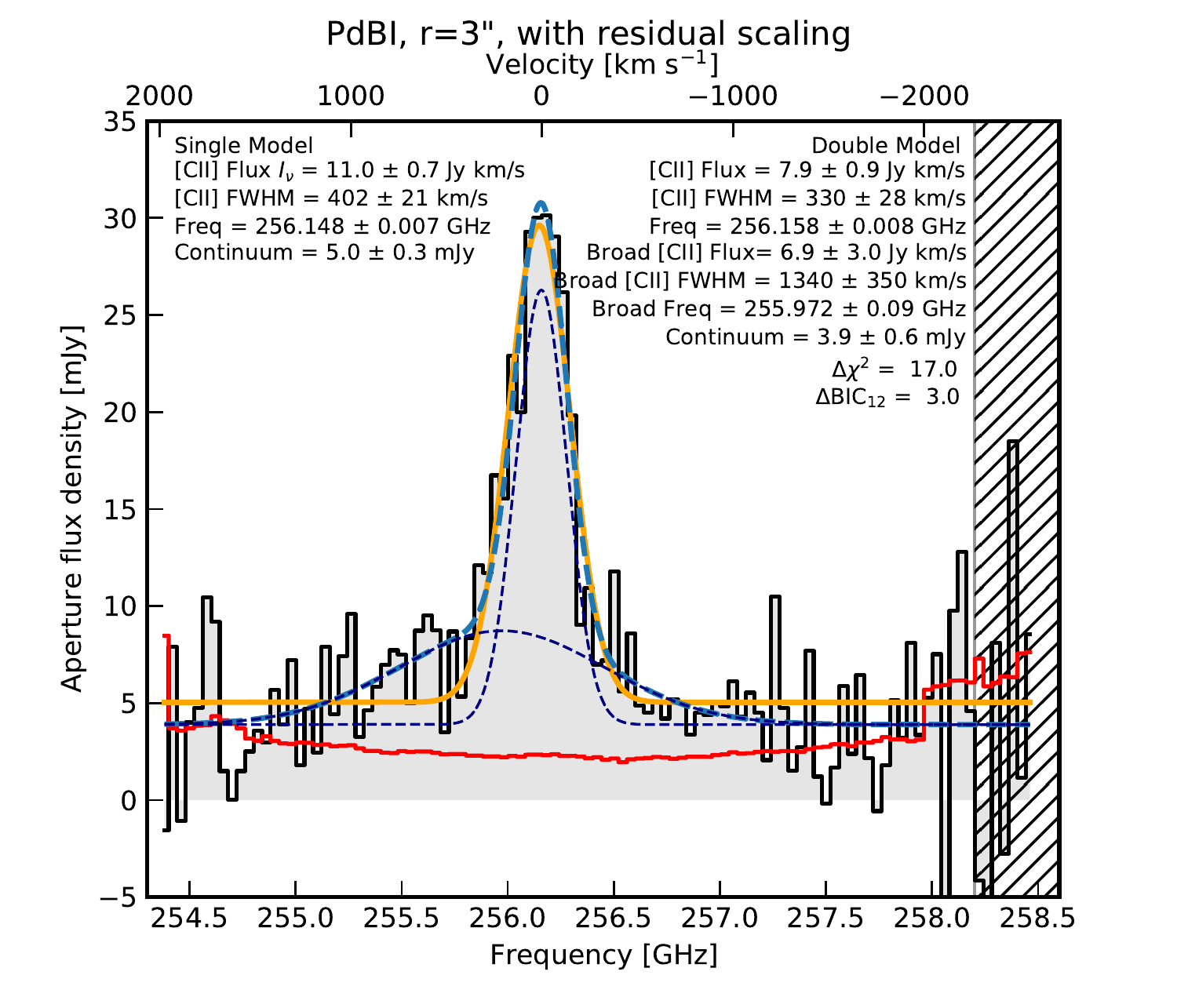}
    \includegraphics[width=0.48\textwidth]{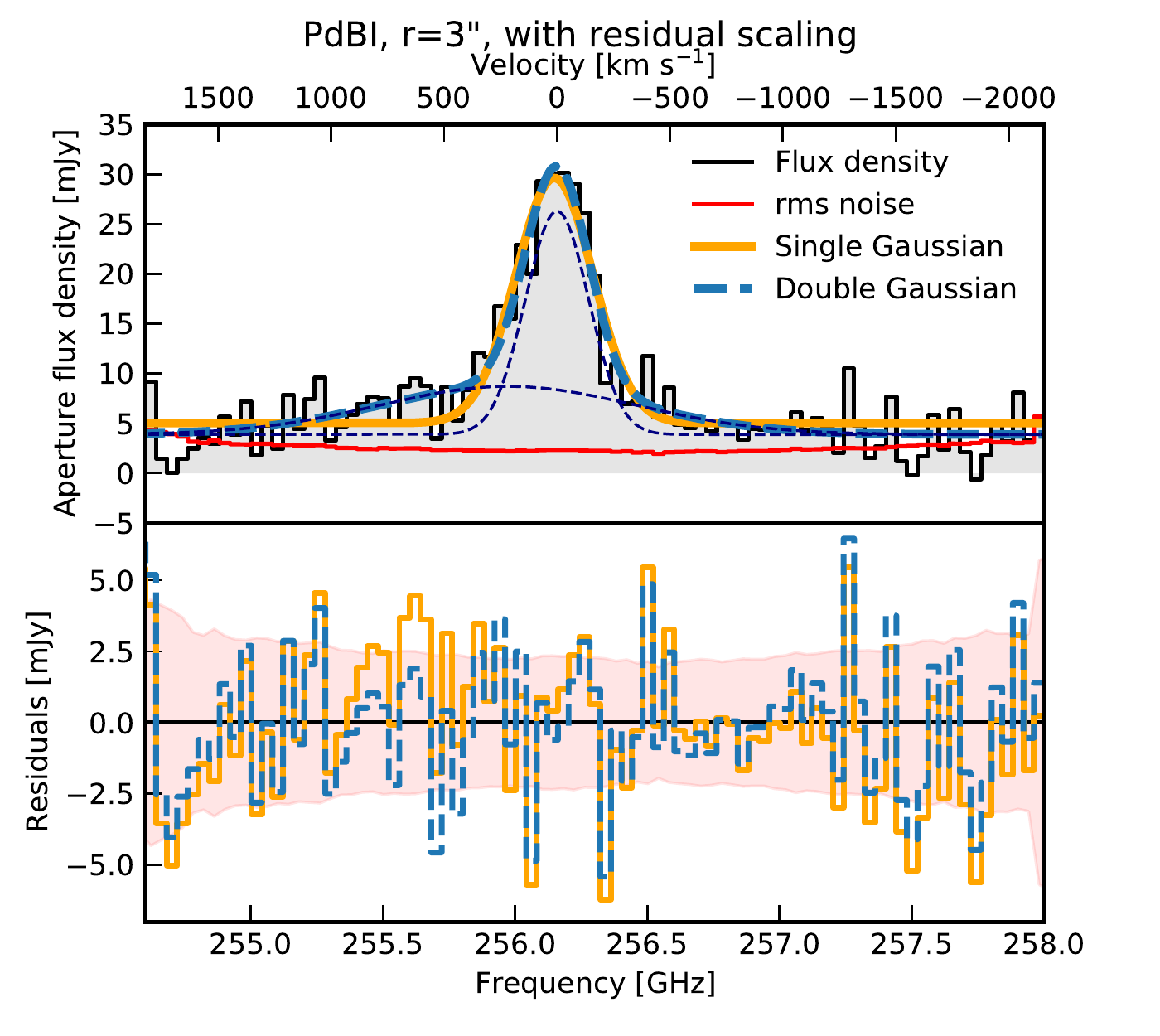}
    \caption{Total spectrum of the \cii\ line (black) presented as in Figure \ref{fig:cii_fit}, using the previous PdBI data, following the methodology used in this paper. Both single-Gaussian and double-Gaussian fits are consistent with that performed on the NOEMA data only (see Fig. \ref{fig:cii_fit}). With a $\Delta\rm{BIC}_{12}=3.0<10$, a broad \cii\ component is only tentatively detected.}
    \label{fig:pdbi_total_cii_spectrum}
\end{figure}

We then show in Figure \ref{fig:pdbi_residual_scaling} different spectra extracted from the previous PdBI dataset with and without residual scaling. It can be clearly seen that the continuum flux at the edges of the band drops significantly in larger apertures when residual scaling is not used. This subtle effect is due to the fact that the continuum is only detected at the $\lesssim 2\sigma$ level in most line-free channels (e.g. at the edges of the band). When cleaning down to a typical threshold $2\sigma$, there is in fact no or little flux above that threshold and therefore no or few clean components are added to the final “clean" map (which is the sum of the clean Gaussian components and the residual map which does not have units of “clean beam" but “dirty beam") for channels that only contain continuum emission. As a consequence, these channels are dominated by the residual map which imprints the dirty beam pattern. With large apertures, e.g. $r>3"$ in this case, one starts to integrate over negative sidelobes of the synthesized beam, decreasing the integrated flux significantly. This effect does not play a dominant role where the \cii\ line is present, as in those channels the flux is dominated by the actual “clean" components. This can then lead, when combined with an under-subtracted continuum, to the illusion of a strong broad component in the final spectrum. However, with the residual scaling approach taken into account, this broad component is less significant (although still tentatively detected, see Fig. \ref{fig:pdbi_total_cii_spectrum}) and the results become compatible with the new NOEMA data.

\begin{figure}
    \centering
    \includegraphics[width=0.5\textwidth]{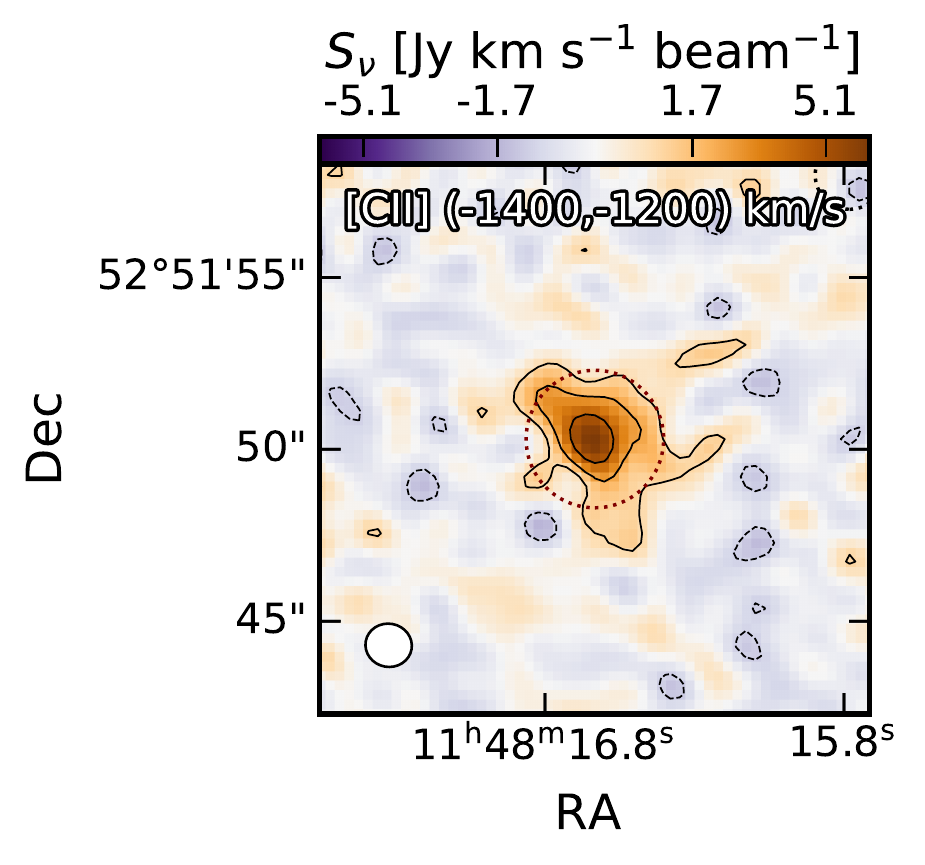}
    \caption{\cii line map integrated over $(-1400, +1200)\ \kms\ $ using the previous PdBI data. The (dashed) black contours contours are at the $(-2,2,4,8) \sigma$ level, where $\sigma$ is the rms computed directly from the map with $\sigma$-clipping. The extended \cii\ structure reported in \citet[][]{Cicone2015} is recovered, albeit at a lower significance level (see main text for details).}
    \label{fig:pdbi_cii_structure}
\end{figure}

We now comment on the \cii\ structure reported in \citet[][]{Cicone2015}. We show in Figure \ref{fig:pdbi_cii_structure} the collapsed \cii\ channels in the range of $(-1400,+1200)\ \kms$ based on the previous PdBI data only. We find a similar structure as previously reported, but at a different significance level (in our case not exceeding the $3\sigma$ level in the extended regions). We therefore speculate that the rms might have been underestimated by the earlier GILDAS/MAPPING software release, which would explain the numerous $-6$ and $-3\sigma$ regions in the previously published \cii\ map.

\begin{figure}
    \centering
    \includegraphics[width=0.51\textwidth]{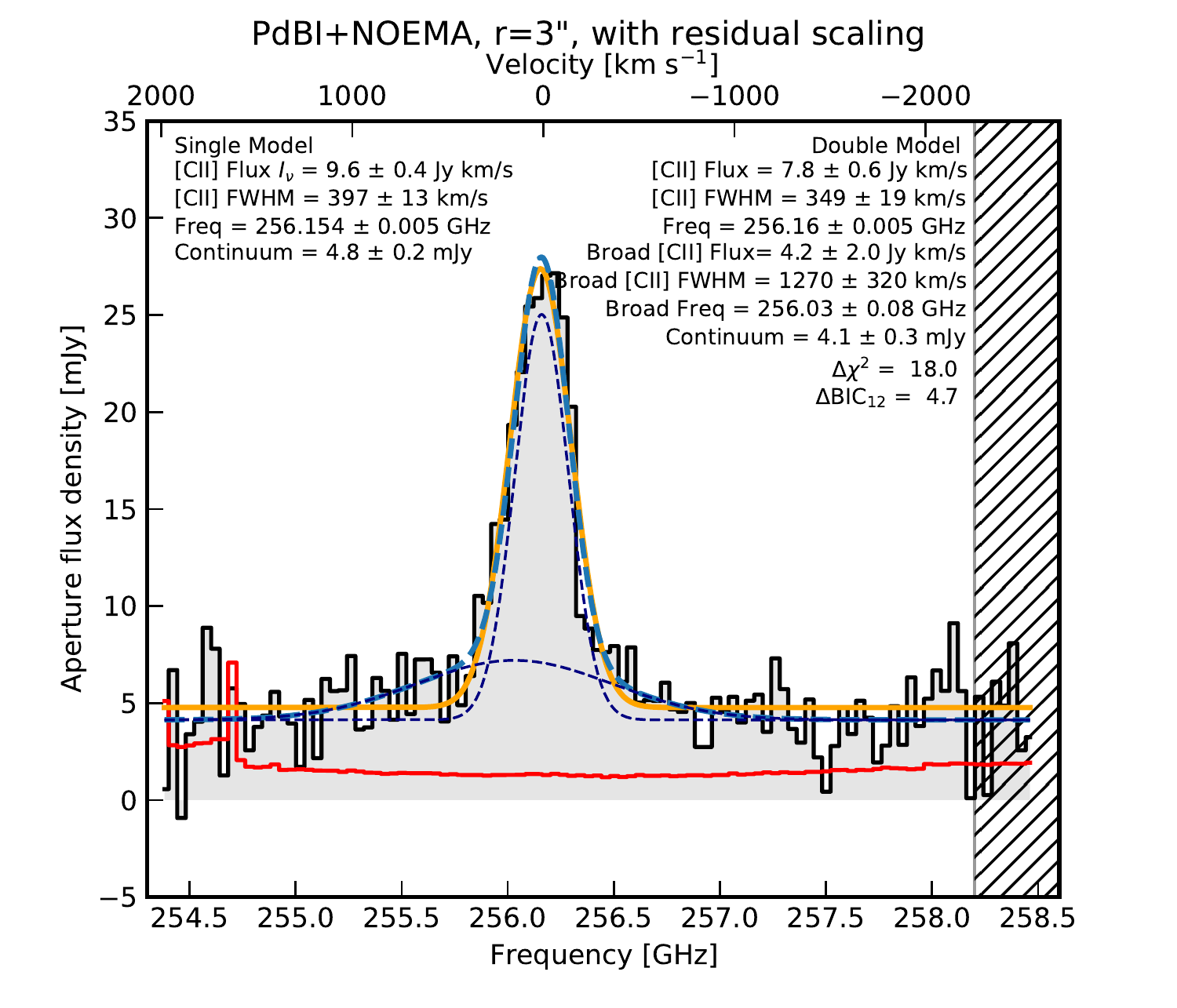}
    \includegraphics[width=0.48\textwidth]{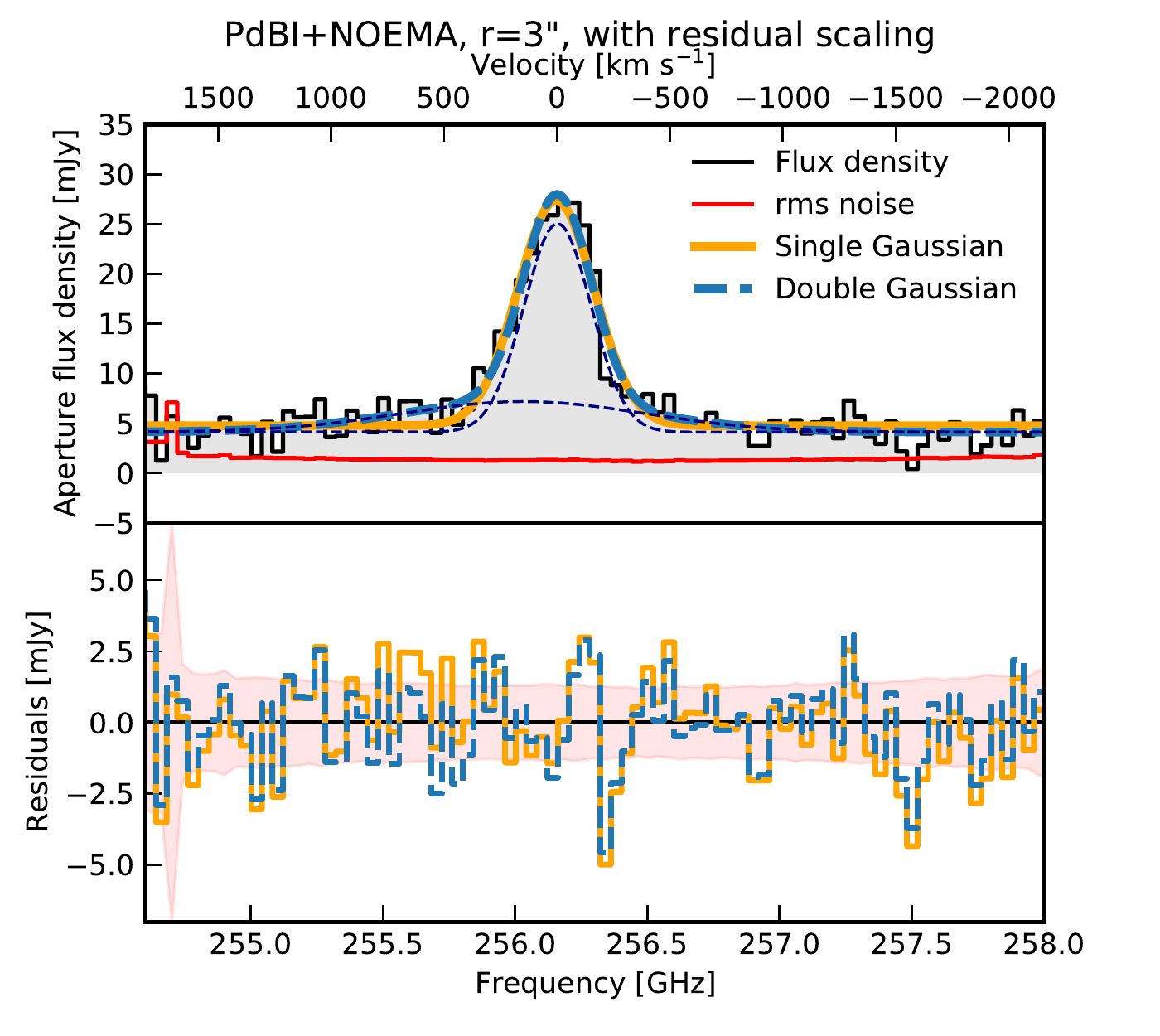}
    \caption{Total spectrum of the \cii\ line (black) presented as in Figure \ref{fig:cii_fit}, but using the merged PdBI and NOEMA dataset. Both single-Gaussian and double-Gaussian fits are consistent with that performed on the NOEMA data only (see Fig. \ref{fig:cii_fit}). With a $\Delta\rm{BIC}_{12}=4.7<10$, a broad \cii\ component is only tentatively detected in the merged PdBI and NOEMA dataset.}
    \label{fig:merged_cii_spectrum}
\end{figure}

We conclude this appendix by presenting the results of our analysis of the merged PdBI and NOEMA \cii\ spectrum. We show in Figure \ref{fig:merged_cii_spectrum} the total and continuum-subtracted spectra of the merged \cii data. Although the statistical analysis of the spectrum extracted from this combined dataset continues to prefer a single component fit, there is evidence for a detection of a small flux excess in the red--shifted \cii\ line wing which might indicate the presence of weak outflow or an unresolved companion as in J0305-3150 \citet[see][]{Venemans2017,Venemans2019}.

\section{Channel and moments maps of the \cii\ emission}
\label{app:velocity}
\begin{figure}
    \centering
    \includegraphics[height= 6cm,trim={0.25cm 0 0 2.1cm},clip]{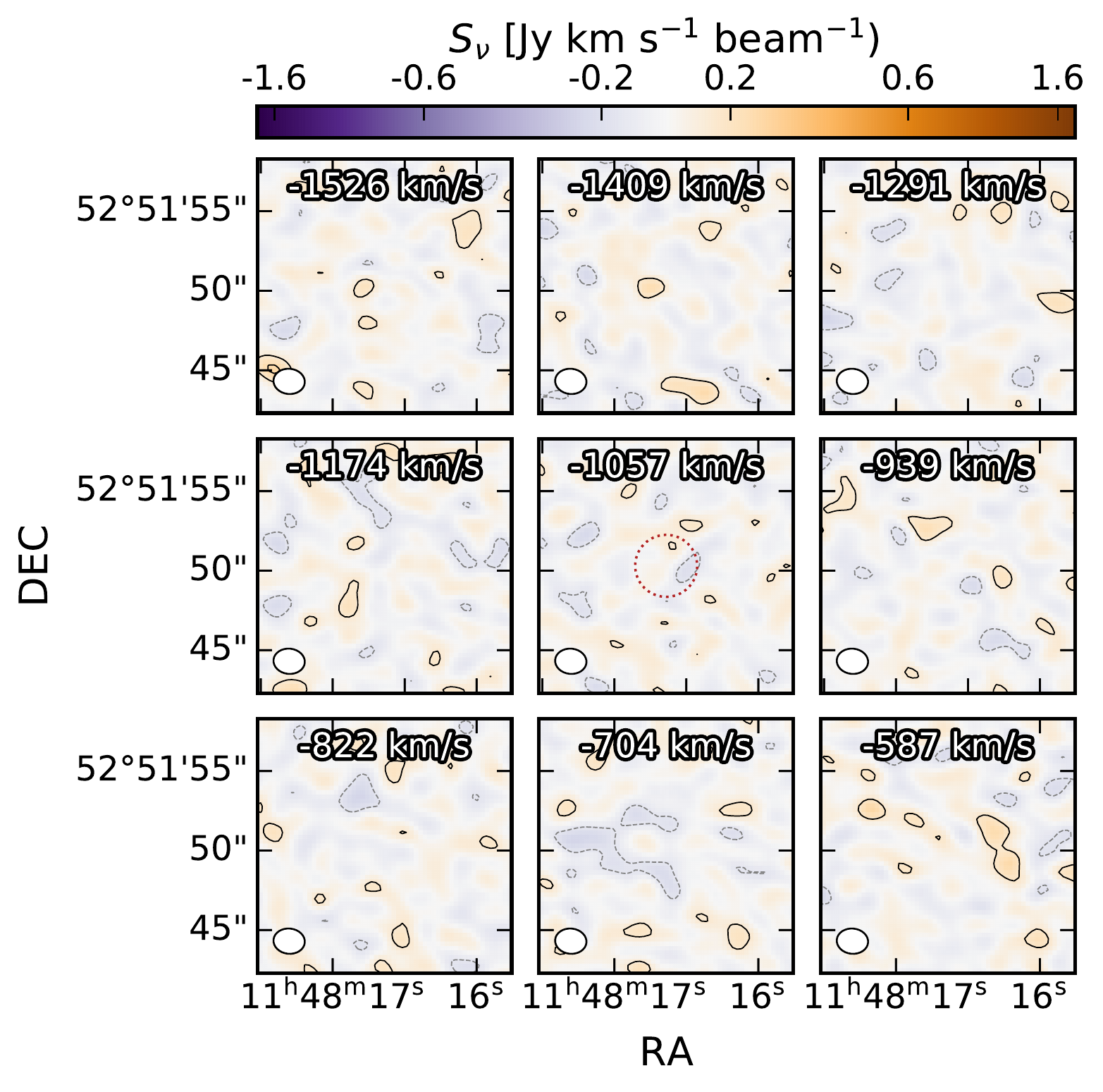} 
    \includegraphics[height= 6.8cm,trim={3.56cm 0 0 0},clip]{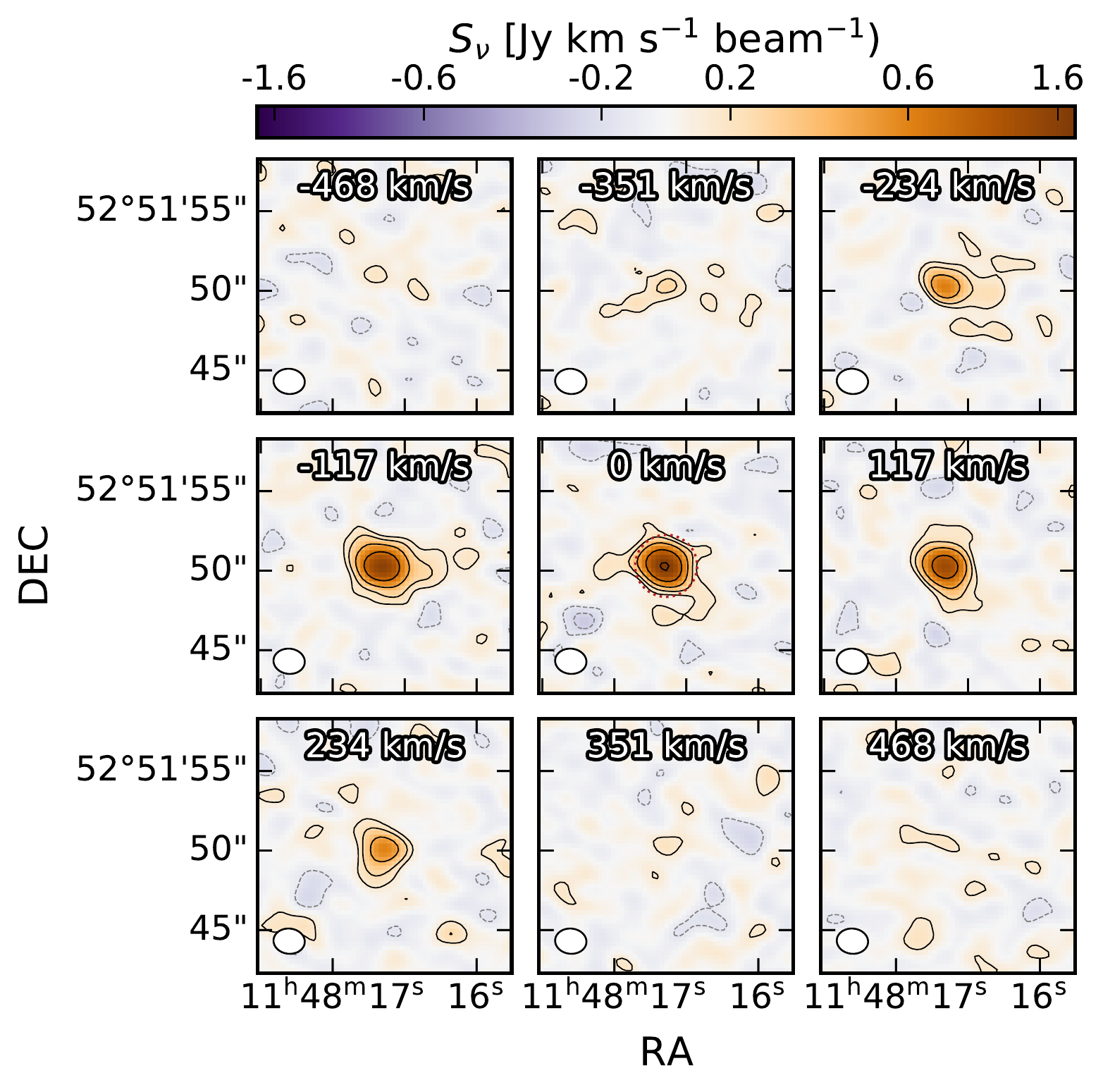} 
    \includegraphics[height= 6cm,trim={3.56cm 0 0 2.1cm},clip]{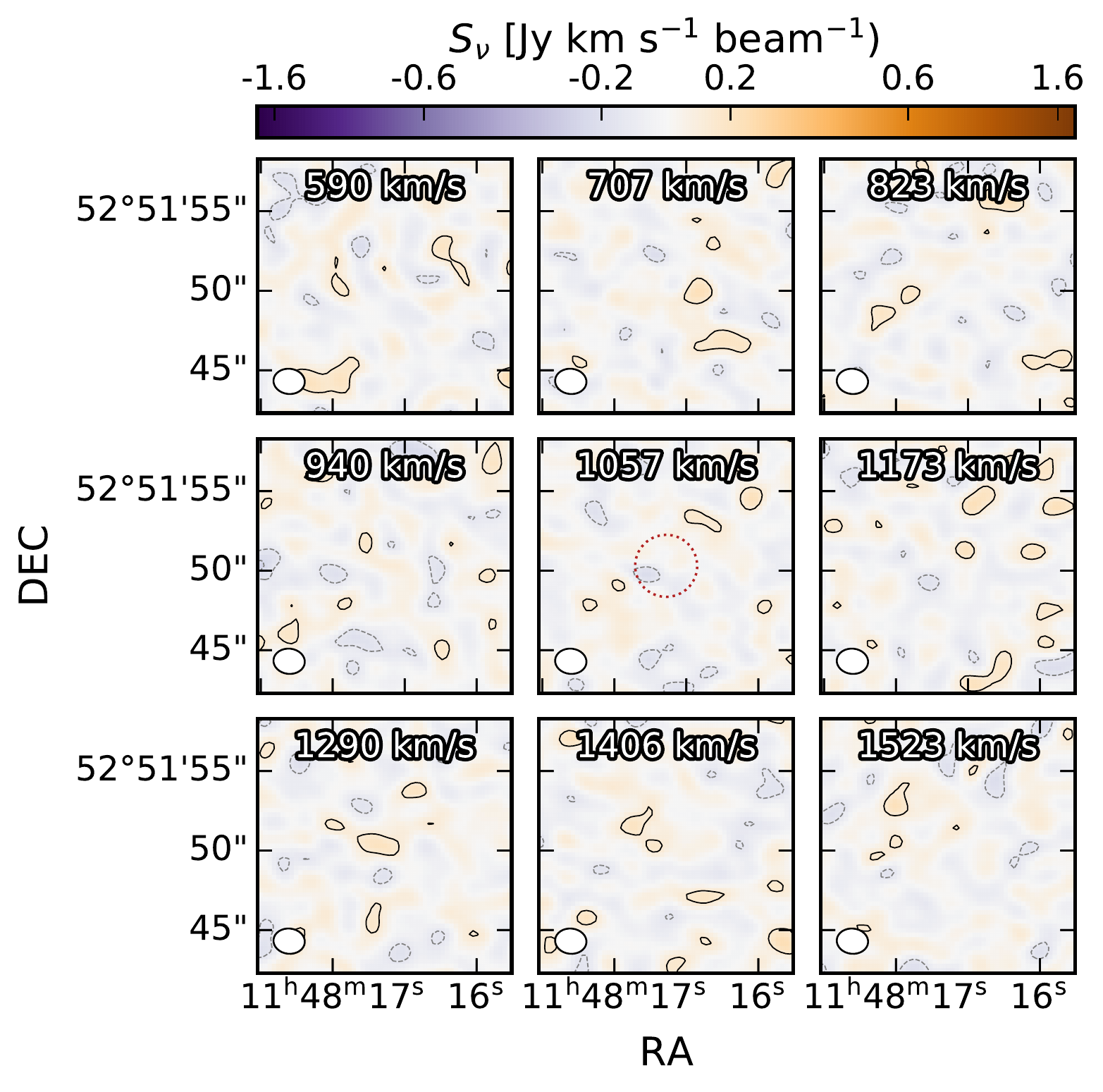}
    \caption{Channel map of the \cii\ line of J1148+5251, in channels of $117\, \kms$. The contours are logarithmic $(-8,-4,-2,2,4,8,16,32)\sigma$ (rms). The colour scaling is log--linear, the threshold being at $3\sigma$ (rms). The colour scaling is log--linear, the threshold being at $3\sigma$ (rms). }
    \label{fig:cii_channels}
\end{figure}

\begin{figure*}
    \centering
    \includegraphics[width=0.8\textwidth]{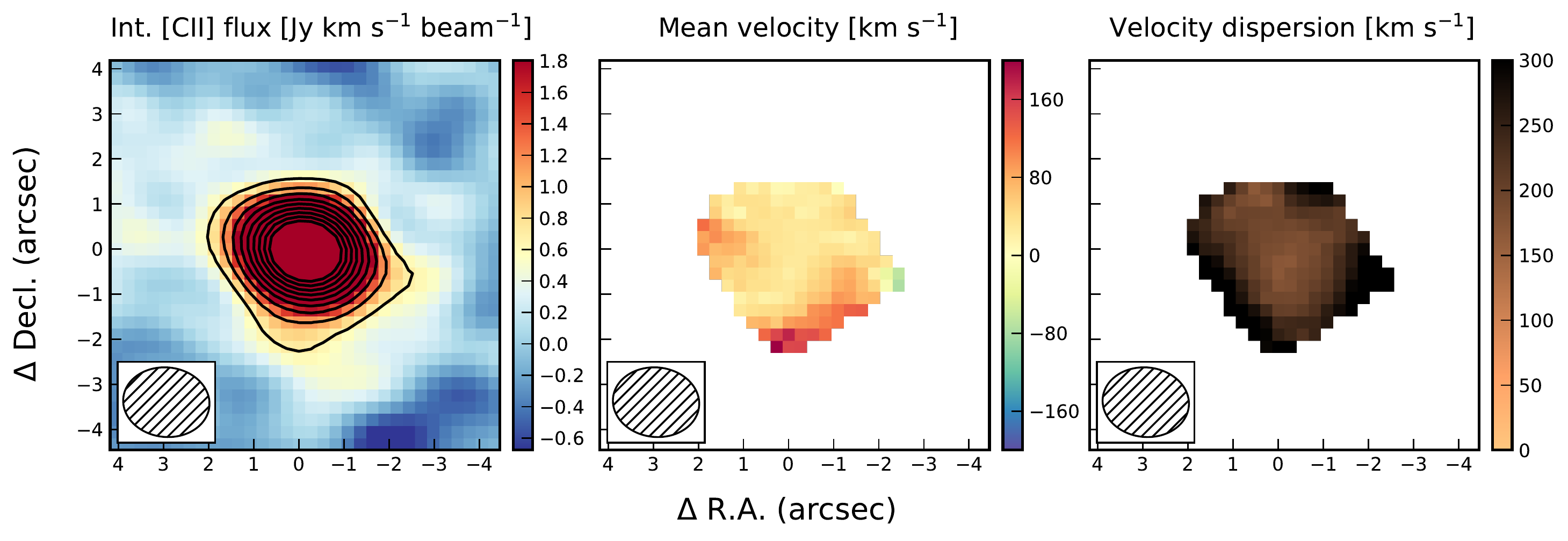}
    \caption{Moment maps of the \cii\ emission. The integrated velocity and velocity dispersion are only shown in  pixels at the $3\sigma$ level. The absence of any velocity structure is expected given the $3\sigma$ detected area covers only $\sim 4$ beams.}
    \label{fig:moments_CII}
\end{figure*}

In this appendix, we present additional visualizations of the \cii\ emission velocity structure in J1148+5251. Firstly, we present channel maps in Figure \ref{fig:cii_channels} with a channel width of $117\, \kms$. No significant emission ($>3\sigma$) is detected at large radii or at velocity offsets $>400\, \kms$ from the peak of the \cii\ line \citep[cf.][]{Cicone2015}. Secondly, we present the integrated flux, mean velocity and velocity dispersion (so--called moment maps) in Figure \ref{fig:moments_CII}. These maps are generated using \emph{Qubefit} \citep[][]{qubefit} using all \cii\ voxels detected at SN$>3$ and  standard parameters. We find no kinematic evidence for a bulge--dominated dispersion or a rotating disk model.

\bibliography{bib}{}
\bibliographystyle{aasjournal}

\end{document}